\documentclass[%
 aip,
 sd,%
 amsmath,amssymb,
 reprint,%
]{revtex4-1}

\usepackage{graphicx}
\usepackage{dcolumn}
\usepackage{bm}

\begin{document}

\preprint{AIP/123-QED}

\title[]{Delay dynamics of neuromorphic optoelectronic nanoscale resonators: Perspectives and applications}

\author{B. Romeira}
  \email{bmromeira@ualg.pt.}
\author{J. M. L. Figueiredo}%
\altaffiliation[Current address: ]{Departamento de F\'{i}sica, Faculdade de Ci\^{e}ncias da Universidade de Lisboa, Campo Grande, 1749-016 Lisboa, Portugal.}
\affiliation{
Centro de Electr\'{o}nica, Optoelectr\'{o}nica e Telecomunica\c{c}\~{o}es (CEOT), Departmento de F\'{i}sica, Universidade do
Algarve, Campus de Gambelas, 8005-139, Faro, Portugal
}%

\author{J. Javaloyes}
\affiliation{%
Departament de F\'{i}sica, Universitat de les Illes Balears, C/ Valldemossa km 7.5, 07122, Palma de Mallorca, Spain
}%


\begin{abstract}
With the recent exponential growth of applications using artificial intelligence (AI), the development of efficient and ultrafast brain-like (neuromorphic) systems is crucial for future information and communication technologies. While the implementation of AI systems using computer algorithms of neural networks is emerging rapidly, scientists are just taking the very first steps in the development of the hardware elements of an artificial brain, specifically neuromorphic microchips. In this review article, we present the current state of neuromorphic photonic circuits based on solid-state optoelectronic oscillators formed by nanoscale double barrier quantum well resonant tunneling diodes. We address, both experimentally and theoretically, the key dynamic properties of recently developed artificial solid-state neuron microchips with delayed perturbations and describe their role in the study of neural activity and regenerative memory. This review covers our recent research work on excitable and delay dynamic characteristics of both single and autaptic (delayed) artificial neurons including all-or-none response, spike-based data encoding, storage, signal regeneration and signal healing. Furthermore, the neural responses of these neuromorphic microchips display all the signatures of extended spatio-temporal localized structures (LSs) of light, which are reviewed here in detail. By taking advantage of the dissipative nature of LSs, we demonstrate potential applications in optical data reconfiguration and clock and timing at high-speeds and with short transients. The results reviewed in this article are a key enabler for the development of high-performance optoelectronic devices in future high-speed brain-inspired optical memories and neuromorphic computing.
\end{abstract}

\keywords{Delay dynamics, excitability, lasers, localized structures, nanostructures, neuromorphic, optical memories, optoelectronics, oscillators, resonant tunneling diodes}
\maketitle

\begin{quotation}

Our aim in this article is to provide a review of our recent achievements for readers who wish to study and emulate the biophysics of spiking neurons and dynamic synapses using advanced brain-inspired (neuromorphic) optoelectronic oscillators. In our approach, electronic and photonic elements converge towards one hybrid micro- nano-technology, offering great advantages for implementing ultra-compact, high-speed and low-power artificial brain-inspired microchips. Firstly, after introducing the state of the art of neuromorphic electronic and photonic circuits, we summarize our experimental and theoretical work on neuromorphic optoelectronic resonators using solid-state resonant tunneling diode circuits integrated with high-speed detectors and light sources. These neuromorphic microchips can be activated either optically or electrically featuring a wide range of neuron-like signal outputs - spiking, bursting, periodic and aperiodic mixed mode oscillations - and other phenomena - e.g., coherence resonance. Secondly, we review our work aiming at using these neuromorphic photonic circuits to build robust, flexible and high-speed brain-inspired regenerative optical memories displaying the unique signatures of spatio-temporal localized structures of light.





\end{quotation}

\section{\label{sec:level1}Introduction}




\begin{figure*}
  \includegraphics[width=1.0\textwidth]{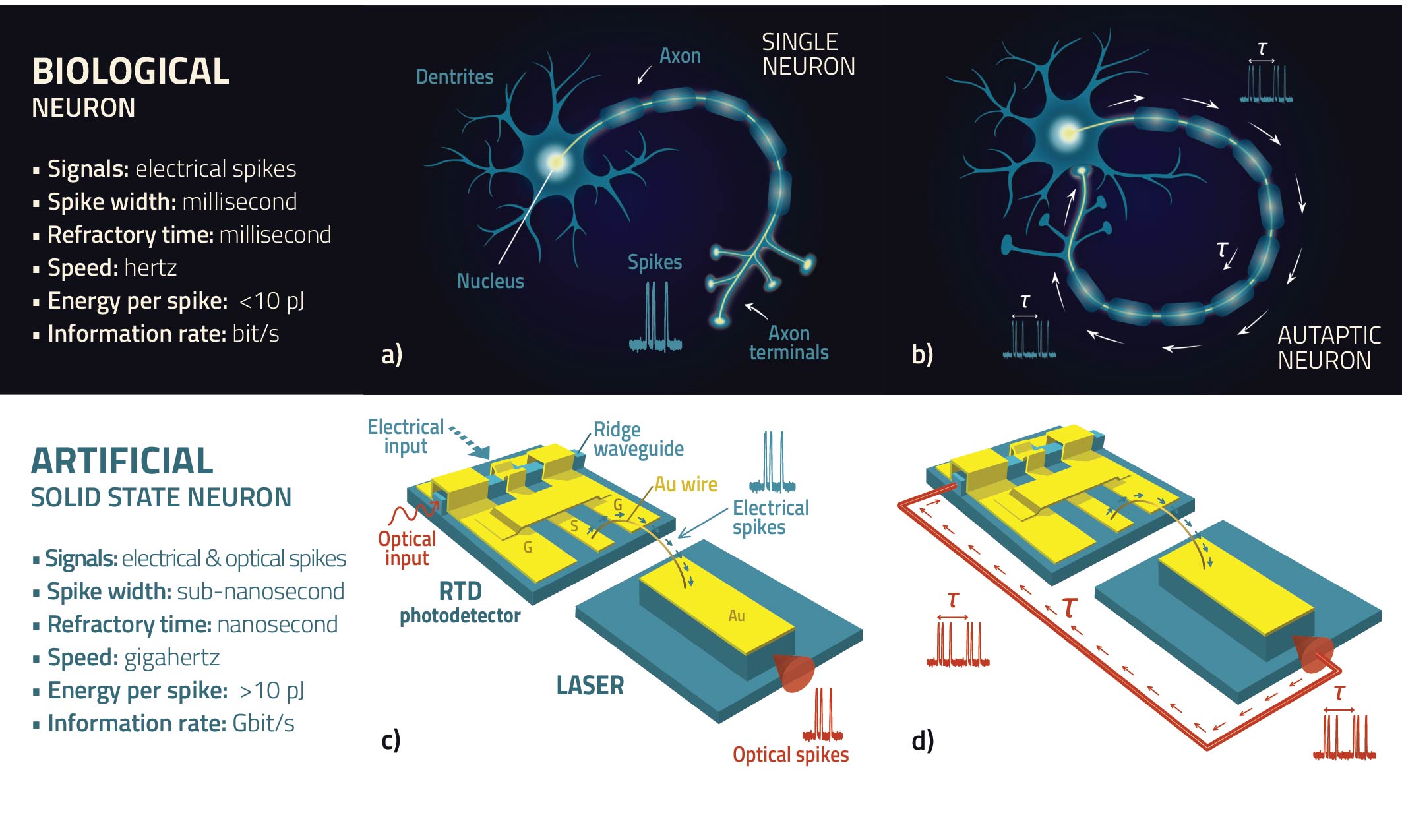}
\caption{a) Artistic view of a biological neuron. b) Representation of a neuron with a self-feedback connection with a time-delay $\tau$ due to the presence of an autapse. c) The solid-state neuron microchip consisting of an RTD-photodetector connected in series with a semiconductor laser diode using a gold (Au) wire bond. Both electrical and optical high-speed excitable spike signals can be activated either electrically or optically. d) Representation of the autaptic neuron microchip. In this configuration, the optical spiking output is re-injected into the RTD-photodetector input after a time-delay $\tau$ due to the propagation in an optical delay line (red trace). The typical performance of our artificial neuron and of a single biological neuron are both detailed in the bottom left and top left, respectively. The energy per spike was calculated using $E=\frac{1}{2}CV_{pp}^2$, assuming $C=245$ pF and $V_{pp}=100$ mV for the case of the biological neuron, while the electrical energy per spike of the artificial neuron assumed a typical device with $C=4.5\,$pF and $V_{pp}=3\,$V. The energy per spike of the artificial neuron can be substantially reduced ($\ll 1\,$pJ) by scaling down the size of the microchip components.}
\label{neuronrtd}       
\end{figure*}

We are currently witnessing an exponential growth of artificial intelligence systems to help humans dealing with highly complex tasks, such as sensing and learning \citep{Jang1993,Rowley1998,Huang2006,Lichtsteiner2008,Shen2016}, needed for the internet of things and to handle with big data. While conventional digital computing has been the engine of the information technology revolution in the past decades, this technology falls far short of the human brain in terms of problem-solving abilities and power consumption. For this reason, neural networks, i.e., collections of artificial neurons mimicking biological brain functions, are currently the focus of much attention. To this end, there is a strong focus not only on the development of deep neural networks using computer algorithms \cite{Hinton2006,Lecun2015,Bengio2013} but also in integrated neuromorphic microchips \cite{Indiveri2011,Merolla2014,Prezioso2015,Shen2016} as the hardware that can reproduce neurotransmission dynamics - the communication between neurons - by interconnecting many artificial neuron-like elements. This communication is encoded in sequences of intensity spikes (the excitable pulses \cite{Lindner2004}) as found in the unique information processing of the brain.

This challenging and highly-interdisciplinary area of research has recently mobilized significant researchers around the world. In the United States, researchers funded by DARPA SyNAPSE developed a one million neuron brain-inspired processor \cite{SyNAPSE}, the TrueNorth chip. In 2012 Intel announced their venture into neuromorphic chip development with a new architecture resulting in the start-up development of the QuarkSE chip. In Stanford University, the Neurogrid \cite{Benjamin2014} implemented a mixed-analog-digital multi-chip system for large-scale neural simulations. In Europe, the FACETS program \cite{Facets} developed a chip with 200,000 neurons and 50 million synaptic connections which in turn has led to the European initiative "The Human Brain Project" launched in 2014 \cite{HumanBrain}. Other examples of projects in brain machine simulation include the SpiNNaker \cite{Spinnaker} and the BrainScaleS \cite{BrainScales} initiatives.

Thus far, the emphasis has been on the development of neuromorphic electronic technologies based on CMOS and/or memristors \cite{Indiveri2006,Jo2010,Indiveri2011,Kim2012,Merolla2014}. Although the impressive advances, these approaches use electronic synapses \cite{Kuzum2013} at kHz speeds, which are difficult to interface with optical technologies for implementation in the context of high-bandwidth optical communications systems. One of the most promising alternatives is to use light-based synapses enabled by neuromorphic photonic integrated chips. This approach takes advantage of energy efficient optical interconnects to achieve low-power neuron-like responses at speeds one billion times faster than neurons ($> 1$ Gb/s). This is also much faster than the electronic-based artificial neurons, making optical neurons excellent candidates to realize the dream of a fully integrated brain-inspired photonic information processor. In photonics, high-speed neuromorphic spiking responses can be achieved using semiconductor lasers\cite{Prucnal2016} and other optoelectronic-based configurations \cite{Romeira2014,Romeira2015}, see a recent review in \cite{Prucnal2017}. Several laser-based neuromorphic systems have been reported allowing operation at telecommunication wavelengths (see for example the recent work of Hurtado et al. \cite{Robertson17}, and references therein), and therefore compatible with current fiber-optic communication systems.

While considerable attention has been dedicated to the realization of optical neurons and network architectures, less attention has been paid to the mechanism of autaptic, i.e. self-feedback, connections in neurons. These synapses between a neuron and a branch of its own axon were reported for the first time more than four decades ago \cite{Loos1972}, and have been found in the neocortex and the hippocampus regions of the brain, among other areas \cite{Flight2009}. Several works suggest that these autaptic neural connections are abundant in specific types of neurons \cite{Tamas1997}, having key implications in synaptic transmission \cite{Bacci2003} and in local feedback neuron regulation \cite{Herrmann2004,Flight2009}. In neural networks, these self-feedback connections may offer energetically effective means for controlling network dynamics towards specific states \cite{Wiles2017,Xu2017}. A number of theoretical studies show that the delays in autaptic inputs affect the bursting behavior and information transfer of individual neurons \cite{Rusin2011,Hashemi2012,Wang2014,WangChaos2014}. Despite the potential applications of autaptic neurons, excluding the work on reservoir computing that uses a single dynamical delayed node as a complex network to perform computation \cite{Appeltant2011}, experimental autaptic neurons remain almost unexplored in the context of neuromorphic microchips.

In this article, we review our recent work on solid-state artificial neurons with autaptic (delayed) perturbations. In Fig. \ref{neuronrtd}, we show a schematic diagram that compares the biological neuron, panels a) and b), with our solid-state neuron, panels c) and d). The biological neuron, panel a), consists of a soma region (cell body nucleus), dendrites (thin structures that arise from the cell body) and axon terminals (a long cellular extension that arises from the cell body). Neurons generate action potentials (electrical spikes), with amplitudes of approximately 100 mV and duration in the range of 0.1-1 ms in their soma. The spikes then propagate through the axon and are transmitted to the next neuron through the synapses consuming less than 10 pJ per spike \cite{Sourikopoulos2017}. The synapses, which are 20-40 nm wide gaps between the axon end and the dendrites, transmit the signal either chemically by releasing neurotransmitters or electrically, depending on the type of the synapse. The typical performance of a biological neuron is detailed in the top left of Fig. \ref{neuronrtd}. In panel b), it is  shown a schematic representation of an autaptic neuron with a self-feedback connection with a time-delay $\tau$ due to the presence of an autapse. These type of connections provide additional control of the synaptic activity of neurons.

The artificial neuron, panel c), consists of a neuromorphic optoelectronic microchip \cite{Romeira2013a} formed by two key components: a nanoscale double barrier quantum well (DBQW) resonant tunneling diode photodetector (RTD-PD) \cite{Romeira2010,Romeira2013b} in an optical ridge waveguide, and a laser diode (LD). Both components operate at telecommunications wavelengths ($\sim$1.55$\,\mu$m). The typical electrical active area of the RTD-PDs used in our work is around 400 $\mu$m$^{2}$ and the ridge mesas of the commercial lasers have typical areas of $\sim 300$ $\mu$m$^{2}$. Although the footprint of both devices is currently much larger than the nanoscale dimensions of the DBQW nanostructure (around 10 nm), our neuromorphic integrated system could potentially be reduced to dimensions (excluding the electrical contacts) of only a few micrometers taking advantage of recent advances in the nanofabrication of nanoscale light sources (nanoLEDs \cite{Huang2014,Dolores-Calzadilla2017} and nanolasers \cite{Hill2014}) with wavelength and sub-wavelength scale dimensions. Furthermore, RTD devices with areas below 25 $\mu$m$^{2}$ have been reported already by several groups, e.g. \cite{Asada2008,Wang2013}, although for different applications, namely terahertz (THz) oscillators. This would bring several advantages since the energy per spike of the artificial neuron could be substantially reduced ($\ll 1\,$pJ) by scaling down both RTD-PD and LD components.

\begin{figure}
  \centering
  \includegraphics[width=3.0in]{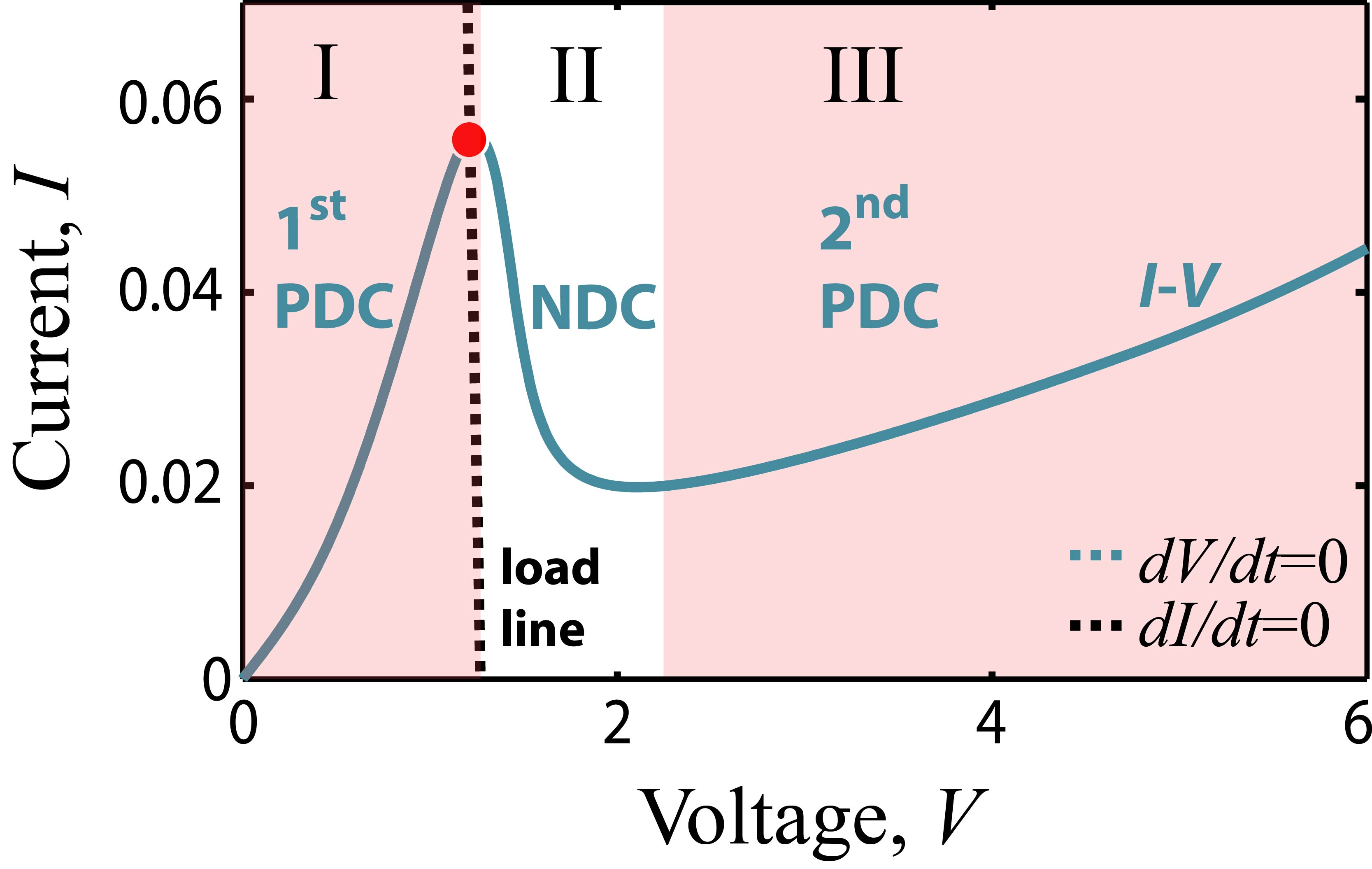}\\
  \caption{Typical antisymmetric nonlinear current-voltage characteristic (solid curve) of a resonant tunneling diode showing the regions of negative (II) and positive (I and III) differential conductance. Also shown is the load line (dashed curve) which defines the operating point of the RTD-based neuromorphic optoelectronic circuit.}\label{rtdiv}
\end{figure}

In the neuromorphic integrated system, the RTD-PD component provides a non-monotonic current-voltage ($I-V$) characteristic with a nonlinear region of negative differential conductance (NDC), Fig. \ref{rtdiv}. Depending on the dc bias point, that is, the intersection point between the load line and the nonlinear $I-V$ curve, the device can be operated in various dynamical regimes, in which the optoelectronic delayed feedback plays different roles:

\begin{enumerate}
  \item \textbf{Bistability:} the load line can intersect the nonlinear $I-V$ curve in one of its two positives slopes, also called positive differential conductance (PDC) regions (I and II in Fig. \ref{rtdiv}). Within either the two PDCs, the device remains at a steady-state but if the load line intersects both the PDCs, bistability is achieved. In this bistable regime, the noise stemming from e.g. thermal fluctuations in the circuit can induce transitions between the two stable states and a phenomenon of stochastic resonance is observed, as reported in \cite{Romeira2014b}.
  \item \textbf{Self-sustained oscillations:} in the NDC, region II in Fig. \ref{rtdiv}, the system operates as a high-frequency nonlinear self-sustained oscillator \cite{Slight2008}. In this situation, the quality of the high-speed radio-frequency signals can benefit from the effect of the delayed optoelectronic feedback \cite{Romeira2011,Romeira2013}, enabling ultra-low phase noise self-sustained oscillations.
  \item \textbf{Excitable dynamics:} when the bias point is set at the border between the first PDC and the NDC, region I in Fig. \ref{rtdiv}, neural, type II, excitable dynamics is achieved (see Section IV-A for details on the physical principle of excitability). Multi-pulse excitable bursting is obtained when biased at the border between the NDC and the second PDC, region III in Fig. \ref{rtdiv}. Depending on the perturbation and load line, periodic and aperiodic mixed mode oscillations can also be obtained under a proper external modulation (see section IV-B). In the excitable dynamic regime, the effect of delayed optoelectronic feedback leads to self-regeneration of the fired excitable pulses enabling regenerative memory operation, as discussed in detail in Section IV-C.
  \item \textbf{Neural inhibition dynamics:} although not discussed in this review paper, our system could be used to obtain neural inhibition dynamics \cite{Mesaritakis2016,Robertson17}. In this case, the bias point is set in the NDC and close to the peak, region II in Fig. \ref{rtdiv}, where oscillations are obtained. In order to get a static response, that is inhibition, an external voltage can be applied (using e.g. a negative pulse voltage) to switch the operation point to the first PDC region, region I in Fig. \ref{rtdiv}. The same principle can be used in the case the bias point is set in the NDC region but closer to the valley. In this situation, a positive pulse voltage is needed to switch the operation point to the second PDC region, region III in Fig. \ref{rtdiv}.
\end{enumerate}


In this review article, we address the neural spike dynamic characteristics detailed above in the excitable dynamic regime (point 3) combined with a mechanism of time-delayed feedback. This corresponds to the solid-state autaptic neuron configuration represented schematically in Fig.~\ref{neuronrtd}d). It consists of an optical delay line inserted off-chip \cite{Romeira2016}, typically using a low loss optical fiber, with a time delay of $\tau$. The delay line provides a mechanism of re-injection of the fired optical pulses, analogous to the autaptic neuron shown in Fig.~\ref{neuronrtd}b). This scheme is exploited for applications in brain-inspired temporal buffer memories enabling writing and storage of information as light intensity pulses. Notably, the neural spike response in the autaptic configuration displays all the signatures of spatio-temporal localized structures (LSs) of light (see a recent review in \cite{YG-JPA-17}). As discussed in Section IV-E, by taking advantage of the dissipative nature of the LSs, robust and flexible data reconfiguration and clock and timing can be achieved in our neuromorphic chips at high-speeds and with short transients.





\section{Nanoscale resonant tunneling diode}

The quantum properties of charge transport and the nonlinearity of nanoscale RTDs have been exploited in a large variety of applications including THz communications \cite{Oshima2016}, THz imaging \cite{Miyamoto2016}, ultrashort pulse generators \cite{Kamegai2008}, highly sensitive light detectors \cite{Romeira2013b,Pfenning2015,Pfenning2016}, single-photon detectors and photon counting,\cite{Blakesley2005,Li2008,Weng2015} ultrafast memories and switches \cite{Shimizu2015}, and emulation of neural excitable responses \cite{Romeira2013b,Romeira2014b,Klofai2015,Romeira2016}. Here, we give particular attention to the mechanism of charge transport based on quantum resonant tunneling that provides the N-shaped negative differential conductance in RTDs, a key property enabling excitable neural responses in nanoscale RTDs.

\begin{figure}
  \centering
  \includegraphics[width=3.4in]{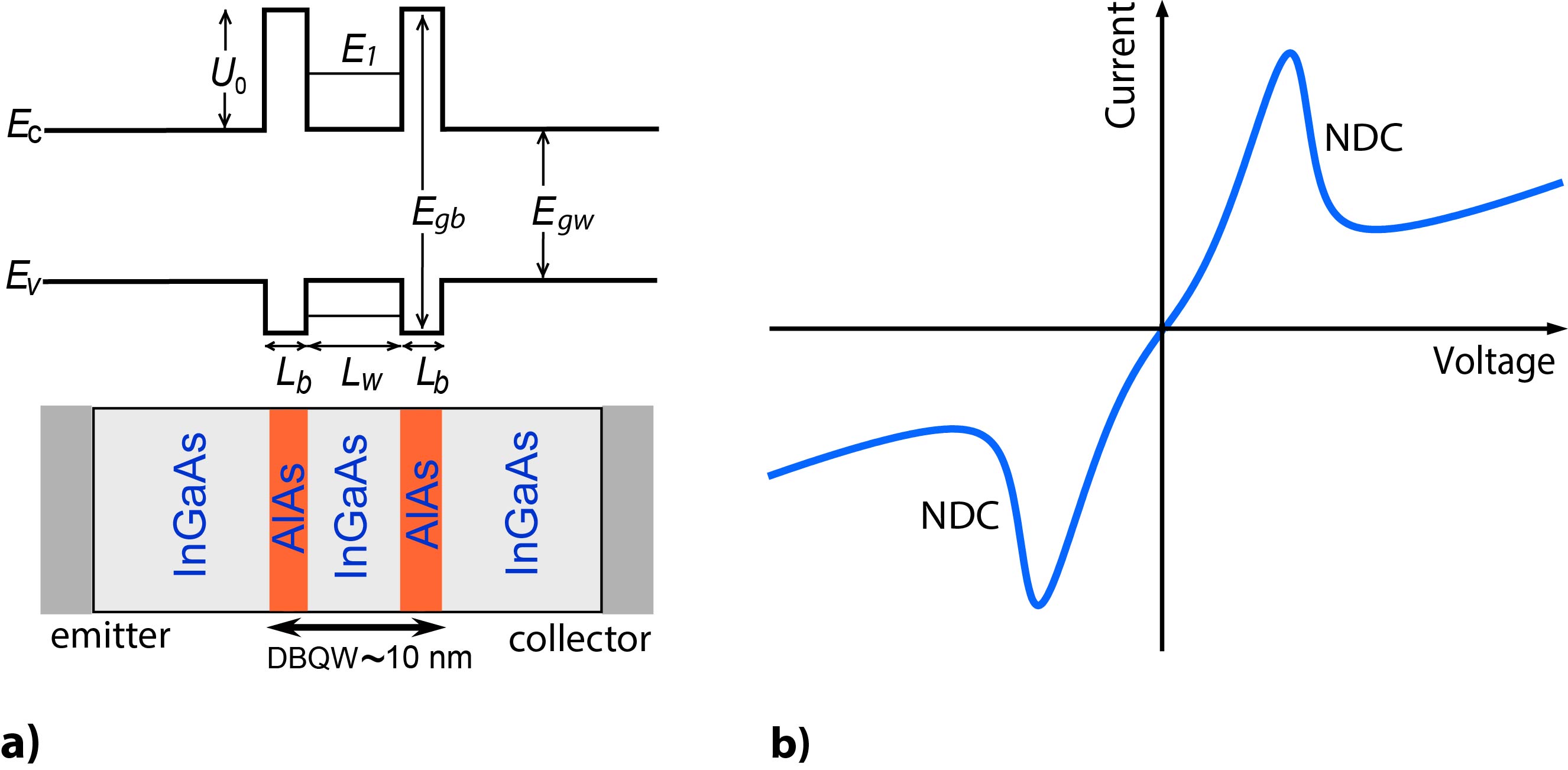}\\
  \caption{a) Schematic energy diagram (top) of the lowest conduction band, $E_c$, and the highest valence band, $E_v$, of a double barrier quantum well nanostructure formed by InGaAs/AlAs semiconductor compounds (bottom). $E_{gb}$ and $E_{gw}$ represent the energy bandgaps of the barriers and quantum well, respectively, and $U_0$ is the potential barrier height. b) Typical antisymmetric N-shaped current-voltage characteristic of a DBQW-RTD nanostructure at room-temperature showing the regions of negative differential conductance. Reproduced with permission from Ph.D. thesis, Universidade do Algarve (2012). Copyright 2012 Bruno Romeira.}\label{dbqw}
\end{figure}

Figure~\ref{dbqw}a) shows a schematic example of the energy diagram of a resonant tunneling diode formed by InGaAs/AlAs compound semiconductors. Its nanostructure consists of a low energy band-gap semiconductor with energy $E_{gw}$, typically a quantum well ranging from 5 nm to 10 nm wide, surrounded by two thinner layers of higher energy band-gap semiconductor barriers (with energy $E_{gb}$), typically ranging from 1.5 nm to 5 nm wide, both sandwiched between lower energy band-gap materials, usually the quantum well material. When both sides are terminated by highly doped semiconductor layers for electrical connection (the emitter and the collector contacts), the nanostructure is called a resonant tunneling diode. These type of low-dimensional nanostructures have attracted a large attention because of the pronounced region of NDC that appears in their $I-V$ characteristics over a wide voltage range, Fig. \ref{dbqw}b).


\subsection{Double barrier quantum well nanostructure}

\begin{figure}
  \centering
  \includegraphics[width=3.4in]{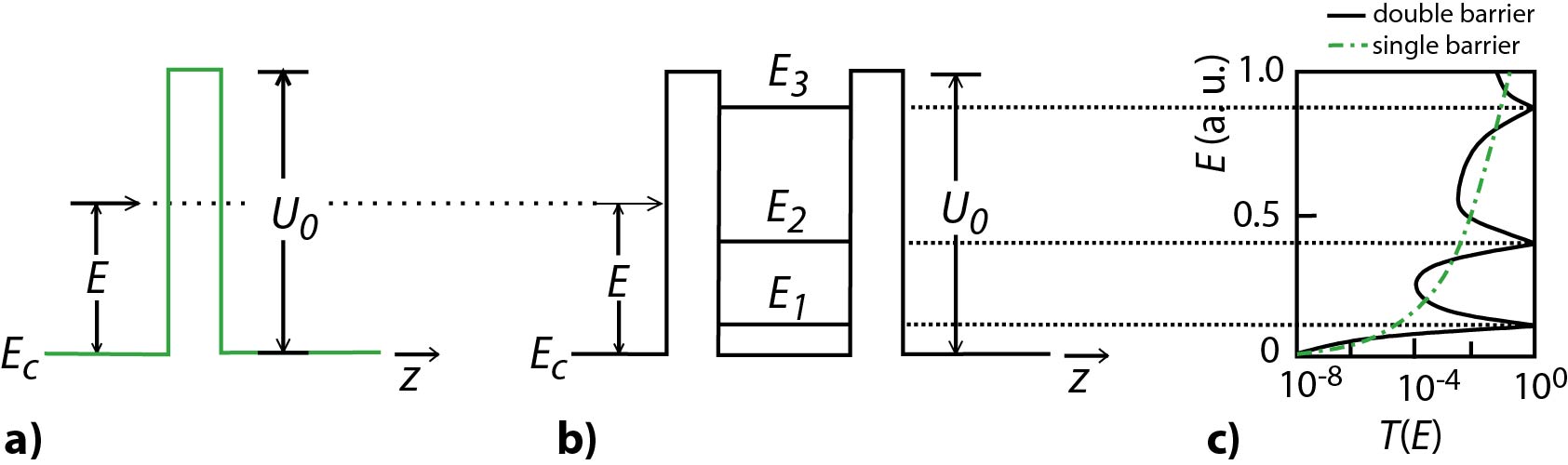}\\
  \caption{Schematic diagram showing a) a single barrier, b) a symmetric double barrier quantum well nanostructure, and c) the corresponding transmission coefficients, $T(E)$, as a function of incident carrier energy, $E$. $U_0$ is the potential barrier height, and $E_1-E_3$ are the resonant energy levels. Reproduced with permission from Ph.D. thesis, Universidade do Algarve (2012). Copyright 2012 Bruno Romeira.}\label{dbqw2}
\end{figure}

The carrier flow through a double barrier quantum well resonant tunneling diode (DBQW-RTD) is fundamentally different from that of a single barrier because the DBQW structure acts as filter to charge carrier energy distribution by controlling the number of carriers that can take part in the conduction through the resonant levels. In the scheme of Fig. \ref{dbqw2}, it is shown a comparison between the transmission coefficient, $T(E)$ (panel c)), for a single barrier, panel a), and a symmetric double barrier quantum well, panel b), as a function of the incident charge carrier energy, $E$. For the case of a symmetric double barrier, because of the finite height of the energy barriers, the allowed energy states in the well region become quasi-bound (or resonant states) rather than bound states. As a consequence, tunneling of charge carriers through the barriers is strongly enhanced (reaching unity), black line in Fig. \ref{dbqw2}c), for incident energies that equal the nanostructure resonant energies. The transmission coefficients in the double barrier quantum well case are much higher than the transmission coefficients of a single barrier at the same energy values of the resonant levels, see green dashed-point trace in Fig \ref{dbqw2}c). The transmission coefficient lobs broaden with increasing energy because the barriers become more transparent. The carrier transmission coefficient maxima shown in Fig. \ref{dbqw2}c) give rise to a current-voltage characteristic with regions of maximum and minimum conduction, resulting in an N-shaped change of the current flow as a function of the applied voltage.

\subsection{Negative differential conductance}

\begin{figure}
  \centering
  \includegraphics[width=2.8in]{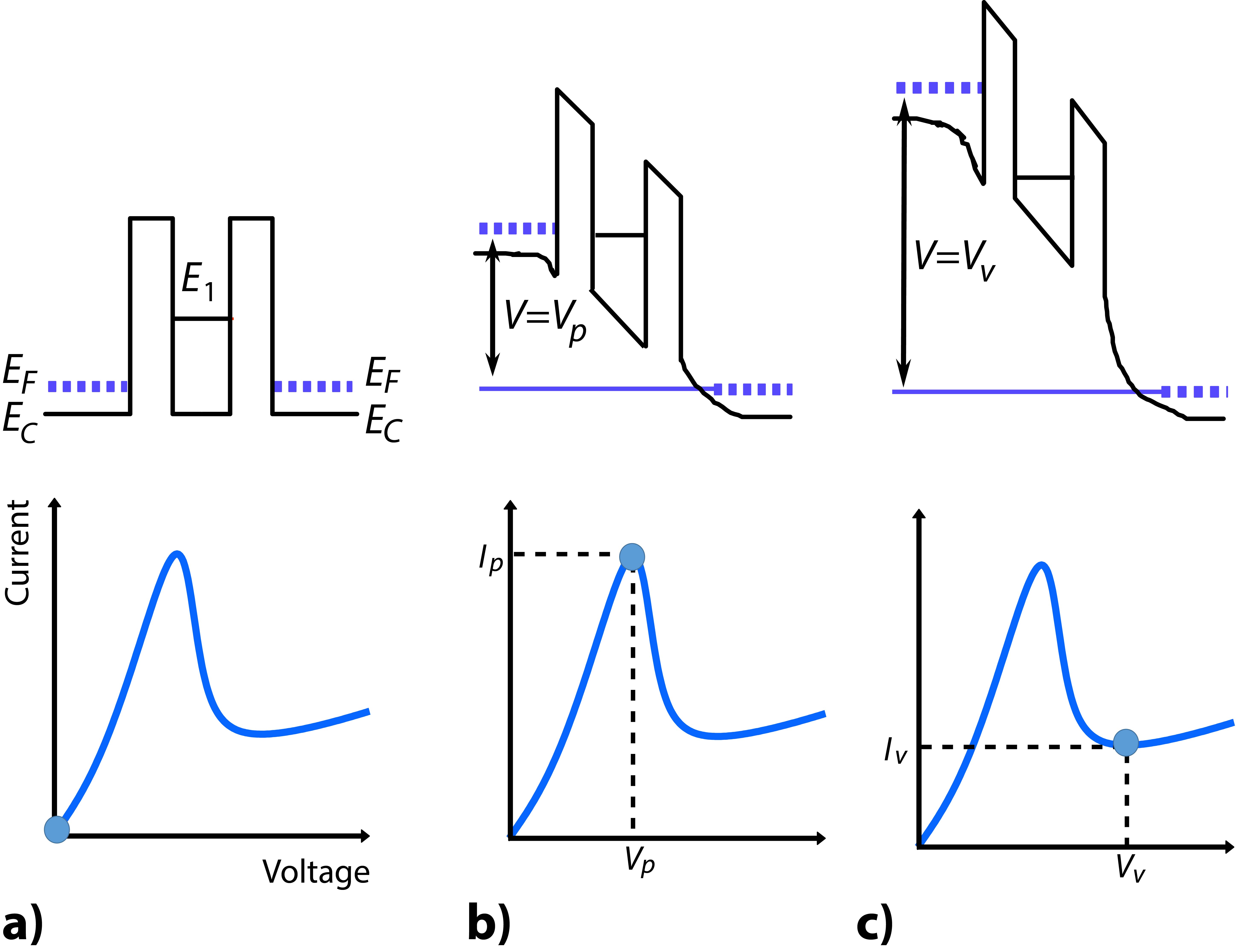}\\
  \caption{a)-c) Lowest conduction band profile under applied voltage (top), and N-shaped negative differential conductance current-voltage characteristic (bottom). Reproduced with permission from Ph.D. thesis, Universidade do Algarve (2012). Copyright 2012 Bruno Romeira.}\label{iv}
\end{figure}

The N-shaped RTD $I-V$ characteristic can be understood with the help of the lowest conduction band profile. Figure \ref{iv} shows a schematic of the lowest conduction band profiles for an $n$-type DBQW-RTD at zero volt, panel a), at the peak voltage (resonance), panel b), and at the valley voltage (off resonance), panel c). When the applied bias is small, i.e., $V \ll V_{p}$ (where $V_p$ is the peak voltage, also referred as resonance voltage), the conduction band profile is not much affected, remaining almost flat, see Fig. \ref{iv}a). The first resonant level is well above the emitter's Fermi level, and very low charge flows. As the voltage is increased, the energy of the first resonant level is moved downwards to the emitter's Fermi level, leading to an almost linear increase of the current with the voltage. The current increases until a local maximum is reached at $I_{p}$, ideally, at $V \simeq 2E_{n=1}/q$, that is, when the overlap between the emitter's Fermi sea energy and the transmission coefficient around the first resonant level reaches a local maximum, see Fig.~\ref{iv}b) upper panel, corresponding to the first positive differential conductance region. A further increase in the applied voltage pulls the first resonant level towards the bottom and into the forbidden band-gap, where there are no longer carriers available to efficiently cross the DBQW. This leads to a sharp current decrease, giving rise to the first NDC portion of the $I-V$ characteristic. At a given voltage, called the valley voltage $V_{v}$ ($V_{v}>V_{p}$), the current reaches a local minimum $I_{v}$, Fig. \ref{iv}c). An additional increase of the bias voltage will further lift up the emitter's Fermi level and tunneling through higher resonant levels or carrier transport above the barriers will lead to a new current increase.

\subsection{Li\'{e}nard oscillator model}\label{sec_lienard}

\begin{figure}
  \centering
  \includegraphics[width=2.8in]{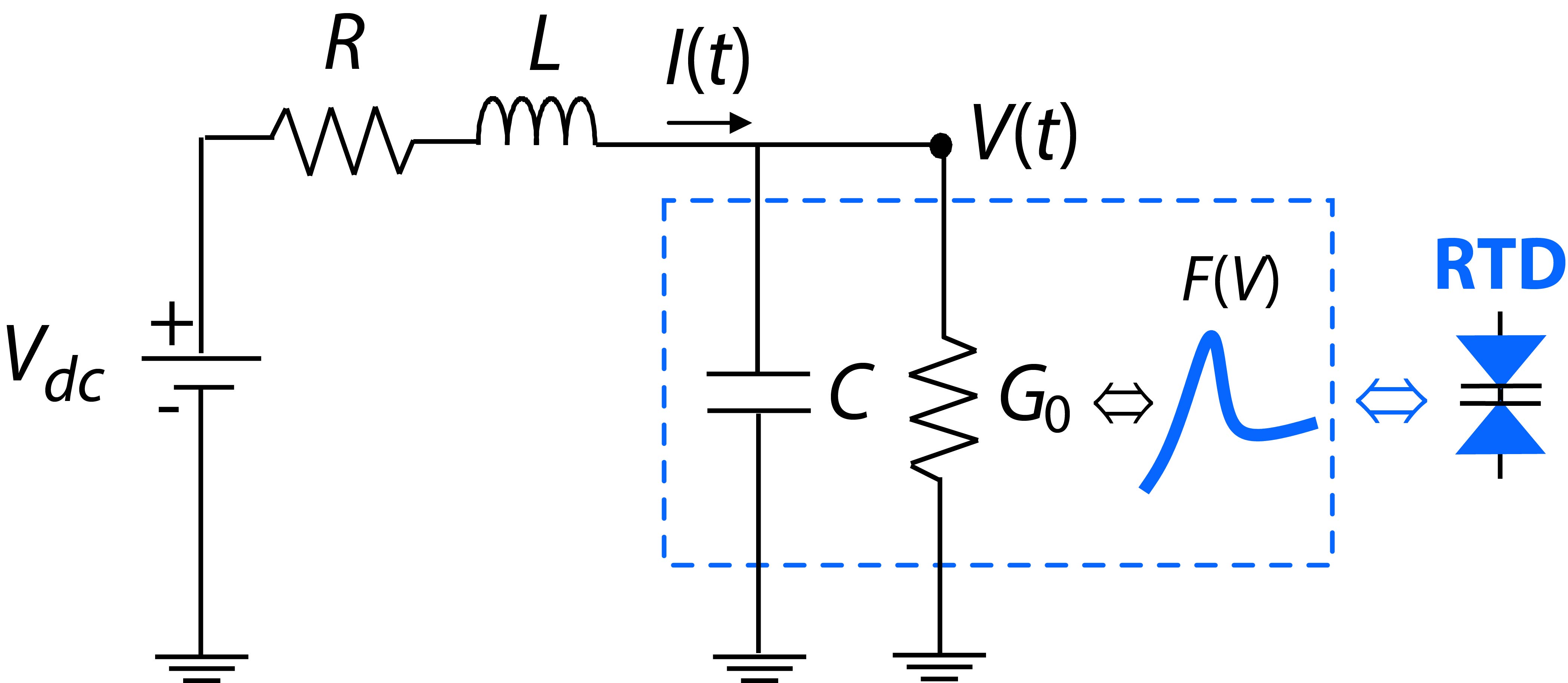}\\
  \caption{Equivalent lumped electrical circuit schematic of the nanoscale RTD nonlinear oscillator. Reproduced with permission from Ph.D. thesis, Universidade do Algarve (2012). Copyright 2012 Bruno Romeira.}\label{rtd_circuit}
\end{figure}

Using the theory of nonlinear differential equations employed to describe oscillator systems, here we formulate the nonlinear dynamic model that describes the dynamic characteristics of resonant tunneling diode resonators. The RTD is represented by an equivalent parallel circuit describing the RTD's conductance $G_{0}$ and its capacitance $C$ as shown in Fig.~\ref{rtd_circuit}. The RTD conductance and the diode capacitance correspond to the static NDC conductance and the RTD's emitter-collector capacitance, respectively. We use a voltage dependent current source, $I=F(V)$ to model the nonlinear $I-V$ characteristic.  The function $F(V)$ describes the experimental N-shaped $I-V$ characteristic and can be either fitted with a polynomial function or using a more detailed equation \cite{Schulman1996}. Since the RTD current lags behind the applied voltage, an inductance in series with the RTD conductance should be considered in the circuit model. This inductance $L_{qw}$ is related with the resonant state given by $L_{qw}=\tau_{d}/G_{0}$, where $\tau_{d}$ is the resonance lifetime. However, in a real circuit operating at moderate frequencies $< 10$ GHz, $L_{qw}$ is negligible, that is, much smaller that the circuit's inductance, Fig. \ref{rtd_circuit}.

By applying Kirchhoff's rules (using Faraday's law) to the circuit of Fig. \ref{rtd_circuit}, the voltage $V(t)$ across the capacitance $C$ and the current $I(t)$ through the inductor $L$ are given by the following system of two first-order differential equations:

\begin{eqnarray}
\mu\frac{dV(t)}{dt} & = & I(t) - F(V)-I_{n}-I_{ph} \label{eq:voltage-norm}\\
\mu^{-1}\frac{dI(t)}{dt} & = &   V_{dc}+V_{ac} \sin(\omega_{in} t)-R I(t)-V(t) \label{eq:current-norm}
\end{eqnarray}
where $V_{dc}$ is the dc bias voltage and $L$ describes the circuit parasitics from the transmission line and wire connections. Lastly, $R$ is the series resistance related to the highly doped bulk regions on either side of the DBQW structure and the external lead resistances associated to the contacts, e.g. wires and other bias circuit components. We also include in Eq.~(\ref{eq:voltage-norm}) an external electrical perturbation $V_{ac}\sin(2\pi f_{in}t)$ describing a sinusoidal modulation, where $V_{ac}$ represents the amplitude and $f_{in}$ its frequency. For purposes of numerical simulation time was rescaled by the RTD natural frequency $\omega_{0}=1/\sqrt{LC}$ so that $\mu=\sqrt{C/L}$, $f_{in}=\frac{\omega_{in}}{\omega_{0} 2\pi}$.

In Eqs. (\ref{eq:voltage-norm})-(\ref{eq:current-norm}), when $V_{AC}=0$, the system is called an autonomous Li\'{e}nard oscillator \cite{Hale}. The Li\'{e}nard's model describes a wide range of physical nonlinear dynamical systems (see \cite{A.LinsW.Melo1977,Hale} and the references therein). More recently, we have demonstrated the Li\'{e}nard system describes the rich dynamics of RTD optoelectronic oscillators, specifically the generation of self-sustained oscillations at GHz speeds \cite{Slight2008,Romeira2013}. The Li\'{e}nard RTD oscillator subjected to a time-dependent external force ($V_{ac}\neq0$) provides an additional degree of freedom and a wide range of additional dynamical regimes besides self-sustained oscillations can emerge. Foremost of these dynamical regimes include synchronization \cite{Figueiredo2008,Romeira2009a}, chaos \cite{Romeira2008,Romeira2009}, and excitability \cite{Romeira2013a}. Lastly, the photo-current, $I_{ph}$, and current noise, $I_{n}$, sources included in Eq.~(\ref{eq:voltage-norm}) describe the Li\'{e}nard oscillator under optical injection \cite{Romeira2010} and perturbed by current noise \cite{Romeira2013}, respectively. The photo-detection characteristics of the RTD are analyzed in detail in Section III-A.

\section{Solid-state optoelectronic neuron resonator}

\begin{figure*}
  \includegraphics[width=1.0\textwidth]{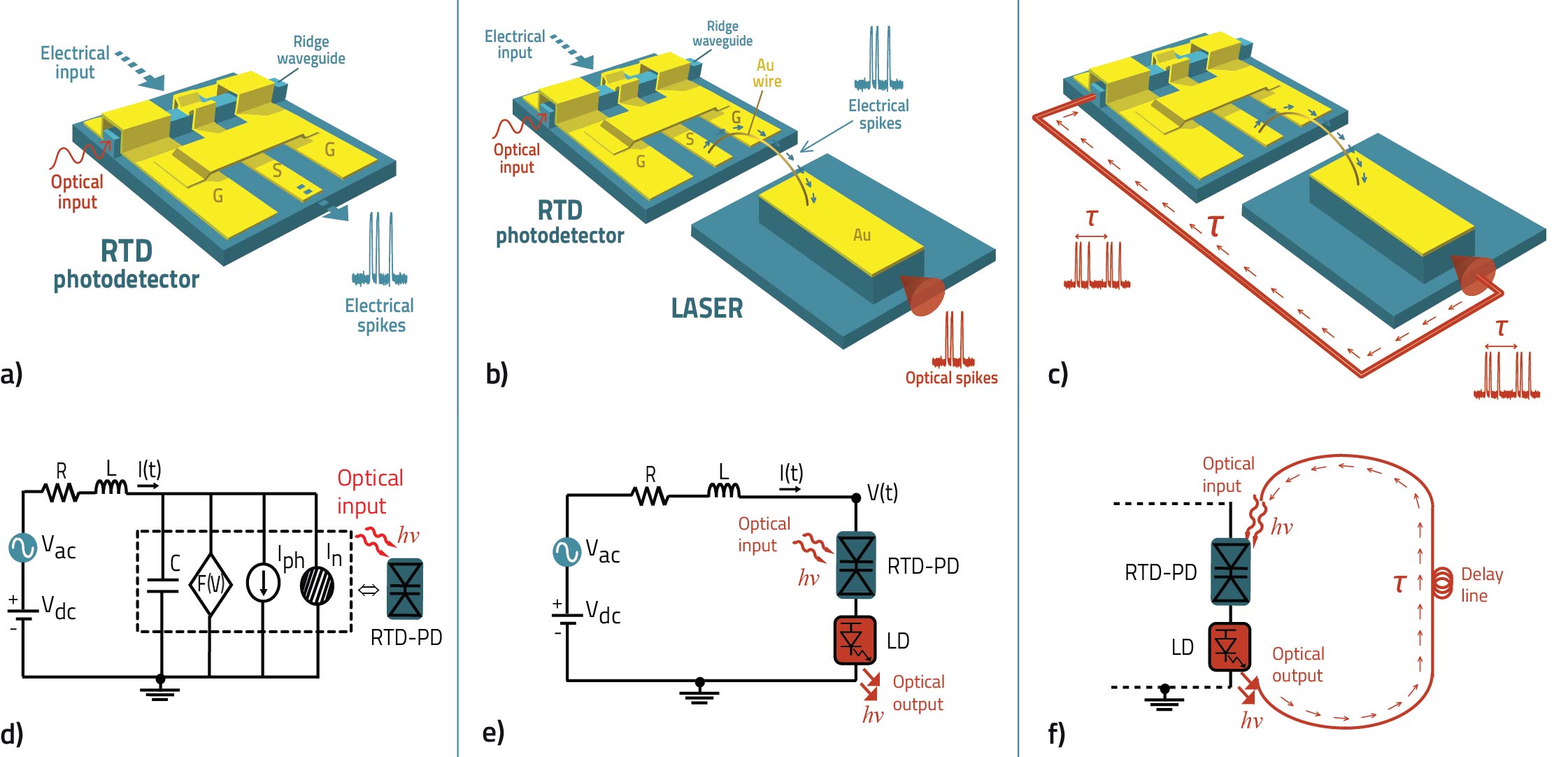}
\caption{a) Schematic of the solid-state neuron microchip consisting of the RTD-photodetector (RTD-PD). b) Schematic of the solid-state optoelectronic neuron microchip consisting of an RTD-PD connected in series with a semiconductor laser diode using a gold (Au) wire bond. Both electrical and optical high-speed excitable spike signals can be activated either electrically or optically. c) Schematic of the autaptic neuron microchip. In this configuration, the optical spiking output is re-injected into the RTD-photodetector input after a time-delay $\tau$ due to the propagation in an optical delay line (red trace). Schematic representation of the equivalent lumped electrical circuit of the d) RTD-PD microchip, e) RTD-PD-LD microchip, f) autaptic microchip.}
\label{neuronrtd_circuit}       
\end{figure*}

In this section we present our work on recent developed solid-state optoelectronic neuron microchips formed by resonant tunneling diode oscillators. In our approach, electronic and photonic components are hybrid integrated in a single chip enabling electro-optical neuromorphic functionalities. Figure \ref{neuronrtd_circuit} shows a schematic representation of the individual and hybrid integrated microchips, panels a)-c), and their respective equivalent circuits, panels d)-f) used to model the experimental circuits. In what follows, we describe, both experimentally and theoretically, the implemented optoelectronic circuits for neuromorphic applications.

\subsection{RTD-photodetector}
The key component of our solid-state neuron microchips is the RTD-photodetector, Fig. \ref{neuronrtd_circuit}a). It consists of a DBQW embedded in an optical ridge waveguide and operates as a waveguide photodetector for incoming light at wavelengths with energy close to or above the waveguide core bandgap energy (the typical operation of our devices is $\sim 1.55 \mu$m). Its typical epi-layer structure is grown by molecular beam epitaxy in a Varian Gen II system on an InP substrate and consists of a 2-nm-thick AlAs barriers separated by a 6-nm-wide InGaAs layer, embedded in a $1$ $\mu$m thick ridge waveguide which corresponds to the photoconductive semiconductor layers. The ridge waveguide consists of a unipolar InAlAs/In$_{0.53}$Ga$_{0.42}$Al$_{0.05}$As/InP (for more detail see \cite{Figueiredo2001,Romeira2013b}). As represented schematically in Fig. \ref{neuronrtd_circuit}a), such configuration enables control of the neuron microchip using both light and electrical signals.

The dynamics of the RTD-PD is analyzed considering the lumped electrical circuit of Fig. \ref{neuronrtd_circuit}d). This circuit is equivalent to the Li\'{e}nard oscillator circuit analyzed in section II-C, Fig. \ref{rtd_circuit}, but assuming an additional photocurrent, $I_{ph}$, and current noise, $I_{n}$, sources (already described in Eq. (\ref{eq:voltage-norm}). The photocurrent source, $I_{ph}$, (see below) describes the photogenerated current in response to an optical modulated signal. The current noise source, $I_{n}$, describes the random noise processes in the oscillator and detector, specifically thermal and shot noises, and are all modelled as Gaussian noise processes (and are described in section III-C).



In Eq. (\ref{eq:voltage-norm}), the RTD-PD photo-generated current, $I_{ph}$, in response to a optical modulated signal, $P(\lambda)$, is given by:

\begin{equation}\label{eq:photocurrent}
    I_{ph}=\eta_{ph} \frac{e \lambda}{h c} P(\lambda)
\end{equation}
where $\lambda$ is the wavelength, $e$ is the electric charge unit, $h$ and $c$ are the Planck constant and the speed of light in the vacuum, respectively, and $\eta_{ph}$ is the waveguide photodetector quantum efficiency given by:

\begin{equation}\label{eq:quantum-efficiency}
    \eta_{ph} = \kappa (1-R_{ref})(1-e^{-\alpha \gamma_{ph} \Lambda})
\end{equation}
where $\kappa$ is the light coupling factor, $R_{ref}$ is the waveguide facet reflectivity, $\alpha$ is the waveguide core absorption coefficient, $\gamma_{ph}$ is the overlap integral of the electric and the optical fields, and $\Lambda$ is the active waveguide PD length. Typical values can be found in.\cite{Romeira2013b}

Although the $I-V$ characteristic of the RTD-PD is a function of both voltage and optical power,\cite{Romeira2013b} considering the low responsivity of the photodetectors analyzed ($< 0.25$ A/W), in this work we assume the static $I-V$ curve in dark conditions. In the cases where the optical injection changes substantially the $I-V$ characteristic, refinements of the model can be included, namely by taking into account the photoconductivity and charge accumulation effects in double-barrier RTD structures (see for example the work of \cite{Coelho2004}).

\subsection{RTD-laser diode}




In the RTD-laser diode configuration, Fig.~\ref{neuronrtd_circuit}b), the RTD-PD is connected in series with a laser diode in a hybrid optoelectronic integrated circuit. The typical laser devices used in our work consisted of an InGaAsP multi-quantum-well active region operating in continuous-wave at $\sim$1550 nm, with 6 mA threshold current, up to 10 mW optical output power, and a modulation bandwidth of $\sim10$ GHz. Connecting a laser diode in series does not substantially change the overall shape of the nonlinear $I-V$ characteristic of the RTD. As detailed in \cite{Slight2008} it just shifts the peak and valley regions to higher voltages (a shift of around 0.84 V, i.e., the voltage drop across the laser $p-n$ junction), while the current values are unchanged. In fact, the nonlinearity of the RTD-PD remains in the laser output. The RTD-PD provides a dynamical bias current control for the laser diode enabling the RTD-LD to operate as a neuromorphic optoelectronic oscillator featuring high-speed response and quadruple electronic and optical input/output functionalities. Moreover, this configuration can be readily adapted to include other types of laser devices (e.g. VCSELs), and on-chip monolithic integration of both RTD-PD and LD components can also be implemented \cite{Slight2007}.

In order to describe the dynamic behavior of the laser diode employed in our neuromorphic microchips, we use a single-mode rate-equation model. The rate equations for the photon, $s(t)$, and carrier, $n(t)$, densities in the semiconductor laser active region are given by:

\begin{eqnarray}
  \frac{dn(t)}{dt} &=& \eta_i \frac{I(t)}{q V_{a}}-\frac{n(t)}{\tau_{n}}-g_{0}(n(t)-n_{0})s(t)\label{eq:laser1} \\
  \frac{ds(t)}{dt} &=& \Gamma g_{0}(n(t)-n_{0})S(t)+\Gamma \beta_{sp} \frac{n(t)}{\tau_{n}}-\frac{s(t)}{\tau_{p}}\label{eq:laser2} \\
  P_{opt} &=&  \frac{\eta_{ext} V_{a} s(t) h c}{\Gamma \tau_{p} \lambda_c}\label{eq:laser3}
\end{eqnarray}

\noindent where $I(t)$ is the modulated current provided by the RTD-PD (given by Li\'enard's model, Eqs. (\ref{eq:voltage-norm})-(\ref{eq:current-norm})), plus the dc current, $V_{a}$ is the volume of the active region, $\tau_{n}$ and $\tau_{p}$ are the spontaneous carrier lifetime and the photon lifetime, respectively; the spontaneous emission factor, $\beta_{sp}$, is the fraction of the spontaneous emission that is coupled to the lasing mode; $n_{0}$ is the carrier density at transparency; $g_{0}$ is the differential gain which includes the effect of gain compression; $\Gamma$ is the optical confinement factor; $\lambda_c$ is the emission wavelength; $\eta_{ext}$ is the external efficiency; and $P_{opt}$ is the laser optical output power. For simplicity of analysis, the nonradiative recombination processes were neglected in the rate equations model.




\subsection{Autaptic neuron resonator}

Here we describe how to operate and model the neuromorphic optoelectronic microchip described previously as an autaptic neuron. In this configuration, Fig.~\ref{neuronrtd_circuit}c)-f), an optical delay line with time-delay $\tau$ is inserted off-chip (typically a low loss optical fiber) providing a mechanism of re-injection of the optical output, analogous to an autaptic biological neuron,  Fig.~\ref{neuronrtd_circuit}c)-f). As detailed in Section IV, this scheme is explored for applications in brain-inspired temporal buffer memories enabling writing and storage of information as light intensity pulses. In order to fully model this neuromorphic optoelectronic resonator system, we consider the set of Eqs. (\ref{eq:voltage-norm})- (\ref{eq:current-norm}) describing the RTD-photodetector and the rate-equation model (\ref{eq:laser1})- (\ref{eq:laser3}) describing the laser diode in a delayed feedback configuration. We then obtain the following dimensionless coupled delay differential equations (DDE) system describing the dynamic characteristics of the solid-state autaptic neuron:

\begin{eqnarray}
  \mu\frac{dV(t)}{dt} &=& I(t) - F(V)-\chi \xi(t)-\eta S(t-\tau) \label{eq:voltage norm}\\
  \mu^{-1}\frac{dI(t)}{dt} &=& V_{dc}+V_{ac}\sin(\omega_{in}t)-R I(t)-V(t) \label{eq:current norm}\\
  \tau'_{n}\frac{dN(t)}{dt} &=& \frac{I(t)}{I_{th}} - N(t) - \frac{N(t) - \delta}{1 - \delta}\frac{S(t)}{1+\epsilon S(t)} \label{eq:electron-norm}\\
  \tau'_{p}\frac{dS(t)}{dt}) &=& \frac{N(t)-\delta}{1-\delta}\frac{S(t)}{1+\epsilon S(t)}-S(t)+\beta_{sp} N(t) \label{eq:photon-norm}
\end{eqnarray}
where $N(t)$ and $S(t)$ are the dimensionless carrier and photon densities, respectively. The variables were rescaled as $n(t)=N(t) N_{th}$ and $s(t)=S(t)S_{0}$, where $S_{0}=\Gamma(\tau_{p}/\tau_{n})N_{th}$ and $N_{th}=n_{0}+(\Gamma g_{0} \tau_{p})^{-1}$ is the threshold carrier density; time is normalized to the characteristic $LC$ resonant tank frequency, $\omega_{0}=(\sqrt{LC})^{-1}$, hence $\tau=\omega_0 t$. The parameter $\eta S(t-\tau_{d})$ is the optical feedback, where $\eta$ is the feedback strength and $\tau$ is the time delay with respect to the dimensionless time $t$. The feedback strength parameter, $\eta$, depends on RTD-PD detection characteristics, Eq. (\ref{eq:photocurrent}), and the fraction of the laser optical output power $P_{opt}$ re-injected into the delayed feedback loop. The carrier density in the laser Eqs. (\ref{eq:electron-norm})-(\ref{eq:photon-norm}) is normalized to threshold, and $I_{th}$ is the dimensionless laser diode threshold current. The parameters $\tau'_{n}$ and $\tau'_{p}$ come from the time rescaling. Lastly, we model the stochastic processes of the neuromorphic system (e.g. thermal and shot noises) as an effective delta-correlated Gaussian white noise of zero mean $\chi \xi(t)$ \cite{G-BOOK-95}, Eq. (\ref{eq:voltage norm}), where the parameter $\chi$ is the dimensionless variance of the distribution and denotes the noise strength. Typical parameters used in the numerical simulations can be found in \cite{Romeira2013}.

\section{Excitable and delay dynamics}\label{sec_results}


In this section, we review our recent results on the excitable and delay dynamic characteristics of both single and autaptic (self-delayed) solid-state neurons including all-or-none spike response (Subsection A) and mixed mode oscillations (Subsection B), and spike-based data encoding, storage, and signal regeneration (subsection C). In subsection D, we link our experimental implementation of the artificial neuron, based upon the nanoscale nonlinear resonant tunneling diode driving a laser, to the paradigm of neuronal activity, the FitzHugh-Nagumo (FHN) model with delayed feedback. In subsection E, we analyze in more detail our autaptic neuron and disclose the existence of a unique temporal response characteristic of extended spatio-temporal localized structures. As thoroughly discussed in subsection E, the ubiquitous physical properties of these localized structures, namely mutual independence, as demonstrated by their uncorrelated random walk motion in the presence of noise, enables the implementation of robust and flexible regenerative photonic memories by addressing, creating and destroying individual localized pulsed patterns of information.

\subsection{Excitability}

Excitability is a concept originally coined to describe the capacity of living organisms e.g. nerves \cite{HH-JOP-52,HH2-JOP-52} or neurons to respond strongly to a weak external stimulus that overcomes a well defined threshold. If the system is perturbed from its rest state, it may relax back towards its steady state in two different ways. If the perturbation remains below a certain threshold, the relaxation is exponential. Above this threshold, the system has to perform a large orbit that involves the whole phase space topology before relaxing again towards the unique fixed point (the rest state). Such a relaxation consisting of two widely different transient regimes towards a unique attractor defines the so-called excitability phenomenon (typically with a pulsed shape). Notably, this unique pulse (or spike) response depends only on the characteristics of the excitable system at hand and not on the details of the stimulus. During its large excursion in phase space, the system cannot respond to another perturbation which defines the so-called lethargic time (or refractory time), $T_l$, as the temporal extent of the orbit. Well known in physiology, this refractory period corresponds physically to the large amount of energy released during the excitable response and may be understood as the time the system needs to recharge before being able to release another response. In neurons, this period of time occurs during the re-polarization and the hyperpolarization of the membrane potential.


The unique characteristics of the excitable response confer to the data transmission and processing systems based upon this mechanism a high degree of robustness due to their inherent capability of signal reshaping. In the spatial domain, a photosensitive Belousov-Zhabotinsky reaction was used to perform image processing \cite{Kuhnert89}, enabling smoothing and Sobel filter edge detection. Recently, an optical torque wrench \cite{Pedaci10} was employed as a sensing technique based in the excitability and was capable of detecting single perturbation events. In the last two decades, there has been a quest for electro-optical semiconductor excitable systems capable of operating at speeds much faster than the typical slow speeds of neurons. In electronic systems, neuron-like semiconductors \cite{SNJ-JAP-11} have been explored showing moderate operation speeds (20 kHz). In the context of high-speed optical communications, short optical pulses (0.73 ns) were obtained using a monolithic vertical cavity laser with an intracavity saturable absorber \cite{BKY-OL-11}. Another examples of generation of fast optical excitable pulses can be found in \cite{GHR-PRL-07} showing excitability in a quantum dot semiconductor laser with optical injection, and in \cite{SBB-PRL-14} where much faster refractory times were reported using a micropillar laser with saturable absorber.

In what follows, we summarize our results on the artificial solid-state neuron of Figs. \ref{neuronrtd_circuit}b) and e), operating at room temperature and at a telecommunication wavelengths ($\sim$1550 nm) enabling high-speed neural pulse response. Our optoelectronic circuit contains the key ingredients that fulfill the excitability paradigm and thus, the inherent capabilities for being used in the framework of bio-inspired photonic data processing. Specifically, this includes a potential for monolithic integration, an intrinsic high-speed response and quadruple electronic and optical inputs/outputs.

\subsubsection{Spiking and bursting}

In order to operate the neuromorphic microchip of Figs. \ref{neuronrtd_circuit}b)-e) and as an excitable system, the optoelectronic circuit is externally activated using either light or electrical signals and dc biased in a monostable condition (below or above the NDC region), that is, in a unique fixed point of at one of the PDC regions. In these conditions, the circuit is highly sensitive to the external input (using either a pulsed, square or noise signals), emitting fixed-shape spikes (or bursts of spikes), depending on the dc bias point. By choosing a white noise perturbation as the input signal, coherence resonance (CR) phenomena can be activated \cite{Lindner2004}, i. e., a maximum of temporal regularity in the output for a finite noise level.

\begin{figure}[htbp]
\centering\includegraphics[width=3.4in]{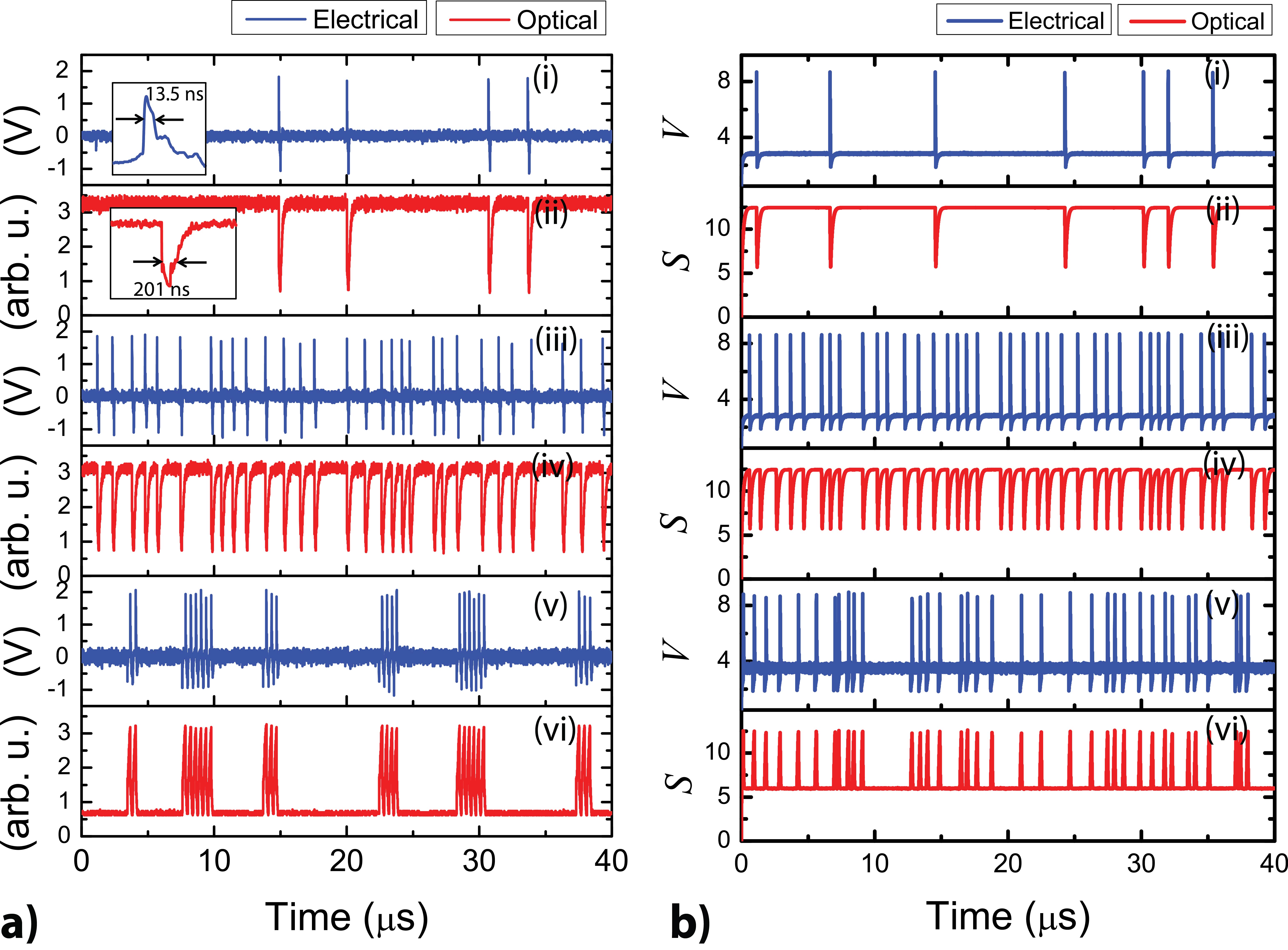}
\caption{a) Experimental time traces of electrically noise induced neuron-like pulsing behavior in an RTD-LD excitable optoelectronic system in both the electrical and the optical domains. The RTD-LD is biased in the first PDC region and is modulated with a noise strength of (i)-(ii) 100 mV; (iii)-(iv) 175 mV. Multi-pulsing bursts when the RTD-LD is biased in the second PDC region and is modulated with a noise strength of 150 mV (v)-(vi). b) Numerical simulation of voltage and photon density $(V,S)$ showing noise induced spike dynamic regimes (i)-(iv) in the first PDC, and (v)-(vi) in the second PDC. The dimensionless noise strength employed in the simulations are: (i)-(ii) $\chi=0.128$; (iii)-(iv) $\chi=0.158$; and (v)-(vi) $\chi=0.310$. Reproduced with permission from Opt. Express 21, 20931 (2013). Copyright 2013 Optical Society of America.
\label{excitable_spikes}}
\end{figure}

Figure \ref{excitable_spikes}a) presents experimental results showing the occurrence of spiking and CR behavior in a neuromorphic circuit activated by electrical noise. For a very weak noise input level, the output maintains its steady (rest) state. For a given noise threshold perturbation, panels (i) and (ii), the circuit is perturbed from its rest state and spiking in both electrical and laser outputs is observed ($V$ and $S$, respectively). The spikes appear random, and the interval between excitations varies substantially. We show in the inset a single pulse event in both the electrical and the optical outputs. The upward voltage pulse event shows a full width half maximum (FWHM) of around 13 ns, inset of panel i). The LD intensity output follows the electrical current switching induced by the RTD with a sequence of downward pulses of identical shape and typical FWHM of $\sim 200$ ns, inset of panel (ii). For a moderate noise level, panels (iii)-(iv), the firing of spikes is more regular and the interspike intervals do not differ much, providing an example of coherence resonance behavior. An analysis of the statistics of the interspike interval (ISI) of the fired spikes allow us to estimate the refractory time to be around 500 ns ($\pm$ 20 ns).

Interestingly, the asymmetric $I-V$ characteristic of our neuromorphic circuit enables two different excitable regimes. As detailed previously, in the first PDC, the response consists in a single isolated peak. But in the second PDC, the temporal response is composed by a succession (or burst) of peaks well separated by the lethargic time. In Fig. \ref{excitable_spikes}a), panels (v) and (vi), is shown multi-pulsing "bursting" behavior when the circuit is dc biased in the second PDC region. As explained next in more detail, in this case the optical pulses are fired in an upward direction since the laser follows the electrical current switching induced by the RTD-LD.

The above observations can be fully explained with the model of Eqs. (\ref{eq:voltage norm})-(\ref{eq:photon-norm}) (see \cite{Romeira2013a} for details on the parameters used). Fig. \ref{excitable_spikes}b) shows a very good agreement between the numerical simulations and the experimental results in both electrical and optical outputs. The white Gaussian noise with amplitude above the excitable threshold activates randomly neuron-like pulses as observed experimentally. The multiple bursting in the second PDC is also predicted by the model due to the asymmetry of the $F(V)$ function.


\begin{figure}[htbp]
\centering\includegraphics[width=3.6in]{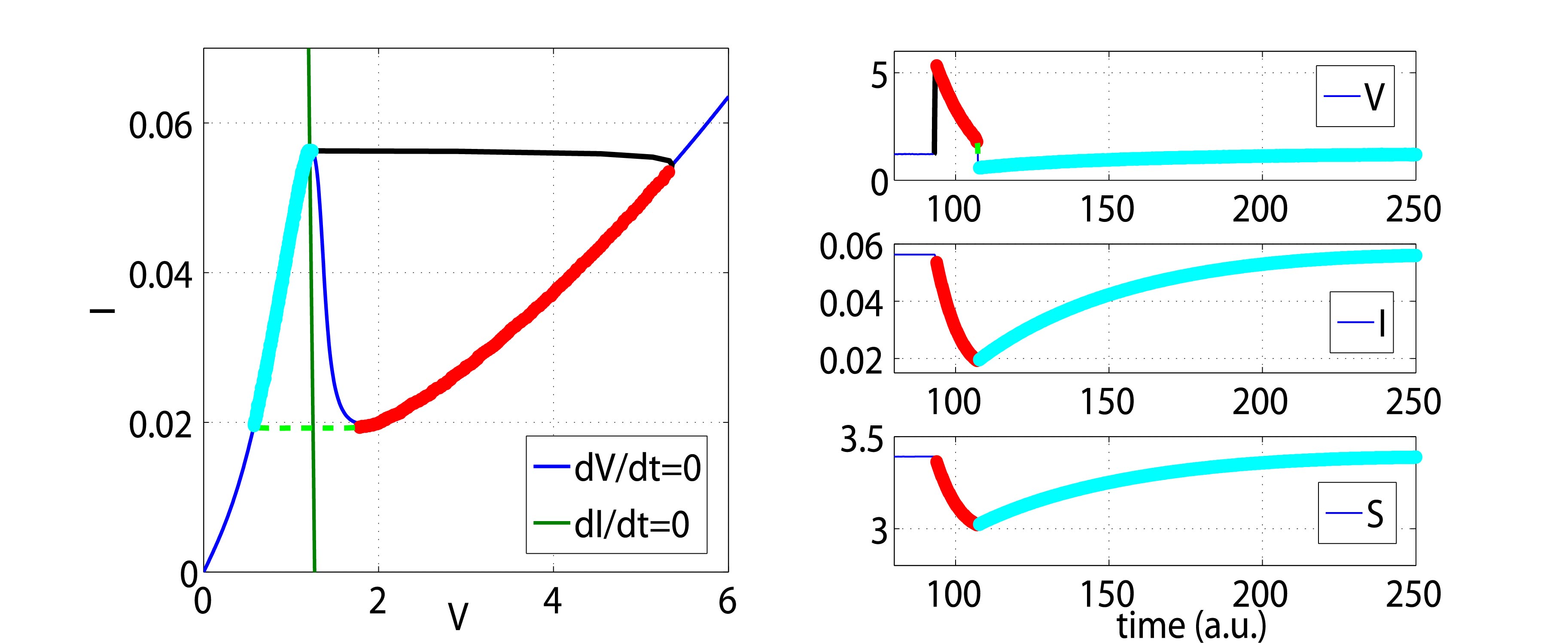}
\caption{Decomposition of the excitable orbit into four stages. The first fast stage corresponds to a sudden rise of the voltage (black line) without variation of the current. The second stage consists in a slow decay of both $V$ and $I$ along the right part of the $F(V)$ nullcline (red line). Next, another fast stage correspond to a voltage drop to the other side of the same nullcline (green dotted line) without variation of the current, finally followed by last slow stage where both $V$ and $I$ recover their initial values. The laser output being modulated by the current, only the slow stages drive its dynamic evolution. Reproduced with permission from Opt. Express 21, 20931 (2013). Copyright 2013 Optical Society of America..\label{orbit}}
\end{figure}

The pulse triggering mechanism is represented schematically in Fig. \ref{orbit} and is explained as follows. The RTD-based autaptic neuron is an example of a slow-fast excitable system in which the fast variable is the voltage and the slow one the current. During the excitable orbit, the two fast stages correspond to a sudden increase and drop of the voltage during which the current does not changes appreciably. These two fast stages are interleaved by two slow stages in which the current evolves along the attracting nullcline as defined by the $F(V)$ function while the voltage follows adiabatically. In the limit $\mu \ll 1$, the two fast stages can be neglected and the lethargic time corresponds mainly to the evolution along the slow stages defined by the two PDC regions. In these regions, the motion is governed by the  equation $ dI/d(\mu t)=V_{dc}-R I - F^{-1}(I)$, that follows from the adiabatic elimination of $V$ using that $\mu^{-1}\gg 1$. Since $F^{-1}\gg \gamma$, the period of the excitable orbit is proportional to $1/\mu=\sqrt{L/C}$ and to the derivative of $F$. Once the scaling of our model by $\omega_{0}=(\sqrt{LC})^{-1}$ is removed, the lethargic time is solely proportional to the inductance of the circuit $L$. An approximate expression of the excitable period $T_l$ can be estimated to be:
\begin{equation}\label{eq: slowmo}
T_{l}/L  =  \frac{1}{F_{2}^{-1}}\ln\frac{V_{dc}-F_{2}^{-1}I_{+}}{V_{dc}-F_{2}^{-1}I_{-}}+\frac{1}{F_{1}^{-1}}\ln\frac{V_{dc}-F_{1}^{-1}I_{-}}{V_{dc}-F_{1}^{-1}I_{+}}
\end{equation}
with $I_{\pm}$ the currents values at the folding points and $F_{1,2}^{-1}$ the inverse of the derivative of $F$ in the first and second PDCs. The estimated inductance of the circuit analyzed in Fig. \ref{excitable_spikes}a) was $L\sim 3$ $\mu$H, explaining the large measured lethargic time ($\sim$500 ns). As shown in the results of subsection C-2, the lethargic time in our neuromorphic microchips can be substantially reduced using a transmission line with lower inductance to connect the RTD and the LD dies.

\subsubsection{High-speed optically induced spike generation}

\begin{figure}
\centering{}\includegraphics[width=3.5in]{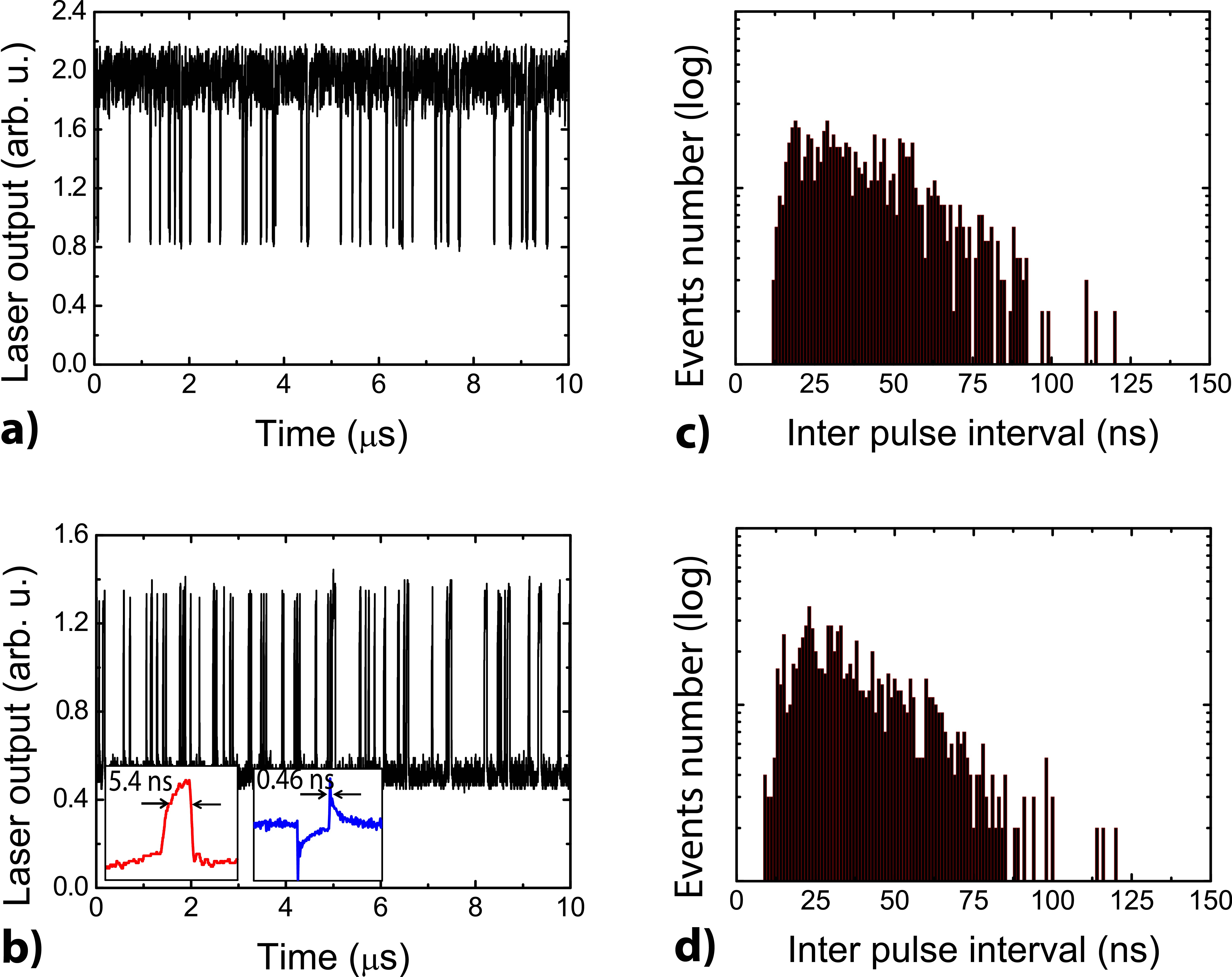}%
\caption{Experimental time traces of the photo-detected laser optical output: (a) electrically noise induced pulsing in the first PDC using a noise amplitude level of 600 mV; (b) optically induced pulsing in the second PDC using an optical power signal of 5.5 mW at $\lambda=1550$ nm and AM modulated with an electrical noise signal with 1.5 V amplitude. Inset: optical and electrical single pulses. Statistic of the times between minima/maxima in the laser output (histogram bin size of 500 ps) when the RTD-LD is biased: (c) in the first PDC region, and (d) in the second PDC region. Reproduced with permission from Opt. Express 21, 20931 (2013). Copyright 2013 Optical Society of America. \label{optical_spikes}}
\end{figure}

Excitability and pulsating behavior can be achieved at much faster speeds as a result of modifications of our neuromorphic optoelectronic circuit. Using a transmission line with $L\sim 8 $~nH to connect the RTD and the LD dies lethargic time can be substantially reduced. Figures \ref{optical_spikes}a) and \ref{optical_spikes}b) show downward and upward optical pulses, respectively, measured in the LD output and triggered by noise. In this case, excitation events were also triggered optically, panel b), by coupling a light signal at 1550 nm to the RTD-PD ridge waveguide using a lensed optical fiber. The optical signal was amplitude modulated (AM) with electrical white noise. The RTD-PD shows a typical responsivity $\sim$0.2 A/W when dc biased in the second PDC region (see section III-A for more details concerning the RTD-PD). In the inset of Fig. \ref{optical_spikes}b), we show the single fired pulse events of both electrical and the optical outputs. The optical pulse shows a FWHM of 5.4 ns, and the upward voltage pulse event shows a FWHM of around 0.46 ns. The measurements were limited by the time of sampling of the oscilloscope (100~ps).



Figures \ref{optical_spikes}c) and d) presents the corresponding ISI distribution. The ISI histogram analysis show the time series measured in the first and second PDCs, showing the typical pulse statistic of the times between minima of the laser output computed using a bin size of 500 ps. We find a statistic with a hard boundary on the left showing the typical exponential behavior of a Kramers escape process \cite{Lindner2004}, displaced by the refractory time of the excitable orbits. This analysis allow us to estimate a refractory time of 12 ns ($\pm$ 500 ps) in the first PDC region, Fig. \ref{optical_spikes}c), and 9 ns ($\pm$ 500 ps) in the second PDC region, Fig. \ref{optical_spikes}d). The lethargic time is more than one order of magnitude lower than the lethargic time measured in the previous neuromorphic circuit analyzed ($\sim$500 ns). Further speed increasing and pulse width reduction is expected with future improvements of the microchips which include reducing the RTD-PD active area (and therefore reduce the intrinsic capacitance $<1$pF), and reducing the series equivalent inductance, here mainly determined by the length of the gold wires used in the electrical connections.


\subsection{Mixed mode oscillations}

Mixed mode oscillations (MMOs) describe trajectories that combine small-amplitude oscillations and large-amplitude oscillations of relaxation type, both recurring in an alternating manner, as compared with the self-sustained relaxation oscillations. Recently, there has been a lot of interest in MMOs that arise due to a generalized canard phenomenon (see \cite{Desroches2012} and references therein). Such MMOs arise in the context of slow-fast systems, similar to our RTD-based neuromorphic system. The small oscillations arise during the passage of the trajectories near a fold, due to the presence of a so-called folded singularity. The dynamics near the folded singularity is transient, yet recurrent: the trajectories return to the neighborhood of the folded singularity by way of a global return mechanism.

\begin{figure}
\centering{}\includegraphics[width=3.5in]{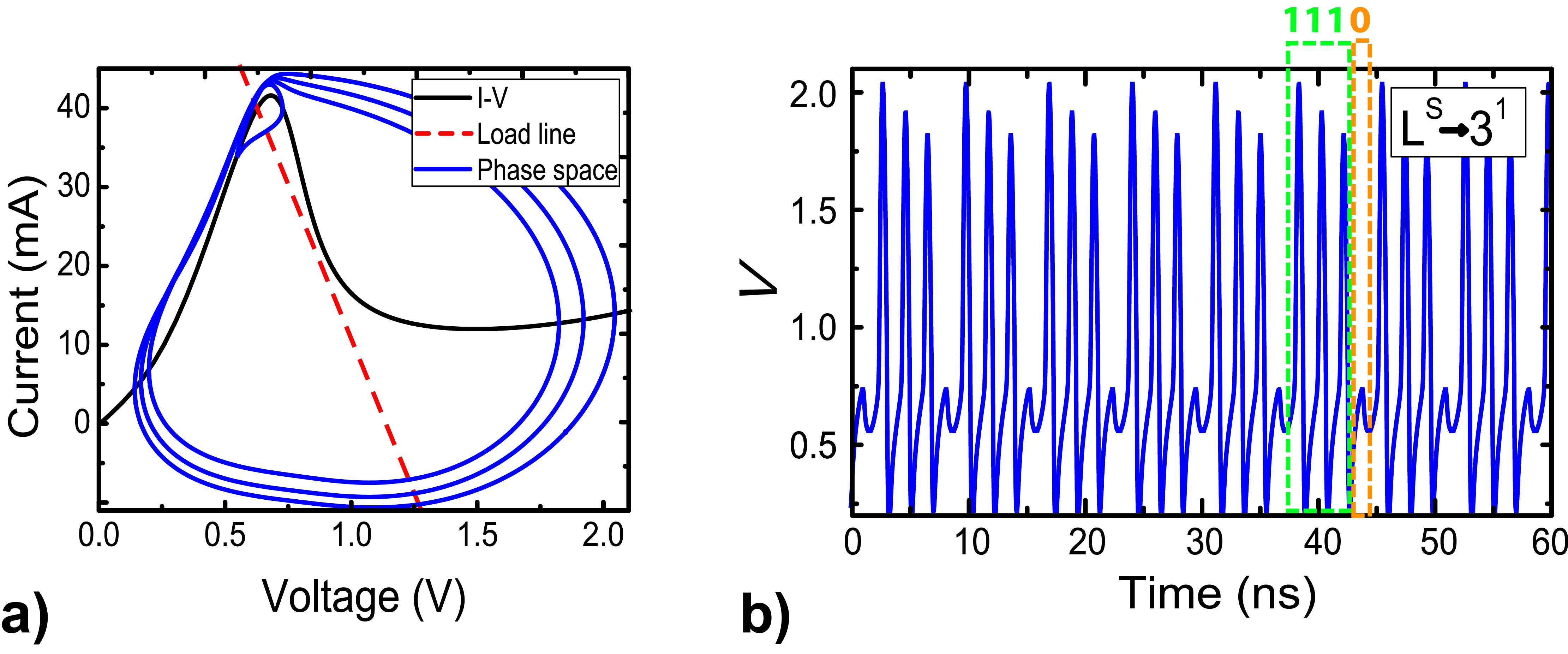}%
\caption{a) Nonlinear $I-V$ curve, $F(V)$, load line, and numerically simulated limit cycle of a $3^{1}$ MMO pattern. b) A typical time series of a numerically simulated $3^{1}$ MMO periodic pattern. Reproduced with permission from Proceedings of the International Conference on Numerical Simulation of Optoelectronic Devices, NUSOD (2014). Copyright 2014 IEEE. \label{mmo_model}}
\end{figure}

In order to understand the dynamics of the MMO pattern generation in the case of our neuromorphic optoelectronic resonator, we discuss the current-voltage phase space diagram displayed in Fig. \ref{mmo_model}a). For the chosen parameters, the load line intersects the $I-V$ in the first PDC region (rest state), but sufficiently close to the NDC region where self-sustained relaxation oscillations can occur. By choosing a periodic sinusoidal modulation, when the driving amplitude is below a given level of $V_{ac}$, the dynamics is only perturbed around the fix point, as shown in the upper left corner of the phase-space diagram. This corresponds to the small amplitude oscillations (binary '0'). If we increase the driving amplitude above a threshold value, oscillations of large amplitude (binary '1') can occur, corresponding to a large excursion in the limit cycle. Using this mechanism, MMO patterns of large and small ($L^{S}$) amplitude oscillations in the electrical domain are achieved. A typical waveform is displayed in Fig. \ref{mmo_model}b), showing a periodic $L^{S}\rightarrow 3^{1}$ MMO sequence with large amplitude signals of $V$ followed by one small amplitude oscillation in the other state. The two-state level operation can be associated with a binary encoding $\{0, 1\}$. The $L^{S}$ electrical current patterns directly modulate the laser diode, enabling identical MMO dynamics to occur in the LD intensity optical output.

\begin{figure}
  \centering
  \includegraphics[width=3.0in]{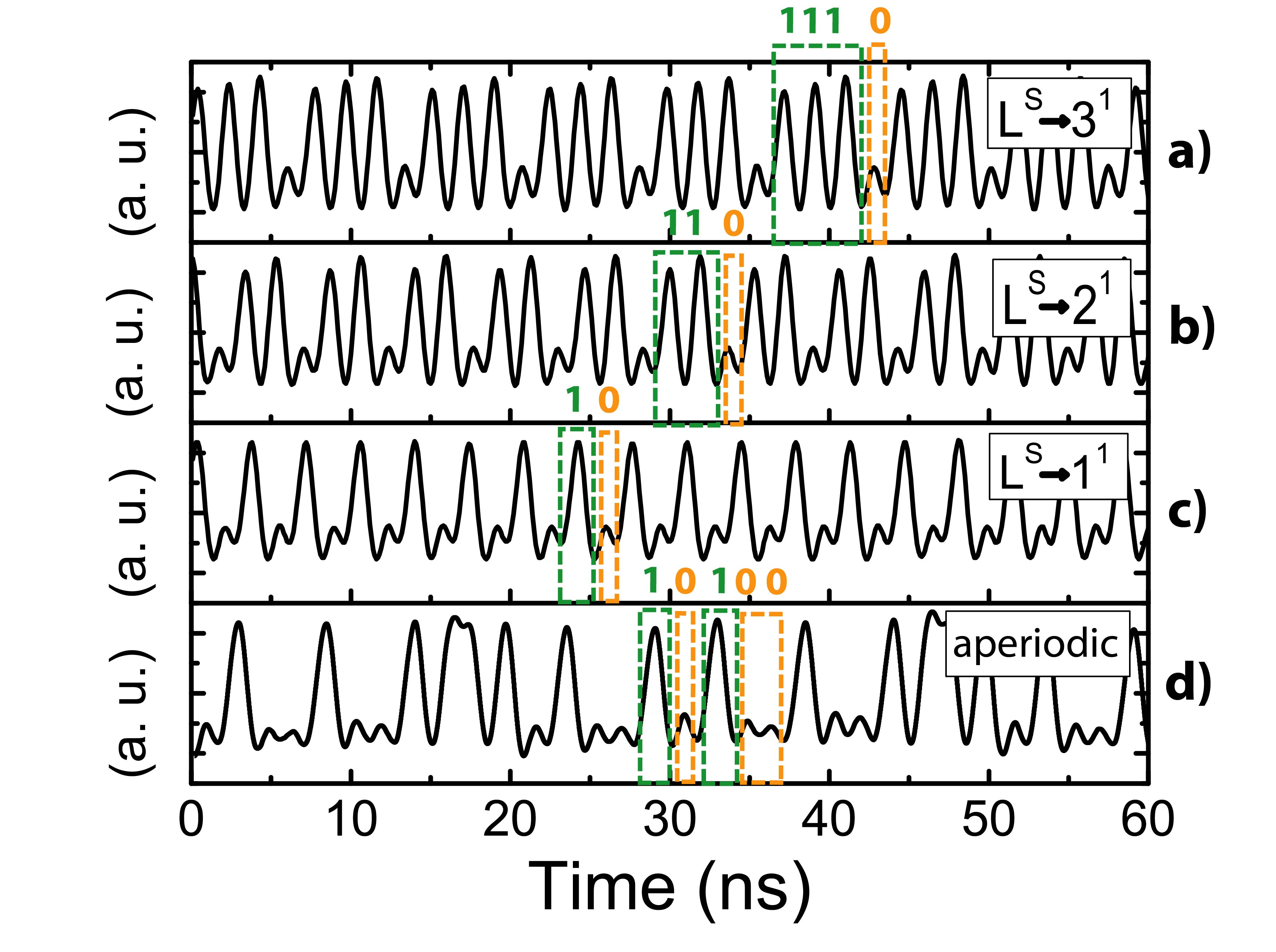}\\
  \caption{Experimental time traces of periodic and aperiodic MMOs triggered by an external electrical periodic signal with an amplitude of 281 mV. The different MMO patterns are activated as a function of the injected frequency: (a) 0.544 GHz, (b) 0.564 GHz, (c) 0.586 GHz, and (d) 0.581 GHz. Reproduced with permission from Proceedings of the International Conference on Numerical Simulation of Optoelectronic Devices, NUSOD (2014). Copyright 2014 IEEE.}\label{mmo}
\end{figure}

In Fig. \ref{mmo} is shown typical experimental time traces of MMOs patterns using a neuromorphic circuit similar to the one discussed in section IV-A2. In this experiment, in order to operate the circuit as a two-state level MMO pattern generator, the bias point was selected in the second PDC region. Depending on the driving frequency and amplitude of the external control ac signal (assuming a fixed dc bias point and circuit parameters) either periodic or aperiodic intermittent MMO patterns can be triggered. In Fig. \ref{mmo} we measured the response of the neuromorphic circuit to sinusoidal electrical signals in the range of 0.53 GHz to 0.63 GHz, that is, close to the natural oscillation frequency of the circuit, and recorded the laser photo-detected output using an oscilloscope. The results show that changing the periodic external signal will trigger an MMO sequence periodic of $L^{S}\rightarrow 3^{1} \rightarrow 2^{1} \rightarrow  1^{1}$, upon varying the frequency parameter. Between periodic MMOs, more complex patterns can be achieved including quasi-periodic and chaotic MMOs. In Fig. \ref{mmo}d) we show an example of an aperiodic sequence displaying random MMOs.

The MMO pattern generation can be used as an efficient switching method to modulate the laser intensity output for applications in pattern and random bit generation. Considering that small external perturbations (in frequency or amplitude) substantially change the dynamical characteristics of the MMO patterns, this can have novel applications in data encoding.

\subsection{Regenerative photonic memories}

Buffering of optical signals is desirable to avoid congestion of information traffic and realize efficient optical interconnects. Recently, all-optical buffer memories have been proposed based on slow-light delay lines, or using the Kerr nonlinearity on a standard silica optical fiber \cite{Leo2010} providing functionalities such as all-optical storage and reshaping. However, for realistic applications an optical buffer has to be compact for on-chip integration, which rules out most existing schemes, as they are not easily scalable to a millimeter size footprint and the writing process is often complex and costly. Therefore, configurations that combine the robustness of semiconductor nanoelectronic and optical devices with the wide-bandwidth of photonics elements offer significant advantages because they can provide small size, high-speed and low cost alternatives to the all-optical buffers currently proposed.

In this section, we review our work on regenerative photonic memories using the autaptic configuration introduced in Fig. 6c) and f). In this configuration, an optical delay line with time delay $\tau$ is inserted off-chip (in this case a low-loss single-mode optical fiber) providing a mechanism of re-injection of the fired excitable pulses analyzed in section IV-A, a mechanism analogous to an autaptic neuron. Our regenerative memory operates under the physical principle of excitable regeneration in which a pulse re-circulates indefinitely in an optical fiber delayed feedback loop enabling robust regenerative signal buffering and the potential for logical operations. The work on the solid-state autaptic neuron follows the strong interest that has been devoted during the past two decades on the effects of communication time delays in biological systems \cite{Buri2003,Stepan2009}, and how they can influence the synchronization dynamics between distant coupled neurons. For instance, it was shown that dynamical systems mimicking coupled neurons exhibit stable periodic pulsating regimes \cite{Yacomotti2002,Scholl2009,Kelleher2010,Weicker2014}, instead of a stable steady-state regime. Following the seminal work of Ikeda\cite{Ikeda1979}, the concept of pattern memorization in time delayed bistable systems was addressed in opto-electronic systems, see for instance \cite{Neyer1982,Aida1992}. Departing from these works, our work involves an excitable element, the RTD-PD resonator, as the nonlinear node. Because of this crucial difference, our approach benefits from the robustness and self-healing properties which are typical of neural signals and do not exist in bistable systems. Specifically, the excitable response of the nonlinear node guarantees a strong and well-defined all-or-nothing pulse response, almost identical in shape and duration to any supra-threshold incoming signal, enabling the reshaping and healing functionality of incoming signals which can be employed in the following critical telecommunication buffer functions: writing, storage, reshaping/healing, and XOR operation.


\begin{figure}
  \centering
  \includegraphics[width=3.3in]{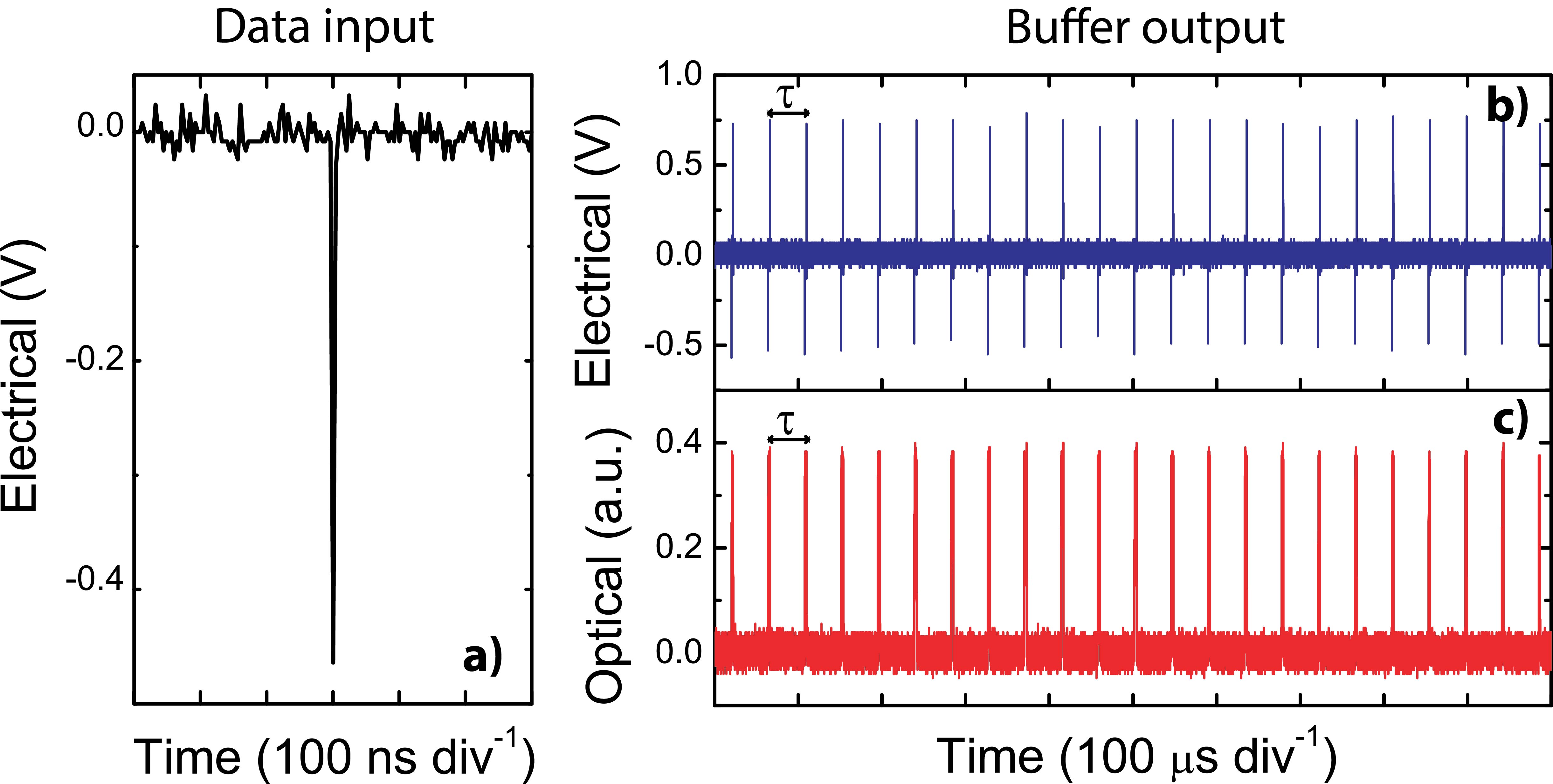}\\
  \caption{Experimental recorded time traces of the regenerative memory output showing writing and storage of a single-bit electrical pulse in a $\tau=46\,\mu$s cavity round-trip time. (Left) single bit data input; (b) electrical and (c) laser photo-detected optical outputs.}\label{one_bit_buffer}
\end{figure}

Figure \ref{one_bit_buffer} presents a typical example of experimental time traces showing the writing and storage of the regenerative memory using electrical injection of a single pulse bit, panel a). The binary-coded data streams were generated using an arbitrary function generator. For purposes of demonstration and experimental convenience, the data signals were injected electrically although the signals could be also injected optically, taking advantage of the optical input port of the RTD-PD. By operating our solid-state autaptic neuron in an excitable regime (close to the NDC region), we are able to store and regenerate optical bits of information in the fiber, the empty region signalling the "0" bits and the excitable optoelectronic pulses the "1". When these bits are re-injected into the RTD-PD they trigger the generation of a new excitable cycle. This regenerative mechanism occurs after each round trip in the fiber which is extremely robust since even if the bit sequence is strongly deteriorated, the all-or-none nonlinear response of the excitable RTD-PD enables the signal regeneration. In the experiments we used a low-loss optical fiber loop with a $46 \,\mu$s cavity round-trip time, $\tau$. The time $\tau$ was chosen in order that the memory buffer is much larger than the typical excitable lethargic time, $T_{l}$, of the RTD-PD-LD excitable system (in this case $T_{l}\sim$500 ns). In the results of Fig. \ref{one_bit_buffer}b), the regeneration lasted more than $10^{4}$ time round-trips, only limited by the acquisition time of the oscilloscope employed. Real-time measurements suggest that storage of a data stream can be achieved for several minutes without using any temperature or vibration controllers of both optical fiber and optoelectronic circuit.


\begin{figure}
  \centering
  \includegraphics[width=3.3in]{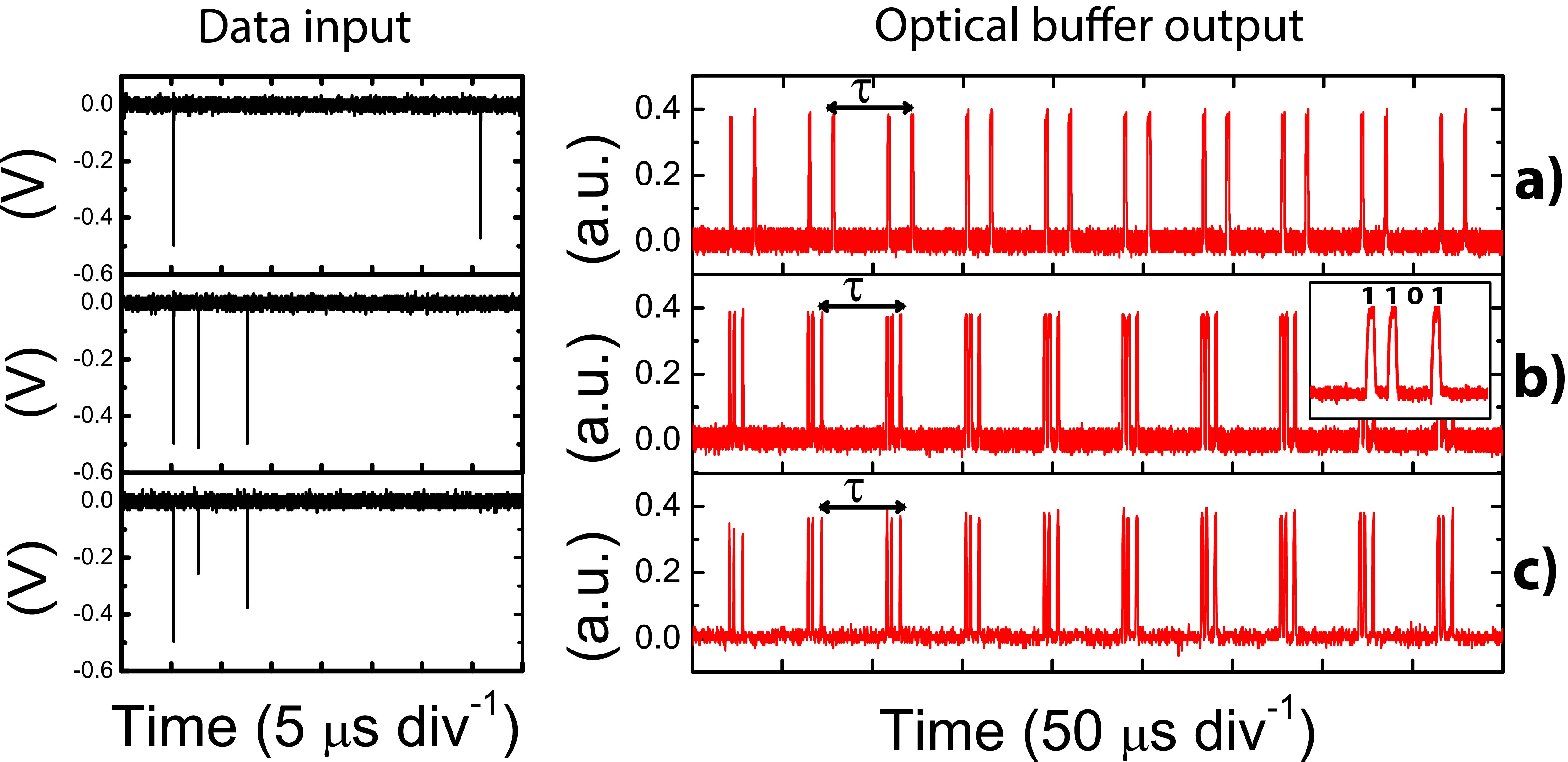}\\
  \caption{Experimental recorded time traces of the regenerative memory output showing writing and storage of binary-coded data streams in a $\tau=46\,\mu$s cavity round-trip time. (Left) data streams input, and (right) laser photo-detected optical buffer output. Sequence of (a) 2-bit (11), (b) 4-bit (1101), and (c) and 4-bit (1101) binary-coded data streams.}\label{bit_buffer}
\end{figure}

\begin{figure}
  \centering
  \includegraphics[width=3.0in]{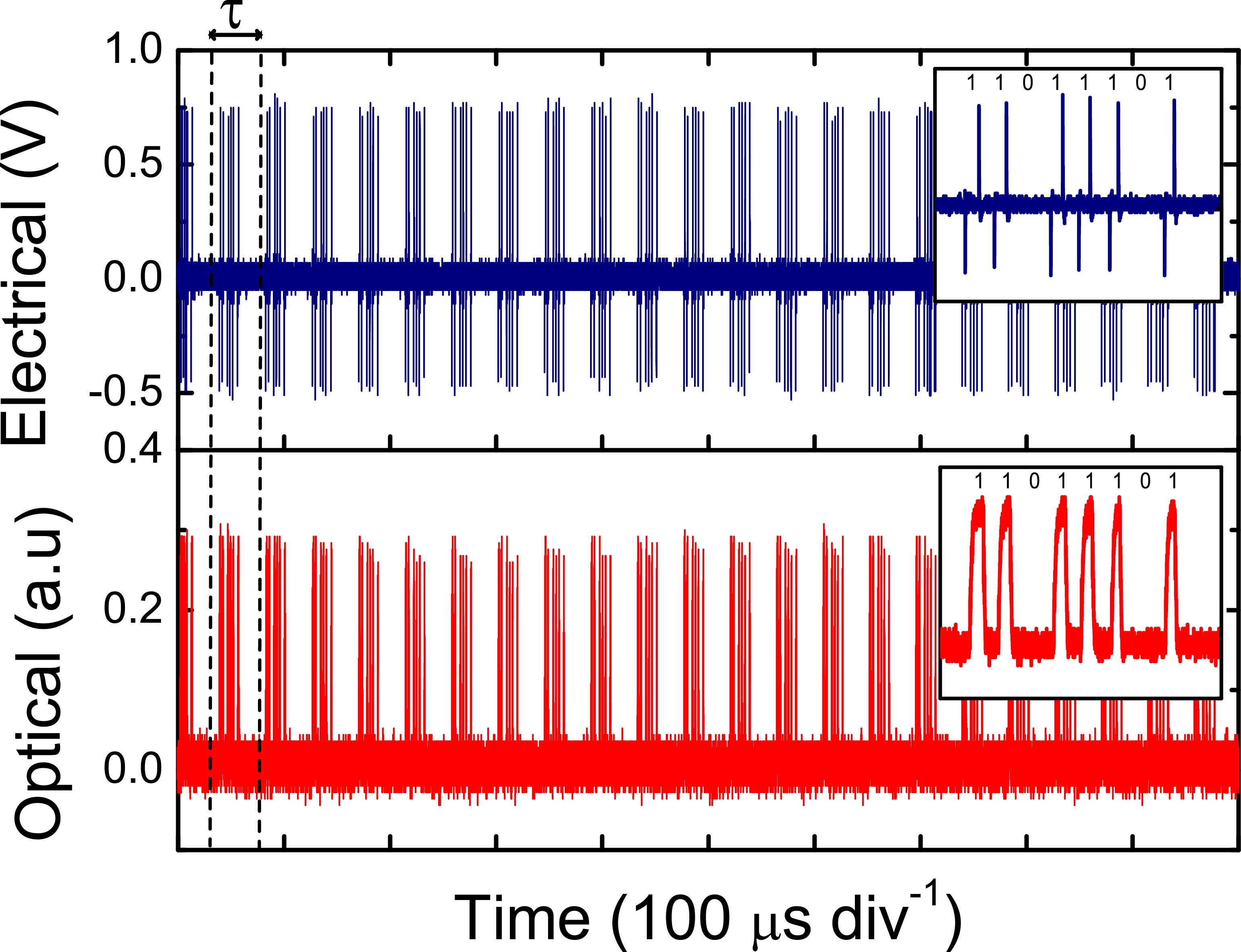}\\
  \caption{Experimental recorded time traces of the regenerative memory showing storage of an 8-bit binary-coded data stream (11011101) in a $\tau=46\,\mu$s cavity round-trip time.}\label{8_bit_buffer}
\end{figure}

In order to assess the robustness of the writing and storage process, a wide range of temporal bit patterns were tested. Figure \ref{bit_buffer} shows an example of complete regeneration using two-bit (11), panel a), and four-bit (1101), panel b), patterns. The bits must be separated by the lethargic time $T_{l}$ of the excitable system, which eventually defines the maximal bit rate as $T_{l}^{-1}$. The lethargic time induces a repulsion between nearest bits of data when they get too close, thereby ensuring signal integrity. Moreover, choosing a bit sequence where the amplitude of the bits are not evenly distributed, panel c), the system is able to perform single pass healing by restoring and self-adjusting the received bits to a fixed amplitude. Thus, the unique characteristics of the excitable response of our neuromorphic system enables the implementation of novel types of regenerative memories almost insensitive (in a certain range) to the exact shape or amplitude of the incoming data signals. As presented in Fig. \ref{8_bit_buffer} showing complete regeneration using a more complex pattern of 8-bits (11011101), the writing and storage process of the regenerative memory is extremely robust. In this situation, the regeneration was stable within the ms range.

It is noteworthy that the writing process of the regenerative memory can be scalable to multi-gigahertz operation, only limited by the laser diode frequency response. The information can be written with a single addressing pulse, either using electrical or optical incoming data, and can be stored in standard low-loss single mode silica optical fibres, which is very beneficial for practical applications.


\subsection{Time-delay FitzHugh-Nagumo model}

A precise modeling of the experimental situation can be achieved within
the framework of a dynamical model employing a Li\'{e}nard equation.
However, by assuming that the excitable response is slower than the
relaxation oscillation frequency of the laser, one can adiabatically
eliminate the laser intensity ($S$) that becomes slaved to the current
($I$) of the RTD. By expanding the nonlinear characteristic of the
RTD at the center of the negative differential resistance, denoting
$\delta V$ and $\delta I$ the deviation of the voltage and of the current,
neglecting the asymmetry of the NDC region, i.e. we assume it is symmetric,
one may reduce exactly the underlying physical model to the FHN model
with delayed feedback, making a complete link with our time-delayed
neuromorphic photonic system and the paradigm of excitability.
The time delayed FHN model reads

\begin{eqnarray}
\dot{\delta V} & = & \delta V-\frac{\delta V^{3}}{3}-\delta I+\eta\left[\delta I\left(t-\tau\right)-\delta I\right],\label{eq:FHN1}\\
\dot{\delta I} & = & \varepsilon\left(\beta+\delta V\right).\label{eq:FHN2}
\end{eqnarray}

The stiffness parameters $\varepsilon=\mu^2$ denotes the ratio of the time
scale governing the slow and the fast variables while $\beta$ is the effective bias parameter.
We choose $\beta>0$ without loss of generality. The influence of the delayed re-injection
of light, proportional to the current in the RTD, is taken into account
by the delayed term in Eq.~\ref{eq:FHN1}. The amplitude of the delayed
feedback is denoted $\eta$. For the sake of convenience we use the
so-called form of non invasive feedback \cite{P-PLA-92}. As such,
the steady states of the FHN model are unchanged by the presence of
feedback. If not otherwise stated the parameters are $\varepsilon=0.05$,
$\eta=0.18$ and $\tau=500$. We included white Gaussian noise of
variable amplitude $\xi$ to model the stochastic processes occurring
in our experimental neuromorphic photonic oscillator.

\subsection{Spatio-temporal localized structures}

The stored information in the optical delay line of the solid-state autaptic neuron is
composed of coherent photon packets that possess all the properties of localized
structures in delayed and spatially extended dynamical systems \cite{Giacomelli1996}.
Localized structures have been widely observed in nature such as in granular
media \cite{Umbanhowar1996}, gas discharges \cite{Astrov2001}, reaction-diffusion
systems \cite{Lee1994}, fluids \cite{Wu1994}, and convective systems \cite{Moses1987}.
Since the observation of LSs in semiconductor microcavities \cite{Barland2002}, LSs
analogues based in lasers have been exploited in periodically modulated and autonomous
delayed dynamical systems \cite{Marino2014,MJB-PRL-14,Garbin2015,MJB-NAP-15,JAH-PRL-15} paving the way to novel methods
of information storage, as proposed theoretically in \cite{Coullet2004}.

\begin{figure}
\centering{}\includegraphics[bb=35bp 0bp 420bp 320bp,clip,width=0.8\columnwidth]{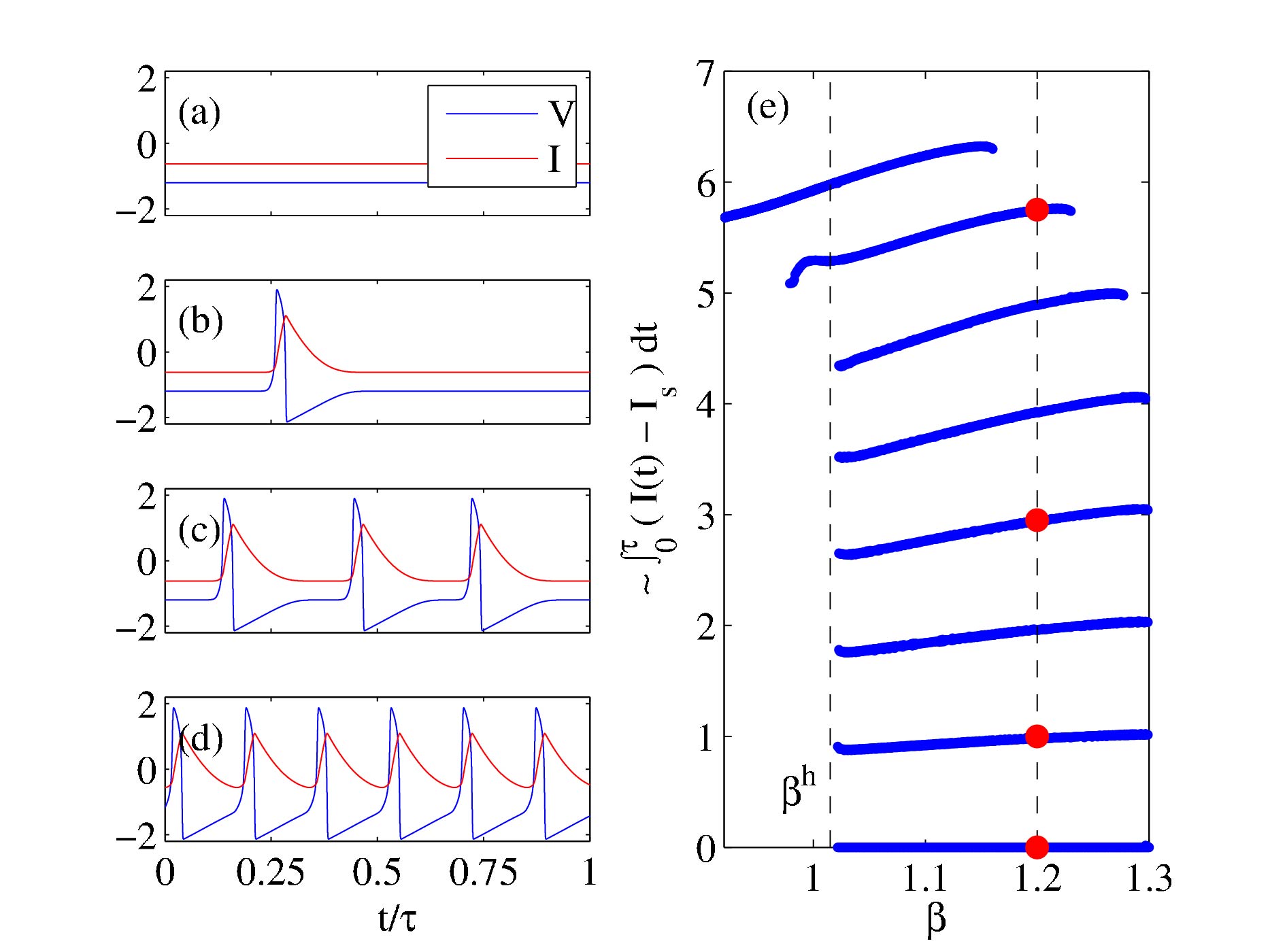}
\caption{Temporal time traces over a single period for various states of the
regenerative memory (a-d) and multi-stability diagram of the coexisting
solutions (e). In (e) we represent some norm of the solutions as the
integral over one period of the deviation of the slow variable ($I-I_{s}$)
yielding upward pulses with a zero background. This integral is normalized
to a value of $1$ when there is a single pulse at $\beta=1.3$. All
the localized solutions becomes unstable close to $\beta^{h}\sim1.018$
where the background get destabilized trough an Andronov-Hopf bifurcation.
Reproduced with permission from Scientific Reports 6, 19510 (2016). Licensed under CC BY.}
\label{diagLS}
\end{figure}

\begin{figure*}
  \includegraphics[width=0.8\textwidth]{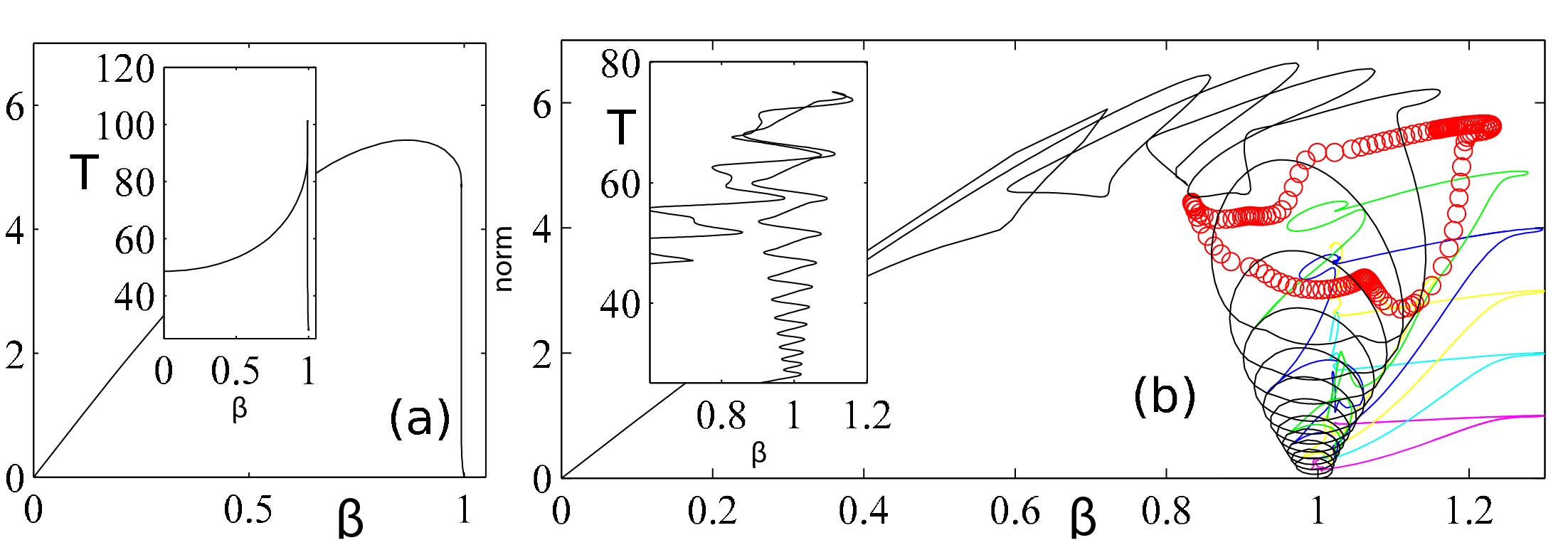}
\caption{Bifurcation theoretical analysis of formation of multiple states of memory. Left: Amplitude of the periodic solutions and variation of the period
along the branches of (inset) for the FHN system without feedback
i.e. $\eta=0$. Right: Same diagram with $\eta=0.18$ the colors correspond
to the branches of solutions with 1,2,...7, equi-spaced localized
structures. Reproduced with permission from Scientific Reports 6, 19510 (2016). Licensed under CC BY.}
\label{ddebif}       
\end{figure*}


We exemplify in Fig.~\ref{diagLS}a-d) various occurrences of such
periodic regimes whose period are close to $\tau$ and that are composed
of 0, 1, 3 and 6 bits of information as embedded excitable responses
within the time delay $\tau$. We stress in Figure~\ref{diagLS}e)
that all these regimes coexist between themselves for a wide range
of the bias parameter $\beta$. Because these isolated temporal patterns
are bistable with the uniform state and are also independent of the
boundary conditions, i.e. the time delay value, and are attractors
of the dynamics, they can be considered as the equivalent of Localized
Structures in time delayed systems.

Interestingly, each branch of solution
corresponds to a well defined number of temporal LSs. However, what
is hidden in such a projection is that for a given number of bits,
i.e. a given branch, an infinity of different arrangements and relative
distances exists. The storage capacity of our regenerative memory
can be understood intuitively. Since the temporal extension of the
excitable orbit is defined by its lethargic time $T_{l}$, the maximal
amount of elements that can be stored in the time delay is the integer
closest to $N\sim\tau/T_{l}$.

The full bifurcation diagram of the multi-LSs solutions was obtained
with DDE-BIFTOOL \cite{DDEBT} and is depicted in Fig.~\ref{ddebif}b.
The continuation of such solutions proved to be particularly challenging numerically
and we were only able to study the equidistant multi-LSs solutions, see \cite{Romeira2016}
for more details.
We notice first that the simple bifurcation scenario found without
feedback in  Fig.~\ref{ddebif}a changes dramatically with $\eta\neq0$.
The dominant periodic branch that corresponds to the canard blow-up (in black) develops
a large number of folds as apparent in Fig.~\ref{ddebif}b. We stress
that only the upper part of this folded branch is stable and that
at, e.g. at $\beta=1.1$, it corresponds to the solution with a maximal
number of LS within the time delay, i.e. the trace depicted in Fig.~\ref{diagLS}d
with $N=7$. By analogy with the terminology of spatially extended
systems we denote this particular temporal trace without empty regions
as the fully developed pattern. We stress that in this regime, all
the $N=7$ pulses are interacting via their tails and can hardly be
considered as independent. For instance, erasing an individual pulse
would result in a smooth rearrangement of the other $N=6$ remaining
LSs in order to minimize their residual repulsive interactions.
The other branches represented in colors in Fig.~\ref{ddebif}b correspond to
solution with $N\in[1,6]\,$LSs. We refer the reader to \cite{Romeira2016} for the
details regarding this analysis.

\begin{figure}
\centering{}\includegraphics[clip,width=0.8\columnwidth]{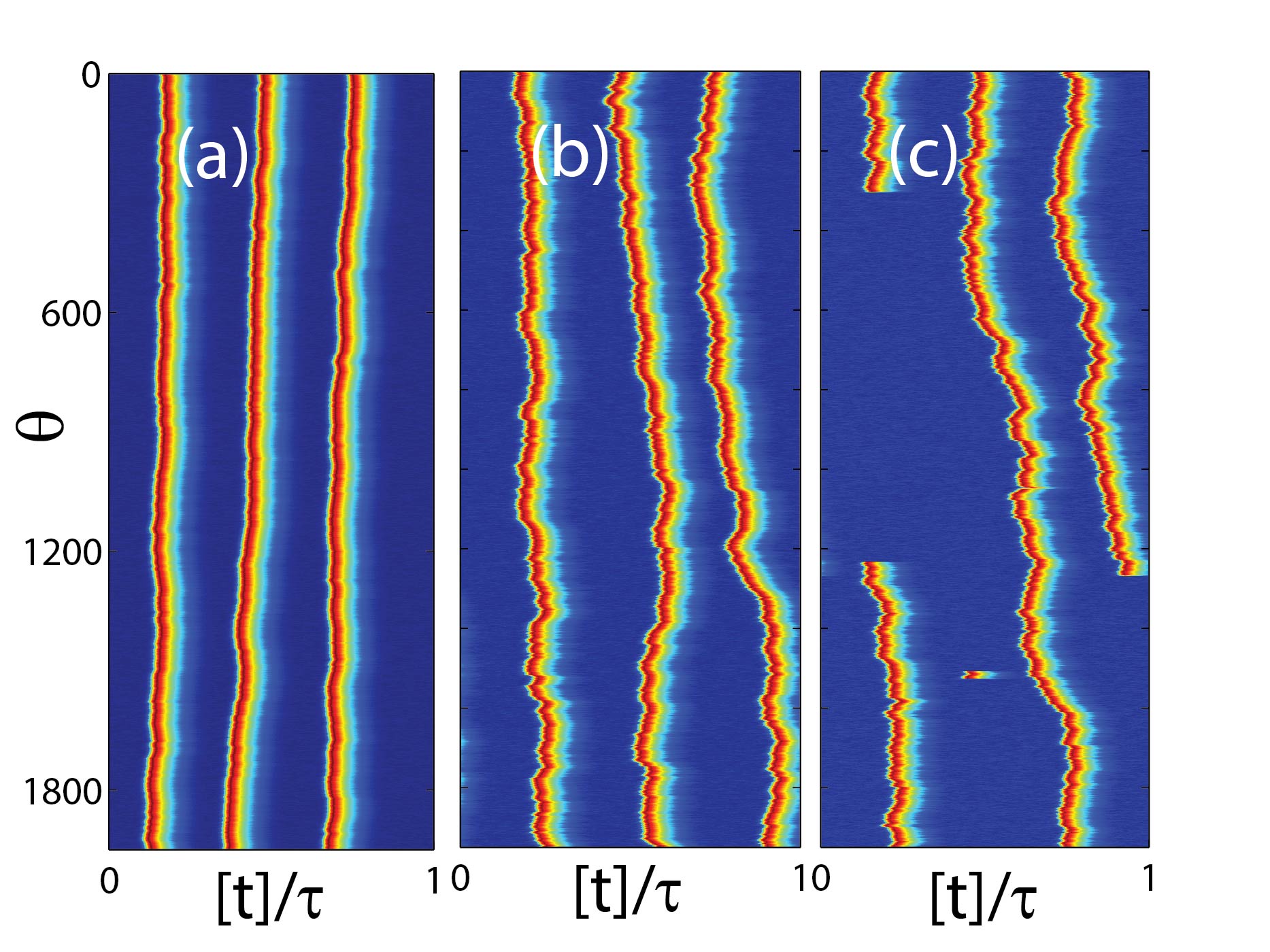}\caption{ The space-time plots represent the evolution of three LSs for an increasing
level of noise. The slow variable represented is $\delta I$. In panel a)
$\xi=10^{-3}$ and the noise induced drift motion is barely visible
over more than $\theta=2000$ periods. In panel b) the uncorrelated
random walk is more visible since $\xi=5\times10^{-3}$ while in c)
the huge level of noise $\xi=8\times10^{-3}$ is capable of inducing
annihilation and nucleation of the LS. \label{FigSpaceTime} }
\end{figure}

As another proof of the mutual independence of the temporal LSs, one can study
their relative motion in presence of noise. For that purpose, we use a two
dimensional pseudo-spatial representation in which the relative motion of the
various LSs are best observed. In this co-moving reference frame, the horizontal
axis is a space-like coordinate that allows to localize the position of the pulses
within a given round-trip while the vertical coordinate corresponds to the slow temporal
evolution of the system over many round-trips, see \cite{Romeira2016} for details.
The mutual independence of these pulses is demonstrated for instance by their uncorrelated
random motion in the presence of noise. We also show the nucleation and annihilation
process in Fig.~\ref{FigSpaceTime}c) simply by including an extremely large amount of noise.
One also notice here a property of the utmost importance:
the almost total absence of transients: the localized bits of information
can be perfectly written and erased in a single round-trip, at variance
with e.g. the results of \cite{LCK-NAP-10,MJB-PRL-14,CJM-PRA-16} where transients
representing tens of round-trip are necessary before a stabilization
of the waveform. From an application point of view, our time-delayed
neuromorphic photonic memory presents the extraordinary advantage
to allow writing and erasing information at a rate comparable to the
nominal reading rate.

\section{Perspectives and applications}
The growing of artificial intelligence and the recognition by the semiconductor industry that the Moore's law is near its end is signaling a clear urgency for disruptive scientific concepts and technologies to keep up with the demands of low energy consumption and ultrafast computing and communications systems. Neuromorphic technologies combined with optics offer great promise for implementing compact, low power consumption artificial brain-inspired computing systems with real-time learning abilities.

Here, we have reviewed our recent work covering photonic memory circuits inspired by the biophysics of neurons. They consist of an integrated high-speed nanoscale nonlinear resonator and photo-detector, the resonant tunneling diode photo-detector (RTD-PD), a laser diode (LD) and an optical fiber, operating at telecommunications wavelengths ($\sim$1550 nm). This optoelectronic configuration forms a new type of high-speed nonlinear neuromorphic photonic resonator microchip, which emulates the biophysics of real spiking neurons and dynamic synapses. The neuron-inspired photonic memory was demonstrated by sending a sequence of random bits of information that triggered the RTD-PD electro-optical spiking excitable response. Excitability is a nonlinear mechanism that not only governs the connections between neurons but also commands the rhythmic action of cardiac cells. Then, by delaying the excitable response using an optical fiber loop before re-injecting back into the RTD-PD input, we have demonstrated a regenerative memory as a result of the time-delayed feedback of the spiking nonlinear response of the system onto itself.

The impacts of this work are twofold. First, in our experiments we were able to write, reshape and store photon packets of information using conventional optical fibers at speeds much faster than the typical responses of biological neurons. The results are very promising for the development of disruptive brain-inspired high bit-rate (Gb/s) optical communications systems. Second, we have demonstrated that the stored information in the optical fiber is composed of coherent photon packets that possess all the properties of localized structures. Such states are abundant in nature and can be found in systems like granular media, semiconductors, fluids, and optical cavities. Their ubiquitous physical properties, namely mutual independence, as demonstrated by their uncorrelated random walk motion in the presence of noise, enables a robust and flexible memory operation by easily addressing, creating and destroying individual localized pulsed patterns of information.

Exploiting such a peculiar neuron-inspired electro-optical response using photonics is a conceptual breakthrough that has been almost unexplored in the field of optical data communications and could open up new applications, e.g. clock and timing, optical memories and data buffers.

\begin{acknowledgments}
This work is partially supported by the Funda\c{c}\~{a}o para a Ci\^{e}ncia e a Tecnologia (FCT) under the project UID/Multi/00631/2013, by the European Structural and Investment Funds (FEEI) through the Competitiveness and Internationalization Operational Program - COMPETE 2020 and by National Funds through FCT under the project ALG-01-0145-FEDER-016432/POCI-01-0145-FEDER-016432, and by the European Commission under the project iBROW (grant agreement no. 645369). J.J. acknowledges financial support project COMBINA (TEC2015-65212-C3-3-PAEI/FEDER UE) and the Ram\'{o}n y Cajal fellowship. We thank Charles Ironside, Curtin University, for the fruitful discussions on RTD optoelectronic devices, Oreste Piro, Universitat de les Illes Balears, for the useful discussions on excitable systems, and Raquel Lu\'{i}s for the artistic representation of the neuron.
\end{acknowledgments}





\begin{thebibliography}{114}%
\makeatletter
\providecommand \@ifxundefined [1]{%
 \@ifx{#1\undefined}
}%
\providecommand \@ifnum [1]{%
 \ifnum #1\expandafter \@firstoftwo
 \else \expandafter \@secondoftwo
 \fi
}%
\providecommand \@ifx [1]{%
 \ifx #1\expandafter \@firstoftwo
 \else \expandafter \@secondoftwo
 \fi
}%
\providecommand \natexlab [1]{#1}%
\providecommand \enquote  [1]{``#1''}%
\providecommand \bibnamefont  [1]{#1}%
\providecommand \bibfnamefont [1]{#1}%
\providecommand \citenamefont [1]{#1}%
\providecommand \href@noop [0]{\@secondoftwo}%
\providecommand \href [0]{\begingroup \@sanitize@url \@href}%
\providecommand \@href[1]{\@@startlink{#1}\@@href}%
\providecommand \@@href[1]{\endgroup#1\@@endlink}%
\providecommand \@sanitize@url [0]{\catcode `\\12\catcode `\$12\catcode
  `\&12\catcode `\#12\catcode `\^12\catcode `\_12\catcode `\%12\relax}%
\providecommand \@@startlink[1]{}%
\providecommand \@@endlink[0]{}%
\providecommand \url  [0]{\begingroup\@sanitize@url \@url }%
\providecommand \@url [1]{\endgroup\@href {#1}{\urlprefix }}%
\providecommand \urlprefix  [0]{URL }%
\providecommand \Eprint [0]{\href }%
\providecommand \doibase [0]{http://dx.doi.org/}%
\providecommand \selectlanguage [0]{\@gobble}%
\providecommand \bibinfo  [0]{\@secondoftwo}%
\providecommand \bibfield  [0]{\@secondoftwo}%
\providecommand \translation [1]{[#1]}%
\providecommand \BibitemOpen [0]{}%
\providecommand \bibitemStop [0]{}%
\providecommand \bibitemNoStop [0]{.\EOS\space}%
\providecommand \EOS [0]{\spacefactor3000\relax}%
\providecommand \BibitemShut  [1]{\csname bibitem#1\endcsname}%
\let\auto@bib@innerbib\@empty
\bibitem [{\citenamefont {Jang}(1993)}]{Jang1993}%
  \BibitemOpen
  \bibfield  {author} {\bibinfo {author} {\bibfnamefont {J.~S.~R.}\
  \bibnamefont {Jang}},\ }\href {\doibase 10.1109/21.256541} {\enquote
  {\bibinfo {title} {{ANFIS: adaptive-network-based fuzzy inference system}},}\
  } (\bibinfo {year} {1993})\BibitemShut {NoStop}%
\bibitem [{\citenamefont {Rowley}, \citenamefont {Baluja},\ and\ \citenamefont
  {Kanade}(1998)}]{Rowley1998}%
  \BibitemOpen
  \bibfield  {author} {\bibinfo {author} {\bibfnamefont {H.~A.}\ \bibnamefont
  {Rowley}}, \bibinfo {author} {\bibfnamefont {S.}~\bibnamefont {Baluja}}, \
  and\ \bibinfo {author} {\bibfnamefont {T.}~\bibnamefont {Kanade}},\ }\href
  {\doibase 10.1109/34.655647} {\enquote {\bibinfo {title} {{Neural
  network-based face detection}},}\ } (\bibinfo {year} {1998})\BibitemShut
  {NoStop}%
\bibitem [{\citenamefont {Huang}, \citenamefont {Zhu},\ and\ \citenamefont
  {Siew}(2006)}]{Huang2006}%
  \BibitemOpen
  \bibfield  {author} {\bibinfo {author} {\bibfnamefont {G.-B.}\ \bibnamefont
  {Huang}}, \bibinfo {author} {\bibfnamefont {Q.-Y.}\ \bibnamefont {Zhu}}, \
  and\ \bibinfo {author} {\bibfnamefont {C.-K.}\ \bibnamefont {Siew}},\
  }\bibfield  {title} {\enquote {\bibinfo {title} {{Extreme learning machine:
  Theory and applications}},}\ }\href {\doibase
  http://dx.doi.org/10.1016/j.neucom.2005.12.126} {\bibfield  {journal}
  {\bibinfo  {journal} {Neurocomputing}\ }\textbf {\bibinfo {volume} {70}},\
  \bibinfo {pages} {489--501} (\bibinfo {year} {2006})}\BibitemShut {NoStop}%
\bibitem [{\citenamefont {Lichtsteiner}, \citenamefont {Posch},\ and\
  \citenamefont {Delbruck}(2008)}]{Lichtsteiner2008}%
  \BibitemOpen
  \bibfield  {author} {\bibinfo {author} {\bibfnamefont {P.}~\bibnamefont
  {Lichtsteiner}}, \bibinfo {author} {\bibfnamefont {C.}~\bibnamefont {Posch}},
  \ and\ \bibinfo {author} {\bibfnamefont {T.}~\bibnamefont {Delbruck}},\
  }\href {\doibase 10.1109/JSSC.2007.914337} {\enquote {\bibinfo {title} {{A
  128x128 120 dB 15 us Latency Asynchronous Temporal Contrast Vision
  Sensor}},}\ } (\bibinfo {year} {2008})\BibitemShut {NoStop}%
\bibitem [{\citenamefont {Shen}\ \emph {et~al.}(2017)\citenamefont {Shen},
  \citenamefont {Harris}, \citenamefont {Skirlo}, \citenamefont {Prabhu},
  \citenamefont {Baehr-Jones}, \citenamefont {Hochberg}, \citenamefont {Sun},
  \citenamefont {Zhao}, \citenamefont {Larochelle}, \citenamefont {Englund},\
  and\ \citenamefont {Soljacic}}]{Shen2016}%
  \BibitemOpen
  \bibfield  {author} {\bibinfo {author} {\bibfnamefont {Y.}~\bibnamefont
  {Shen}}, \bibinfo {author} {\bibfnamefont {N.}~\bibnamefont {Harris}},
  \bibinfo {author} {\bibfnamefont {S.}~\bibnamefont {Skirlo}}, \bibinfo
  {author} {\bibfnamefont {M.}~\bibnamefont {Prabhu}}, \bibinfo {author}
  {\bibfnamefont {T.}~\bibnamefont {Baehr-Jones}}, \bibinfo {author}
  {\bibfnamefont {M.}~\bibnamefont {Hochberg}}, \bibinfo {author}
  {\bibfnamefont {X.}~\bibnamefont {Sun}}, \bibinfo {author} {\bibfnamefont
  {S.}~\bibnamefont {Zhao}}, \bibinfo {author} {\bibfnamefont {H.}~\bibnamefont
  {Larochelle}}, \bibinfo {author} {\bibfnamefont {D.}~\bibnamefont {Englund}},
  \ and\ \bibinfo {author} {\bibfnamefont {M.}~\bibnamefont {Soljacic}},\
  }\bibfield  {title} {\enquote {\bibinfo {title} {{Deep learning with coherent
  nanophotonic circuits}},}\ }\href@noop {} {\bibfield  {journal} {\bibinfo
  {journal} {Nature Photonics}\ }\textbf {\bibinfo {volume} {11}},\ \bibinfo
  {pages} {441–446} (\bibinfo {year} {2017})}\BibitemShut {NoStop}%
\bibitem [{\citenamefont {Hinton}\ and\ \citenamefont
  {Salakhutdinov}(2006)}]{Hinton2006}%
  \BibitemOpen
  \bibfield  {author} {\bibinfo {author} {\bibfnamefont {G.~E.}\ \bibnamefont
  {Hinton}}\ and\ \bibinfo {author} {\bibfnamefont {R.~R.}\ \bibnamefont
  {Salakhutdinov}},\ }\bibfield  {title} {\enquote {\bibinfo {title} {{Reducing
  the Dimensionality of Data with Neural Networks}},}\ }\href {\doibase
  10.1126/science.1127647} {\bibfield  {journal} {\bibinfo  {journal}
  {Science}\ }\textbf {\bibinfo {volume} {313}},\ \bibinfo {pages} {504--507}
  (\bibinfo {year} {2006})}\BibitemShut {NoStop}%
\bibitem [{\citenamefont {LeCun}, \citenamefont {Bengio},\ and\ \citenamefont
  {Hinton}(2015)}]{Lecun2015}%
  \BibitemOpen
  \bibfield  {author} {\bibinfo {author} {\bibfnamefont {Y.}~\bibnamefont
  {LeCun}}, \bibinfo {author} {\bibfnamefont {Y.}~\bibnamefont {Bengio}}, \
  and\ \bibinfo {author} {\bibfnamefont {G.}~\bibnamefont {Hinton}},\
  }\bibfield  {title} {\enquote {\bibinfo {title} {{Deep learning}},}\ }\href
  {http://dx.doi.org/10.1038/nature14539 http://10.0.4.14/nature14539}
  {\bibfield  {journal} {\bibinfo  {journal} {Nature}\ }\textbf {\bibinfo
  {volume} {521}},\ \bibinfo {pages} {436--444} (\bibinfo {year}
  {2015})}\BibitemShut {NoStop}%
\bibitem [{\citenamefont {Bengio}, \citenamefont {Courville},\ and\
  \citenamefont {Vincent}(2013)}]{Bengio2013}%
  \BibitemOpen
  \bibfield  {author} {\bibinfo {author} {\bibfnamefont {Y.}~\bibnamefont
  {Bengio}}, \bibinfo {author} {\bibfnamefont {A.}~\bibnamefont {Courville}}, \
  and\ \bibinfo {author} {\bibfnamefont {P.}~\bibnamefont {Vincent}},\ }\href
  {\doibase 10.1109/TPAMI.2013.50} {\enquote {\bibinfo {title} {{Representation
  Learning: A Review and New Perspectives}},}\ } (\bibinfo {year}
  {2013})\BibitemShut {NoStop}%
\bibitem [{\citenamefont {Indiveri}\ \emph {et~al.}(2011)\citenamefont
  {Indiveri}, \citenamefont {Linares-Barranco}, \citenamefont {Hamilton},
  \citenamefont {van Schaik}, \citenamefont {Etienne-Cummings}, \citenamefont
  {Delbruck}, \citenamefont {Liu}, \citenamefont {Dudek}, \citenamefont
  {H{\"{a}}fliger}, \citenamefont {Renaud}, \citenamefont {Schemmel},
  \citenamefont {Cauwenberghs}, \citenamefont {Arthur}, \citenamefont {Hynna},
  \citenamefont {Folowosele}, \citenamefont {SA{\"{I}}GHI}, \citenamefont
  {Serrano-Gotarredona}, \citenamefont {Wijekoon}, \citenamefont {Wang},\ and\
  \citenamefont {Boahen}}]{Indiveri2011}%
  \BibitemOpen
  \bibfield  {author} {\bibinfo {author} {\bibfnamefont {G.}~\bibnamefont
  {Indiveri}}, \bibinfo {author} {\bibfnamefont {B.}~\bibnamefont
  {Linares-Barranco}}, \bibinfo {author} {\bibfnamefont {T.}~\bibnamefont
  {Hamilton}}, \bibinfo {author} {\bibfnamefont {A.}~\bibnamefont {van
  Schaik}}, \bibinfo {author} {\bibfnamefont {R.}~\bibnamefont
  {Etienne-Cummings}}, \bibinfo {author} {\bibfnamefont {T.}~\bibnamefont
  {Delbruck}}, \bibinfo {author} {\bibfnamefont {S.-C.}\ \bibnamefont {Liu}},
  \bibinfo {author} {\bibfnamefont {P.}~\bibnamefont {Dudek}}, \bibinfo
  {author} {\bibfnamefont {P.}~\bibnamefont {H{\"{a}}fliger}}, \bibinfo
  {author} {\bibfnamefont {S.}~\bibnamefont {Renaud}}, \bibinfo {author}
  {\bibfnamefont {J.}~\bibnamefont {Schemmel}}, \bibinfo {author}
  {\bibfnamefont {G.}~\bibnamefont {Cauwenberghs}}, \bibinfo {author}
  {\bibfnamefont {J.}~\bibnamefont {Arthur}}, \bibinfo {author} {\bibfnamefont
  {K.}~\bibnamefont {Hynna}}, \bibinfo {author} {\bibfnamefont
  {F.}~\bibnamefont {Folowosele}}, \bibinfo {author} {\bibfnamefont
  {S.}~\bibnamefont {SA{\"{I}}GHI}}, \bibinfo {author} {\bibfnamefont
  {T.}~\bibnamefont {Serrano-Gotarredona}}, \bibinfo {author} {\bibfnamefont
  {J.}~\bibnamefont {Wijekoon}}, \bibinfo {author} {\bibfnamefont
  {Y.}~\bibnamefont {Wang}}, \ and\ \bibinfo {author} {\bibfnamefont
  {K.}~\bibnamefont {Boahen}},\ }\bibfield  {title} {\enquote {\bibinfo {title}
  {{Neuromorphic Silicon Neuron Circuits}},}\ }\href {\doibase
  10.3389/fnins.2011.00073} {\bibfield  {journal} {\bibinfo  {journal}
  {Frontiers in Neuroscience}\ }\textbf {\bibinfo {volume} {5}},\ \bibinfo
  {pages} {73} (\bibinfo {year} {2011})}\BibitemShut {NoStop}%
\bibitem [{\citenamefont {Merolla}\ \emph {et~al.}(2014)\citenamefont
  {Merolla}, \citenamefont {Arthur}, \citenamefont {Alvarez-Icaza},
  \citenamefont {Cassidy}, \citenamefont {Sawada}, \citenamefont {Akopyan},
  \citenamefont {Jackson}, \citenamefont {Imam}, \citenamefont {Guo},
  \citenamefont {Nakamura}, \citenamefont {Brezzo}, \citenamefont {Vo},
  \citenamefont {Esser}, \citenamefont {Appuswamy}, \citenamefont {Taba},
  \citenamefont {Amir}, \citenamefont {Flickner}, \citenamefont {Risk},
  \citenamefont {Manohar},\ and\ \citenamefont {Modha}}]{Merolla2014}%
  \BibitemOpen
  \bibfield  {author} {\bibinfo {author} {\bibfnamefont {P.~A.}\ \bibnamefont
  {Merolla}}, \bibinfo {author} {\bibfnamefont {J.~V.}\ \bibnamefont {Arthur}},
  \bibinfo {author} {\bibfnamefont {R.}~\bibnamefont {Alvarez-Icaza}}, \bibinfo
  {author} {\bibfnamefont {A.~S.}\ \bibnamefont {Cassidy}}, \bibinfo {author}
  {\bibfnamefont {J.}~\bibnamefont {Sawada}}, \bibinfo {author} {\bibfnamefont
  {F.}~\bibnamefont {Akopyan}}, \bibinfo {author} {\bibfnamefont {B.~L.}\
  \bibnamefont {Jackson}}, \bibinfo {author} {\bibfnamefont {N.}~\bibnamefont
  {Imam}}, \bibinfo {author} {\bibfnamefont {C.}~\bibnamefont {Guo}}, \bibinfo
  {author} {\bibfnamefont {Y.}~\bibnamefont {Nakamura}}, \bibinfo {author}
  {\bibfnamefont {B.}~\bibnamefont {Brezzo}}, \bibinfo {author} {\bibfnamefont
  {I.}~\bibnamefont {Vo}}, \bibinfo {author} {\bibfnamefont {S.~K.}\
  \bibnamefont {Esser}}, \bibinfo {author} {\bibfnamefont {R.}~\bibnamefont
  {Appuswamy}}, \bibinfo {author} {\bibfnamefont {B.}~\bibnamefont {Taba}},
  \bibinfo {author} {\bibfnamefont {A.}~\bibnamefont {Amir}}, \bibinfo {author}
  {\bibfnamefont {M.~D.}\ \bibnamefont {Flickner}}, \bibinfo {author}
  {\bibfnamefont {W.~P.}\ \bibnamefont {Risk}}, \bibinfo {author}
  {\bibfnamefont {R.}~\bibnamefont {Manohar}}, \ and\ \bibinfo {author}
  {\bibfnamefont {D.~S.}\ \bibnamefont {Modha}},\ }\bibfield  {title} {\enquote
  {\bibinfo {title} {{A million spiking-neuron integrated circuit with a
  scalable communication network and interface}},}\ }\href
  {http://science.sciencemag.org/content/345/6197/668.abstract} {\bibfield
  {journal} {\bibinfo  {journal} {Science}\ }\textbf {\bibinfo {volume}
  {345}},\ \bibinfo {pages} {668 LP -- 673} (\bibinfo {year}
  {2014})}\BibitemShut {NoStop}%
\bibitem [{\citenamefont {Prezioso}\ \emph {et~al.}(2015)\citenamefont
  {Prezioso}, \citenamefont {Merrikh-Bayat}, \citenamefont {Hoskins},
  \citenamefont {Adam}, \citenamefont {Likharev},\ and\ \citenamefont
  {Strukov}}]{Prezioso2015}%
  \BibitemOpen
  \bibfield  {author} {\bibinfo {author} {\bibfnamefont {M.}~\bibnamefont
  {Prezioso}}, \bibinfo {author} {\bibfnamefont {F.}~\bibnamefont
  {Merrikh-Bayat}}, \bibinfo {author} {\bibfnamefont {B.~D.}\ \bibnamefont
  {Hoskins}}, \bibinfo {author} {\bibfnamefont {G.~C.}\ \bibnamefont {Adam}},
  \bibinfo {author} {\bibfnamefont {K.~K.}\ \bibnamefont {Likharev}}, \ and\
  \bibinfo {author} {\bibfnamefont {D.~B.}\ \bibnamefont {Strukov}},\
  }\bibfield  {title} {\enquote {\bibinfo {title} {{Training and operation of
  an integrated neuromorphic network based on metal-oxide memristors}},}\
  }\href {http://dx.doi.org/10.1038/nature14441 http://10.0.4.14/nature14441
  http://www.nature.com/nature/journal/v521/n7550/abs/nature14441.html{\#}supplementary-information}
  {\bibfield  {journal} {\bibinfo  {journal} {Nature}\ }\textbf {\bibinfo
  {volume} {521}},\ \bibinfo {pages} {61--64} (\bibinfo {year}
  {2015})}\BibitemShut {NoStop}%
\bibitem [{\citenamefont {Lindner}\ \emph {et~al.}(2004)\citenamefont
  {Lindner}, \citenamefont {Garcı́a-Ojalvo}, \citenamefont {Neiman},\ and\
  \citenamefont {Schimansky-Geier}}]{Lindner2004}%
  \BibitemOpen
  \bibfield  {author} {\bibinfo {author} {\bibfnamefont {B.}~\bibnamefont
  {Lindner}}, \bibinfo {author} {\bibfnamefont {J.}~\bibnamefont
  {Garcı́a-Ojalvo}}, \bibinfo {author} {\bibfnamefont {A.}~\bibnamefont
  {Neiman}}, \ and\ \bibinfo {author} {\bibfnamefont {L.}~\bibnamefont
  {Schimansky-Geier}},\ }\bibfield  {title} {\enquote {\bibinfo {title}
  {{Effects of noise in excitable systems}},}\ }\href {\doibase
  http://dx.doi.org/10.1016/j.physrep.2003.10.015} {\bibfield  {journal}
  {\bibinfo  {journal} {Physics Reports}\ }\textbf {\bibinfo {volume} {392}},\
  \bibinfo {pages} {321--424} (\bibinfo {year} {2004})}\BibitemShut {NoStop}%
\bibitem [{\citenamefont {SyNAPSE}()}]{SyNAPSE}%
  \BibitemOpen
  \bibfield  {author} {\bibinfo {author} {\bibnamefont {SyNAPSE}},\ }\bibfield
  {title} {\enquote {\bibinfo {title}
  {{Http://www.artificialbrains.com/darpa-synapse-program}},}\ }\href@noop {}
  {\ }\BibitemShut {NoStop}%
\bibitem [{\citenamefont {Benjamin}\ \emph {et~al.}(2014)\citenamefont
  {Benjamin}, \citenamefont {Gao}, \citenamefont {McQuinn}, \citenamefont
  {Choudhary}, \citenamefont {Chandrasekaran}, \citenamefont {Bussat},
  \citenamefont {Alvarez-Icaza}, \citenamefont {Arthur}, \citenamefont
  {Merolla},\ and\ \citenamefont {Boahen}}]{Benjamin2014}%
  \BibitemOpen
  \bibfield  {author} {\bibinfo {author} {\bibfnamefont {B.~V.}\ \bibnamefont
  {Benjamin}}, \bibinfo {author} {\bibfnamefont {P.}~\bibnamefont {Gao}},
  \bibinfo {author} {\bibfnamefont {E.}~\bibnamefont {McQuinn}}, \bibinfo
  {author} {\bibfnamefont {S.}~\bibnamefont {Choudhary}}, \bibinfo {author}
  {\bibfnamefont {A.~R.}\ \bibnamefont {Chandrasekaran}}, \bibinfo {author}
  {\bibfnamefont {J.~M.}\ \bibnamefont {Bussat}}, \bibinfo {author}
  {\bibfnamefont {R.}~\bibnamefont {Alvarez-Icaza}}, \bibinfo {author}
  {\bibfnamefont {J.~V.}\ \bibnamefont {Arthur}}, \bibinfo {author}
  {\bibfnamefont {P.~A.}\ \bibnamefont {Merolla}}, \ and\ \bibinfo {author}
  {\bibfnamefont {K.}~\bibnamefont {Boahen}},\ }\href {\doibase
  10.1109/JPROC.2014.2313565} {\enquote {\bibinfo {title} {{Neurogrid: A
  Mixed-Analog-Digital Multichip System for Large-Scale Neural Simulations}},}\
  } (\bibinfo {year} {2014})\BibitemShut {NoStop}%
\bibitem [{\citenamefont {FACETS}()}]{Facets}%
  \BibitemOpen
  \bibfield  {author} {\bibinfo {author} {\bibnamefont {FACETS}},\ }\bibfield
  {title} {\enquote {\bibinfo {title} {http://facets.kip.uni-heidelberg.de/},}\
  }\href@noop {} {\ }\BibitemShut {NoStop}%
\bibitem [{\citenamefont {HumanBrain}()}]{HumanBrain}%
  \BibitemOpen
  \bibfield  {author} {\bibinfo {author} {\bibnamefont {HumanBrain}},\
  }\bibfield  {title} {\enquote {\bibinfo {title}
  {https://www.humanbrainproject.eu/},}\ }\href@noop {} {\ }\BibitemShut
  {NoStop}%
\bibitem [{\citenamefont {Spinnaker}()}]{Spinnaker}%
  \BibitemOpen
  \bibfield  {author} {\bibinfo {author} {\bibnamefont {Spinnaker}},\
  }\bibfield  {title} {\enquote {\bibinfo {title}
  {http://www.artificialbrains.com/spinnaker},}\ }\href@noop {} {\
  }\BibitemShut {NoStop}%
\bibitem [{\citenamefont {BrainScales}()}]{BrainScales}%
  \BibitemOpen
  \bibfield  {author} {\bibinfo {author} {\bibnamefont {BrainScales}},\
  }\bibfield  {title} {\enquote {\bibinfo {title}
  {{Nohttps://brainscales.kip.uni-heidelberg.de/index.html}},}\ }\href@noop {}
  {\ }\BibitemShut {NoStop}%
\bibitem [{\citenamefont {Indiveri}, \citenamefont {Chicca},\ and\
  \citenamefont {Douglas}(2006)}]{Indiveri2006}%
  \BibitemOpen
  \bibfield  {author} {\bibinfo {author} {\bibfnamefont {G.}~\bibnamefont
  {Indiveri}}, \bibinfo {author} {\bibfnamefont {E.}~\bibnamefont {Chicca}}, \
  and\ \bibinfo {author} {\bibfnamefont {R.}~\bibnamefont {Douglas}},\ }\href
  {\doibase 10.1109/TNN.2005.860850} {\enquote {\bibinfo {title} {{A VLSI array
  of low-power spiking neurons and bistable synapses with spike-timing
  dependent plasticity}},}\ } (\bibinfo {year} {2006})\BibitemShut {NoStop}%
\bibitem [{\citenamefont {Jo}\ \emph {et~al.}(2010)\citenamefont {Jo},
  \citenamefont {Chang}, \citenamefont {Ebong}, \citenamefont {Bhadviya},
  \citenamefont {Mazumder},\ and\ \citenamefont {Lu}}]{Jo2010}%
  \BibitemOpen
  \bibfield  {author} {\bibinfo {author} {\bibfnamefont {S.~H.}\ \bibnamefont
  {Jo}}, \bibinfo {author} {\bibfnamefont {T.}~\bibnamefont {Chang}}, \bibinfo
  {author} {\bibfnamefont {I.}~\bibnamefont {Ebong}}, \bibinfo {author}
  {\bibfnamefont {B.~B.}\ \bibnamefont {Bhadviya}}, \bibinfo {author}
  {\bibfnamefont {P.}~\bibnamefont {Mazumder}}, \ and\ \bibinfo {author}
  {\bibfnamefont {W.}~\bibnamefont {Lu}},\ }\bibfield  {title} {\enquote
  {\bibinfo {title} {{Nanoscale Memristor Device as Synapse in Neuromorphic
  Systems}},}\ }\href {\doibase 10.1021/nl904092h} {\bibfield  {journal}
  {\bibinfo  {journal} {Nano Letters}\ }\textbf {\bibinfo {volume} {10}},\
  \bibinfo {pages} {1297--1301} (\bibinfo {year} {2010})}\BibitemShut {NoStop}%
\bibitem [{\citenamefont {Kim}\ \emph {et~al.}(2012)\citenamefont {Kim},
  \citenamefont {Gaba}, \citenamefont {Wheeler}, \citenamefont {Cruz-Albrecht},
  \citenamefont {Hussain}, \citenamefont {Srinivasa},\ and\ \citenamefont
  {Lu}}]{Kim2012}%
  \BibitemOpen
  \bibfield  {author} {\bibinfo {author} {\bibfnamefont {K.-H.}\ \bibnamefont
  {Kim}}, \bibinfo {author} {\bibfnamefont {S.}~\bibnamefont {Gaba}}, \bibinfo
  {author} {\bibfnamefont {D.}~\bibnamefont {Wheeler}}, \bibinfo {author}
  {\bibfnamefont {J.~M.}\ \bibnamefont {Cruz-Albrecht}}, \bibinfo {author}
  {\bibfnamefont {T.}~\bibnamefont {Hussain}}, \bibinfo {author} {\bibfnamefont
  {N.}~\bibnamefont {Srinivasa}}, \ and\ \bibinfo {author} {\bibfnamefont
  {W.}~\bibnamefont {Lu}},\ }\bibfield  {title} {\enquote {\bibinfo {title} {{A
  Functional Hybrid Memristor Crossbar-Array/CMOS System for Data Storage and
  Neuromorphic Applications}},}\ }\href {\doibase 10.1021/nl203687n} {\bibfield
   {journal} {\bibinfo  {journal} {Nano Letters}\ }\textbf {\bibinfo {volume}
  {12}},\ \bibinfo {pages} {389--395} (\bibinfo {year} {2012})}\BibitemShut
  {NoStop}%
\bibitem [{\citenamefont {Kuzum}, \citenamefont {Yu},\ and\ \citenamefont
  {Wong}(2013)}]{Kuzum2013}%
  \BibitemOpen
  \bibfield  {author} {\bibinfo {author} {\bibfnamefont {D.}~\bibnamefont
  {Kuzum}}, \bibinfo {author} {\bibfnamefont {S.}~\bibnamefont {Yu}}, \ and\
  \bibinfo {author} {\bibfnamefont {H.-S.~P.}\ \bibnamefont {Wong}},\
  }\bibfield  {title} {\enquote {\bibinfo {title} {{Synaptic electronics:
  materials, devices and applications}},}\ }\href
  {http://stacks.iop.org/0957-4484/24/i=38/a=382001} {\bibfield  {journal}
  {\bibinfo  {journal} {Nanotechnology}\ }\textbf {\bibinfo {volume} {24}},\
  \bibinfo {pages} {382001} (\bibinfo {year} {2013})}\BibitemShut {NoStop}%
\bibitem [{\citenamefont {Prucnal}\ \emph {et~al.}(2016)\citenamefont
  {Prucnal}, \citenamefont {Shastri}, \citenamefont {de~Lima}, \citenamefont
  {Nahmias},\ and\ \citenamefont {Tait}}]{Prucnal2016}%
  \BibitemOpen
  \bibfield  {author} {\bibinfo {author} {\bibfnamefont {P.~R.}\ \bibnamefont
  {Prucnal}}, \bibinfo {author} {\bibfnamefont {B.~J.}\ \bibnamefont
  {Shastri}}, \bibinfo {author} {\bibfnamefont {T.~F.}\ \bibnamefont
  {de~Lima}}, \bibinfo {author} {\bibfnamefont {M.~A.}\ \bibnamefont
  {Nahmias}}, \ and\ \bibinfo {author} {\bibfnamefont {A.~N.}\ \bibnamefont
  {Tait}},\ }\bibfield  {title} {\enquote {\bibinfo {title} {{Recent progress
  in semiconductor excitable lasers for photonic spike processing}},}\ }\href
  {\doibase 10.1364/AOP.8.000228} {\bibfield  {journal} {\bibinfo  {journal}
  {Adv. Opt. Photon.}\ }\textbf {\bibinfo {volume} {8}},\ \bibinfo {pages}
  {228--299} (\bibinfo {year} {2016})}\BibitemShut {NoStop}%
\bibitem [{\citenamefont {Romeira}\ \emph
  {et~al.}(2014{\natexlab{a}})\citenamefont {Romeira}, \citenamefont {Kong},
  \citenamefont {Li}, \citenamefont {Figueiredo}, \citenamefont {Javaloyes},\
  and\ \citenamefont {Yao}}]{Romeira2014}%
  \BibitemOpen
  \bibfield  {author} {\bibinfo {author} {\bibfnamefont {B.}~\bibnamefont
  {Romeira}}, \bibinfo {author} {\bibfnamefont {F.}~\bibnamefont {Kong}},
  \bibinfo {author} {\bibfnamefont {W.}~\bibnamefont {Li}}, \bibinfo {author}
  {\bibfnamefont {J.~M.~L.}\ \bibnamefont {Figueiredo}}, \bibinfo {author}
  {\bibfnamefont {J.}~\bibnamefont {Javaloyes}}, \ and\ \bibinfo {author}
  {\bibfnamefont {J.}~\bibnamefont {Yao}},\ }\bibfield  {title} {\enquote
  {\bibinfo {title} {Broadband chaotic signals and breather oscillations in an
  optoelectronic oscillator incorporating a microwave photonic filter},}\
  }\href {\doibase 10.1109/JLT.2014.2308261} {\bibfield  {journal} {\bibinfo
  {journal} {Journal of Lightwave Technology}\ }\textbf {\bibinfo {volume}
  {32}},\ \bibinfo {pages} {3933--3942} (\bibinfo {year}
  {2014}{\natexlab{a}})}\BibitemShut {NoStop}%
\bibitem [{\citenamefont {Romeira}\ \emph {et~al.}(2015)\citenamefont
  {Romeira}, \citenamefont {Kong}, \citenamefont {Figueiredo}, \citenamefont
  {Javaloyes},\ and\ \citenamefont {Yao}}]{Romeira2015}%
  \BibitemOpen
  \bibfield  {author} {\bibinfo {author} {\bibfnamefont {B.}~\bibnamefont
  {Romeira}}, \bibinfo {author} {\bibfnamefont {F.}~\bibnamefont {Kong}},
  \bibinfo {author} {\bibfnamefont {J.~M.~L.}\ \bibnamefont {Figueiredo}},
  \bibinfo {author} {\bibfnamefont {J.}~\bibnamefont {Javaloyes}}, \ and\
  \bibinfo {author} {\bibfnamefont {J.}~\bibnamefont {Yao}},\ }\bibfield
  {title} {\enquote {\bibinfo {title} {High-speed spiking and bursting
  oscillations in a long-delayed broadband optoelectronic oscillator},}\ }\href
  {http://jlt.osa.org/abstract.cfm?URI=jlt-33-2-503} {\bibfield  {journal}
  {\bibinfo  {journal} {J. Lightwave Technol.}\ }\textbf {\bibinfo {volume}
  {33}},\ \bibinfo {pages} {503--510} (\bibinfo {year} {2015})}\BibitemShut
  {NoStop}%
\bibitem [{\citenamefont {{Paul R. Prucnal and Bhavin J.
  Shastri}}(2017)}]{Prucnal2017}%
  \BibitemOpen
  \bibfield  {author} {\bibinfo {author} {\bibnamefont {{Paul R. Prucnal and
  Bhavin J. Shastri}}},\ }\href@noop {} {\emph {\bibinfo {title} {{Neuromorphic
  Photonics}}}}\ (\bibinfo  {publisher} {CRC Press, Taylor and Francis Group},\
  \bibinfo {address} {Boca Raton, Florida},\ \bibinfo {year}
  {2017})\BibitemShut {NoStop}%
\bibitem [{\citenamefont {Robertson}\ \emph {et~al.}(2017)\citenamefont
  {Robertson}, \citenamefont {Deng}, \citenamefont {Javaloyes},\ and\
  \citenamefont {Hurtado}}]{Robertson17}%
  \BibitemOpen
  \bibfield  {author} {\bibinfo {author} {\bibfnamefont {J.}~\bibnamefont
  {Robertson}}, \bibinfo {author} {\bibfnamefont {T.}~\bibnamefont {Deng}},
  \bibinfo {author} {\bibfnamefont {J.}~\bibnamefont {Javaloyes}}, \ and\
  \bibinfo {author} {\bibfnamefont {A.}~\bibnamefont {Hurtado}},\ }\bibfield
  {title} {\enquote {\bibinfo {title} {Controlled inhibition of spiking
  dynamics in vcsels for neuromorphic photonics: theory and experiments},}\
  }\href {\doibase 10.1364/OL.42.001560} {\bibfield  {journal} {\bibinfo
  {journal} {Opt. Lett.}\ }\textbf {\bibinfo {volume} {42}},\ \bibinfo {pages}
  {1560--1563} (\bibinfo {year} {2017})}\BibitemShut {NoStop}%
\bibitem [{\citenamefont {Loos}\ and\ \citenamefont {Glaser}(1972)}]{Loos1972}%
  \BibitemOpen
  \bibfield  {author} {\bibinfo {author} {\bibfnamefont {H.~V.~D.}\
  \bibnamefont {Loos}}\ and\ \bibinfo {author} {\bibfnamefont {E.~M.}\
  \bibnamefont {Glaser}},\ }\bibfield  {title} {\enquote {\bibinfo {title}
  {{Autapses in neocortex cerebri: synapses between a pyramidal cell's axon and
  its own dendrites}},}\ }\href {\doibase
  http://dx.doi.org/10.1016/0006-8993(72)90189-8} {\bibfield  {journal}
  {\bibinfo  {journal} {Brain Research}\ }\textbf {\bibinfo {volume} {48}},\
  \bibinfo {pages} {355--360} (\bibinfo {year} {1972})}\BibitemShut {NoStop}%
\bibitem [{\citenamefont {Flight}(2009)}]{Flight2009}%
  \BibitemOpen
  \bibfield  {author} {\bibinfo {author} {\bibfnamefont {M.~H.}\ \bibnamefont
  {Flight}},\ }\bibfield  {title} {\enquote {\bibinfo {title}
  {{Neuromodulation: Exerting self-control for persistence}},}\ }\href
  {http://dx.doi.org/10.1038/nrn2637} {\bibfield  {journal} {\bibinfo
  {journal} {Nat Rev Neurosci}\ }\textbf {\bibinfo {volume} {10}},\ \bibinfo
  {pages} {316} (\bibinfo {year} {2009})}\BibitemShut {NoStop}%
\bibitem [{\citenamefont {Tam{\'{a}}s}, \citenamefont {Buhl},\ and\
  \citenamefont {Somogyi}(1997)}]{Tamas1997}%
  \BibitemOpen
  \bibfield  {author} {\bibinfo {author} {\bibfnamefont {G.}~\bibnamefont
  {Tam{\'{a}}s}}, \bibinfo {author} {\bibfnamefont {E.~H.}\ \bibnamefont
  {Buhl}}, \ and\ \bibinfo {author} {\bibfnamefont {P.}~\bibnamefont
  {Somogyi}},\ }\bibfield  {title} {\enquote {\bibinfo {title} {{Massive
  Autaptic Self-Innervation of GABAergic Neurons in Cat Visual Cortex}},}\
  }\href {http://www.jneurosci.org/content/17/16/6352} {\bibfield  {journal}
  {\bibinfo  {journal} {Journal of Neuroscience}\ }\textbf {\bibinfo {volume}
  {17}},\ \bibinfo {pages} {6352--6364} (\bibinfo {year} {1997})}\BibitemShut
  {NoStop}%
\bibitem [{\citenamefont {Bacci}, \citenamefont {Huguenard},\ and\
  \citenamefont {Prince}(2003)}]{Bacci2003}%
  \BibitemOpen
  \bibfield  {author} {\bibinfo {author} {\bibfnamefont {A.}~\bibnamefont
  {Bacci}}, \bibinfo {author} {\bibfnamefont {J.~R.}\ \bibnamefont
  {Huguenard}}, \ and\ \bibinfo {author} {\bibfnamefont {D.~A.}\ \bibnamefont
  {Prince}},\ }\bibfield  {title} {\enquote {\bibinfo {title} {{Functional
  Autaptic Neurotransmission in Fast-Spiking Interneurons: A Novel Form of
  Feedback Inhibition in the Neocortex}},}\ }\href
  {http://www.jneurosci.org/content/23/3/859} {\bibfield  {journal} {\bibinfo
  {journal} {Journal of Neuroscience}\ }\textbf {\bibinfo {volume} {23}},\
  \bibinfo {pages} {859--866} (\bibinfo {year} {2003})}\BibitemShut {NoStop}%
\bibitem [{\citenamefont {Herrmann}\ and\ \citenamefont
  {Klaus}(2004)}]{Herrmann2004}%
  \BibitemOpen
  \bibfield  {author} {\bibinfo {author} {\bibfnamefont {C.~S.}\ \bibnamefont
  {Herrmann}}\ and\ \bibinfo {author} {\bibfnamefont {A.}~\bibnamefont
  {Klaus}},\ }\bibfield  {title} {\enquote {\bibinfo {title} {{Autapse turns
  neuron into oscillator}},}\ }\href {\doibase 10.1142/S0218127404009338}
  {\bibfield  {journal} {\bibinfo  {journal} {International Journal of
  Bifurcation and Chaos}\ }\textbf {\bibinfo {volume} {14}},\ \bibinfo {pages}
  {623--633} (\bibinfo {year} {2004})}\BibitemShut {NoStop}%
\bibitem [{\citenamefont {Wiles}\ \emph {et~al.}(2017)\citenamefont {Wiles},
  \citenamefont {Gu}, \citenamefont {Pasqualetti}, \citenamefont {Parvesse},
  \citenamefont {Gabrieli}, \citenamefont {Bassett},\ and\ \citenamefont
  {Meaney}}]{Wiles2017}%
  \BibitemOpen
  \bibfield  {author} {\bibinfo {author} {\bibfnamefont {L.}~\bibnamefont
  {Wiles}}, \bibinfo {author} {\bibfnamefont {S.}~\bibnamefont {Gu}}, \bibinfo
  {author} {\bibfnamefont {F.}~\bibnamefont {Pasqualetti}}, \bibinfo {author}
  {\bibfnamefont {B.}~\bibnamefont {Parvesse}}, \bibinfo {author}
  {\bibfnamefont {D.}~\bibnamefont {Gabrieli}}, \bibinfo {author}
  {\bibfnamefont {D.~S.}\ \bibnamefont {Bassett}}, \ and\ \bibinfo {author}
  {\bibfnamefont {D.~F.}\ \bibnamefont {Meaney}},\ }\bibfield  {title}
  {\enquote {\bibinfo {title} {{Autaptic Connections Shift Network Excitability
  and Bursting}},}\ }\href {http://dx.doi.org/10.1038/srep44006
  http://10.0.4.14/srep44006
  https://www.nature.com/articles/srep44006{\#}supplementary-information}
  {\bibfield  {journal} {\bibinfo  {journal} {Scientific Reports}\ }\textbf
  {\bibinfo {volume} {7}},\ \bibinfo {pages} {44006} (\bibinfo {year}
  {2017})}\BibitemShut {NoStop}%
\bibitem [{\citenamefont {Xu}\ \emph {et~al.}(2017)\citenamefont {Xu},
  \citenamefont {Ying}, \citenamefont {Jia}, \citenamefont {Ma},\ and\
  \citenamefont {Hayat}}]{Xu2017}%
  \BibitemOpen
  \bibfield  {author} {\bibinfo {author} {\bibfnamefont {Y.}~\bibnamefont
  {Xu}}, \bibinfo {author} {\bibfnamefont {H.}~\bibnamefont {Ying}}, \bibinfo
  {author} {\bibfnamefont {Y.}~\bibnamefont {Jia}}, \bibinfo {author}
  {\bibfnamefont {J.}~\bibnamefont {Ma}}, \ and\ \bibinfo {author}
  {\bibfnamefont {T.}~\bibnamefont {Hayat}},\ }\bibfield  {title} {\enquote
  {\bibinfo {title} {{Autaptic regulation of electrical activities in neuron
  under electromagnetic induction}},}\ }\href
  {http://dx.doi.org/10.1038/srep43452 http://10.0.4.14/srep43452} {\bibfield
  {journal} {\bibinfo  {journal} {Scientific Reports}\ }\textbf {\bibinfo
  {volume} {7}},\ \bibinfo {pages} {43452} (\bibinfo {year}
  {2017})}\BibitemShut {NoStop}%
\bibitem [{\citenamefont {Rusin}\ \emph {et~al.}(2011)\citenamefont {Rusin},
  \citenamefont {Johnson}, \citenamefont {Kapur},\ and\ \citenamefont
  {Hudson}}]{Rusin2011}%
  \BibitemOpen
  \bibfield  {author} {\bibinfo {author} {\bibfnamefont {C.~G.}\ \bibnamefont
  {Rusin}}, \bibinfo {author} {\bibfnamefont {S.~E.}\ \bibnamefont {Johnson}},
  \bibinfo {author} {\bibfnamefont {J.}~\bibnamefont {Kapur}}, \ and\ \bibinfo
  {author} {\bibfnamefont {J.~L.}\ \bibnamefont {Hudson}},\ }\bibfield  {title}
  {\enquote {\bibinfo {title} {{Engineering the synchronization of neuron
  action potentials using global time-delayed feedback stimulation}},}\ }\href
  {\doibase 10.1103/PhysRevE.84.066202} {\bibfield  {journal} {\bibinfo
  {journal} {Phys. Rev. E}\ }\textbf {\bibinfo {volume} {84}},\ \bibinfo
  {pages} {66202} (\bibinfo {year} {2011})}\BibitemShut {NoStop}%
\bibitem [{\citenamefont {Hashemi}, \citenamefont {Valizadeh},\ and\
  \citenamefont {Azizi}(2012)}]{Hashemi2012}%
  \BibitemOpen
  \bibfield  {author} {\bibinfo {author} {\bibfnamefont {M.}~\bibnamefont
  {Hashemi}}, \bibinfo {author} {\bibfnamefont {A.}~\bibnamefont {Valizadeh}},
  \ and\ \bibinfo {author} {\bibfnamefont {Y.}~\bibnamefont {Azizi}},\
  }\bibfield  {title} {\enquote {\bibinfo {title} {{Effect of duration of
  synaptic activity on spike rate of a Hodgkin-Huxley neuron with delayed
  feedback}},}\ }\href {\doibase 10.1103/PhysRevE.85.021917} {\bibfield
  {journal} {\bibinfo  {journal} {Phys. Rev. E}\ }\textbf {\bibinfo {volume}
  {85}},\ \bibinfo {pages} {21917} (\bibinfo {year} {2012})}\BibitemShut
  {NoStop}%
\bibitem [{\citenamefont {Wang}\ \emph {et~al.}(2014)\citenamefont {Wang},
  \citenamefont {Ma}, \citenamefont {Chen},\ and\ \citenamefont
  {Chen}}]{Wang2014}%
  \BibitemOpen
  \bibfield  {author} {\bibinfo {author} {\bibfnamefont {H.}~\bibnamefont
  {Wang}}, \bibinfo {author} {\bibfnamefont {J.}~\bibnamefont {Ma}}, \bibinfo
  {author} {\bibfnamefont {Y.}~\bibnamefont {Chen}}, \ and\ \bibinfo {author}
  {\bibfnamefont {Y.}~\bibnamefont {Chen}},\ }\bibfield  {title} {\enquote
  {\bibinfo {title} {{Effect of an autapse on the firing pattern transition in
  a bursting neuron}},}\ }\href {\doibase
  http://doi.org/10.1016/j.cnsns.2014.02.018} {\bibfield  {journal} {\bibinfo
  {journal} {Communications in Nonlinear Science and Numerical Simulation}\
  }\textbf {\bibinfo {volume} {19}},\ \bibinfo {pages} {3242--3254} (\bibinfo
  {year} {2014})}\BibitemShut {NoStop}%
\bibitem [{\citenamefont {{Hengtong Wang and Longfei Wang and Yueling Chen and
  Yong Chen}}(2014)}]{WangChaos2014}%
  \BibitemOpen
  \bibfield  {author} {\bibinfo {author} {\bibnamefont {{Hengtong Wang and
  Longfei Wang and Yueling Chen and Yong Chen}}},\ }\bibfield  {title}
  {\enquote {\bibinfo {title} {{Effect of autaptic activity on the response of
  a Hodgkin-Huxley neuron}},}\ }\href {\doibase 10.1063/1.4892769} {\bibfield
  {journal} {\bibinfo  {journal} {Chaos: An Interdisciplinary Journal of
  Nonlinear Science}\ }\textbf {\bibinfo {volume} {24}},\ \bibinfo {pages}
  {33122} (\bibinfo {year} {2014})}\BibitemShut {NoStop}%
\bibitem [{\citenamefont {Appeltant}\ \emph {et~al.}(2011)\citenamefont
  {Appeltant}, \citenamefont {Soriano}, \citenamefont {{Van der Sande}},
  \citenamefont {Danckaert}, \citenamefont {Massar}, \citenamefont {Dambre},
  \citenamefont {Schrauwen}, \citenamefont {Mirasso},\ and\ \citenamefont
  {Fischer}}]{Appeltant2011}%
  \BibitemOpen
  \bibfield  {author} {\bibinfo {author} {\bibfnamefont {L.}~\bibnamefont
  {Appeltant}}, \bibinfo {author} {\bibfnamefont {M.~C.}\ \bibnamefont
  {Soriano}}, \bibinfo {author} {\bibfnamefont {G.}~\bibnamefont {{Van der
  Sande}}}, \bibinfo {author} {\bibfnamefont {J.}~\bibnamefont {Danckaert}},
  \bibinfo {author} {\bibfnamefont {S.}~\bibnamefont {Massar}}, \bibinfo
  {author} {\bibfnamefont {J.}~\bibnamefont {Dambre}}, \bibinfo {author}
  {\bibfnamefont {B.}~\bibnamefont {Schrauwen}}, \bibinfo {author}
  {\bibfnamefont {C.~R.}\ \bibnamefont {Mirasso}}, \ and\ \bibinfo {author}
  {\bibfnamefont {I.}~\bibnamefont {Fischer}},\ }\bibfield  {title} {\enquote
  {\bibinfo {title} {{Information processing using a single dynamical node as
  complex system}},}\ }\href {http://dx.doi.org/10.1038/ncomms1476
  http://10.0.4.14/ncomms1476
  https://www.nature.com/articles/ncomms1476{\#}supplementary-information}
  {\bibfield  {journal} {\bibinfo  {journal} {Nature Communications}\ }\textbf
  {\bibinfo {volume} {2}},\ \bibinfo {pages} {468} (\bibinfo {year}
  {2011})}\BibitemShut {NoStop}%
\bibitem [{\citenamefont {Sourikopoulos}\ \emph {et~al.}(2017)\citenamefont
  {Sourikopoulos}, \citenamefont {Hedayat}, \citenamefont {Loyez},
  \citenamefont {Danneville}, \citenamefont {Hoel}, \citenamefont {Mercier},\
  and\ \citenamefont {Cappy}}]{Sourikopoulos2017}%
  \BibitemOpen
  \bibfield  {author} {\bibinfo {author} {\bibfnamefont {I.}~\bibnamefont
  {Sourikopoulos}}, \bibinfo {author} {\bibfnamefont {S.}~\bibnamefont
  {Hedayat}}, \bibinfo {author} {\bibfnamefont {C.}~\bibnamefont {Loyez}},
  \bibinfo {author} {\bibfnamefont {F.}~\bibnamefont {Danneville}}, \bibinfo
  {author} {\bibfnamefont {V.}~\bibnamefont {Hoel}}, \bibinfo {author}
  {\bibfnamefont {E.}~\bibnamefont {Mercier}}, \ and\ \bibinfo {author}
  {\bibfnamefont {A.}~\bibnamefont {Cappy}},\ }\bibfield  {title} {\enquote
  {\bibinfo {title} {{A 4-fJ/Spike Artificial Neuron in 65 nm CMOS
  Technology}},}\ }\href {\doibase 10.3389/fnins.2017.00123} {\bibfield
  {journal} {\bibinfo  {journal} {Frontiers in Neuroscience}\ }\textbf
  {\bibinfo {volume} {11}},\ \bibinfo {pages} {123} (\bibinfo {year}
  {2017})}\BibitemShut {NoStop}%
\bibitem [{\citenamefont {Romeira}\ \emph
  {et~al.}(2013{\natexlab{a}})\citenamefont {Romeira}, \citenamefont
  {Javaloyes}, \citenamefont {Ironside}, \citenamefont {Figueiredo},
  \citenamefont {Balle},\ and\ \citenamefont {Piro}}]{Romeira2013a}%
  \BibitemOpen
  \bibfield  {author} {\bibinfo {author} {\bibfnamefont {B.}~\bibnamefont
  {Romeira}}, \bibinfo {author} {\bibfnamefont {J.}~\bibnamefont {Javaloyes}},
  \bibinfo {author} {\bibfnamefont {C.~N.}\ \bibnamefont {Ironside}}, \bibinfo
  {author} {\bibfnamefont {J.~M.~L.}\ \bibnamefont {Figueiredo}}, \bibinfo
  {author} {\bibfnamefont {S.}~\bibnamefont {Balle}}, \ and\ \bibinfo {author}
  {\bibfnamefont {O.}~\bibnamefont {Piro}},\ }\bibfield  {title} {\enquote
  {\bibinfo {title} {{Excitability and optical pulse generation in
  semiconductor lasers driven by resonant tunneling diode photo-detectors}},}\
  }\href {\doibase 10.1364/OE.21.020931} {\bibfield  {journal} {\bibinfo
  {journal} {Optics Express}\ }\textbf {\bibinfo {volume} {21}},\ \bibinfo
  {pages} {20931} (\bibinfo {year} {2013}{\natexlab{a}})}\BibitemShut {NoStop}%
\bibitem [{\citenamefont {Romeira}\ \emph {et~al.}(2010)\citenamefont
  {Romeira}, \citenamefont {Figueiredo}, \citenamefont {Ironside},
  \citenamefont {Kelly},\ and\ \citenamefont {Slight}}]{Romeira2010}%
  \BibitemOpen
  \bibfield  {author} {\bibinfo {author} {\bibfnamefont {B.}~\bibnamefont
  {Romeira}}, \bibinfo {author} {\bibfnamefont {J.~M.~L.}\ \bibnamefont
  {Figueiredo}}, \bibinfo {author} {\bibfnamefont {C.~N.}\ \bibnamefont
  {Ironside}}, \bibinfo {author} {\bibfnamefont {A.~E.}\ \bibnamefont {Kelly}},
  \ and\ \bibinfo {author} {\bibfnamefont {T.~J.}\ \bibnamefont {Slight}},\
  }\bibfield  {title} {\enquote {\bibinfo {title} {{Optical control of a
  resonant tunneling diode microwave-photonic oscillator}},}\ }\href {\doibase
  10.1109/LPT.2010.2076331} {\bibfield  {journal} {\bibinfo  {journal} {IEEE
  Photonics Technology Letters}\ }\textbf {\bibinfo {volume} {22}},\ \bibinfo
  {pages} {1610--1612} (\bibinfo {year} {2010})}\BibitemShut {NoStop}%
\bibitem [{\citenamefont {Romeira}\ \emph
  {et~al.}(2013{\natexlab{b}})\citenamefont {Romeira}, \citenamefont {Pessoa},
  \citenamefont {Salgado}, \citenamefont {Ironside},\ and\ \citenamefont
  {Figueiredo}}]{Romeira2013b}%
  \BibitemOpen
  \bibfield  {author} {\bibinfo {author} {\bibfnamefont {B.}~\bibnamefont
  {Romeira}}, \bibinfo {author} {\bibfnamefont {L.}~\bibnamefont {Pessoa}},
  \bibinfo {author} {\bibfnamefont {H.}~\bibnamefont {Salgado}}, \bibinfo
  {author} {\bibfnamefont {C.}~\bibnamefont {Ironside}}, \ and\ \bibinfo
  {author} {\bibfnamefont {J.}~\bibnamefont {Figueiredo}},\ }\bibfield  {title}
  {\enquote {\bibinfo {title} {{Photo-detectors integrated with resonant
  tunneling diodes}},}\ }\href {\doibase 10.3390/s130709464} {\bibfield
  {journal} {\bibinfo  {journal} {Sensors (Switzerland)}\ }\textbf {\bibinfo
  {volume} {13}} (\bibinfo {year} {2013}{\natexlab{b}}),\
  10.3390/s130709464}\BibitemShut {NoStop}%
\bibitem [{\citenamefont {Huang}\ \emph {et~al.}(2014)\citenamefont {Huang},
  \citenamefont {Seo}, \citenamefont {Sarmiento}, \citenamefont {Huo},
  \citenamefont {Harris},\ and\ \citenamefont {Brongersma}}]{Huang2014}%
  \BibitemOpen
  \bibfield  {author} {\bibinfo {author} {\bibfnamefont {K.~C.~Y.}\
  \bibnamefont {Huang}}, \bibinfo {author} {\bibfnamefont {M.~K.}\ \bibnamefont
  {Seo}}, \bibinfo {author} {\bibfnamefont {T.}~\bibnamefont {Sarmiento}},
  \bibinfo {author} {\bibfnamefont {Y.}~\bibnamefont {Huo}}, \bibinfo {author}
  {\bibfnamefont {J.~S.}\ \bibnamefont {Harris}}, \ and\ \bibinfo {author}
  {\bibfnamefont {M.~L.}\ \bibnamefont {Brongersma}},\ }\bibfield  {title}
  {\enquote {\bibinfo {title} {{Electrically driven subwavelength optical
  nanocircuits}},}\ }\href@noop {} {\bibfield  {journal} {\bibinfo  {journal}
  {Nature Photonics}\ }\textbf {\bibinfo {volume} {8}},\ \bibinfo {pages}
  {244--249} (\bibinfo {year} {2014})}\BibitemShut {NoStop}%
\bibitem [{\citenamefont {Dolores-Calzadilla}\ \emph
  {et~al.}(2017)\citenamefont {Dolores-Calzadilla}, \citenamefont {Romeira},
  \citenamefont {Pagliano}, \citenamefont {Birindelli}, \citenamefont
  {Higuera-Rodriguez}, \citenamefont {van Veldhoven}, \citenamefont {Smit},
  \citenamefont {Fiore},\ and\ \citenamefont {Heiss}}]{Dolores-Calzadilla2017}%
  \BibitemOpen
  \bibfield  {author} {\bibinfo {author} {\bibfnamefont {V.}~\bibnamefont
  {Dolores-Calzadilla}}, \bibinfo {author} {\bibfnamefont {B.}~\bibnamefont
  {Romeira}}, \bibinfo {author} {\bibfnamefont {F.}~\bibnamefont {Pagliano}},
  \bibinfo {author} {\bibfnamefont {S.}~\bibnamefont {Birindelli}}, \bibinfo
  {author} {\bibfnamefont {A.}~\bibnamefont {Higuera-Rodriguez}}, \bibinfo
  {author} {\bibfnamefont {P.~J.}\ \bibnamefont {van Veldhoven}}, \bibinfo
  {author} {\bibfnamefont {M.~K.}\ \bibnamefont {Smit}}, \bibinfo {author}
  {\bibfnamefont {A.}~\bibnamefont {Fiore}}, \ and\ \bibinfo {author}
  {\bibfnamefont {D.}~\bibnamefont {Heiss}},\ }\bibfield  {title} {\enquote
  {\bibinfo {title} {{Waveguide-coupled nanopillar metal-cavity light-emitting
  diodes on silicon}},}\ }\href@noop {} {\bibfield  {journal} {\bibinfo
  {journal} {Nature Communications}\ }\textbf {\bibinfo {volume} {8}},\
  \bibinfo {pages} {14323} (\bibinfo {year} {2017})}\BibitemShut {NoStop}%
\bibitem [{\citenamefont {Hill}\ and\ \citenamefont {Gather}(2014)}]{Hill2014}%
  \BibitemOpen
  \bibfield  {author} {\bibinfo {author} {\bibfnamefont {M.~T.}\ \bibnamefont
  {Hill}}\ and\ \bibinfo {author} {\bibfnamefont {M.~C.}\ \bibnamefont
  {Gather}},\ }\bibfield  {title} {\enquote {\bibinfo {title} {{Advances in
  small lasers}},}\ }\href {\doibase 10.1038/nphoton.2014.239} {\bibfield
  {journal} {\bibinfo  {journal} {Nature Photonics}\ }\textbf {\bibinfo
  {volume} {8}},\ \bibinfo {pages} {908--918} (\bibinfo {year}
  {2014})}\BibitemShut {NoStop}%
\bibitem [{\citenamefont {Asada}, \citenamefont {Suzuki},\ and\ \citenamefont
  {Kishimoto}(2008)}]{Asada2008}%
  \BibitemOpen
  \bibfield  {author} {\bibinfo {author} {\bibfnamefont {M.}~\bibnamefont
  {Asada}}, \bibinfo {author} {\bibfnamefont {S.}~\bibnamefont {Suzuki}}, \
  and\ \bibinfo {author} {\bibfnamefont {N.}~\bibnamefont {Kishimoto}},\
  }\bibfield  {title} {\enquote {\bibinfo {title} {Resonant tunneling diodes
  for sub-terahertz and terahertz oscillators},}\ }\href
  {http://stacks.iop.org/1347-4065/47/i=6R/a=4375} {\bibfield  {journal}
  {\bibinfo  {journal} {Japanese Journal of Applied Physics}\ }\textbf
  {\bibinfo {volume} {47}},\ \bibinfo {pages} {4375} (\bibinfo {year}
  {2008})}\BibitemShut {NoStop}%
\bibitem [{\citenamefont {Wang}\ \emph {et~al.}(2013)\citenamefont {Wang},
  \citenamefont {Wang}, \citenamefont {Li}, \citenamefont {Romeira},\ and\
  \citenamefont {Wasige}}]{Wang2013}%
  \BibitemOpen
  \bibfield  {author} {\bibinfo {author} {\bibfnamefont {J.}~\bibnamefont
  {Wang}}, \bibinfo {author} {\bibfnamefont {L.}~\bibnamefont {Wang}}, \bibinfo
  {author} {\bibfnamefont {C.}~\bibnamefont {Li}}, \bibinfo {author}
  {\bibfnamefont {B.}~\bibnamefont {Romeira}}, \ and\ \bibinfo {author}
  {\bibfnamefont {E.}~\bibnamefont {Wasige}},\ }\bibfield  {title} {\enquote
  {\bibinfo {title} {28 ghz mmic resonant tunnelling diode oscillator of around
  1mw output power},}\ }\href {\doibase 10.1049/el.2013.1583} {\bibfield
  {journal} {\bibinfo  {journal} {Electronics Letters}\ }\textbf {\bibinfo
  {volume} {49}},\ \bibinfo {pages} {816--818} (\bibinfo {year}
  {2013})}\BibitemShut {NoStop}%
\bibitem [{\citenamefont {Romeira}\ \emph
  {et~al.}(2014{\natexlab{b}})\citenamefont {Romeira}, \citenamefont
  {Av{\'{o}}}, \citenamefont {Javaloyes}, \citenamefont {Balle}, \citenamefont
  {Ironside},\ and\ \citenamefont {Figueiredo}}]{Romeira2014b}%
  \BibitemOpen
  \bibfield  {author} {\bibinfo {author} {\bibfnamefont {B.}~\bibnamefont
  {Romeira}}, \bibinfo {author} {\bibfnamefont {R.}~\bibnamefont {Av{\'{o}}}},
  \bibinfo {author} {\bibfnamefont {J.}~\bibnamefont {Javaloyes}}, \bibinfo
  {author} {\bibfnamefont {S.}~\bibnamefont {Balle}}, \bibinfo {author}
  {\bibfnamefont {C.}~\bibnamefont {Ironside}}, \ and\ \bibinfo {author}
  {\bibfnamefont {J.}~\bibnamefont {Figueiredo}},\ }\bibfield  {title}
  {\enquote {\bibinfo {title} {{Stochastic induced dynamics in neuromorphic
  optoelectronic oscillators}},}\ }\href {\doibase 10.1007/s11082-014-9905-3}
  {\bibfield  {journal} {\bibinfo  {journal} {Optical and Quantum Electronics}\
  }\textbf {\bibinfo {volume} {46}} (\bibinfo {year} {2014}{\natexlab{b}}),\
  10.1007/s11082-014-9905-3}\BibitemShut {NoStop}%
\bibitem [{\citenamefont {Slight}\ \emph {et~al.}(2008)\citenamefont {Slight},
  \citenamefont {Romeira}, \citenamefont {Wang}, \citenamefont {Figueiredo},
  \citenamefont {Wasige},\ and\ \citenamefont {Ironside}}]{Slight2008}%
  \BibitemOpen
  \bibfield  {author} {\bibinfo {author} {\bibfnamefont {T.}~\bibnamefont
  {Slight}}, \bibinfo {author} {\bibfnamefont {B.}~\bibnamefont {Romeira}},
  \bibinfo {author} {\bibfnamefont {L.}~\bibnamefont {Wang}}, \bibinfo {author}
  {\bibfnamefont {J.}~\bibnamefont {Figueiredo}}, \bibinfo {author}
  {\bibfnamefont {E.}~\bibnamefont {Wasige}}, \ and\ \bibinfo {author}
  {\bibfnamefont {C.}~\bibnamefont {Ironside}},\ }\bibfield  {title} {\enquote
  {\bibinfo {title} {{A Li{\'{e}}nard oscillator resonant tunnelling
  diode-laser diode hybrid integrated circuit: Model and experiment}},}\ }\href
  {\doibase 10.1109/JQE.2008.2000924} {\bibfield  {journal} {\bibinfo
  {journal} {IEEE Journal of Quantum Electronics}\ }\textbf {\bibinfo {volume}
  {44}} (\bibinfo {year} {2008}),\ 10.1109/JQE.2008.2000924}\BibitemShut
  {NoStop}%
\bibitem [{\citenamefont {Romeira}\ \emph {et~al.}(2011)\citenamefont
  {Romeira}, \citenamefont {Seunarine}, \citenamefont {Ironside}, \citenamefont
  {Kelly},\ and\ \citenamefont {Figueiredo}}]{Romeira2011}%
  \BibitemOpen
  \bibfield  {author} {\bibinfo {author} {\bibfnamefont {B.}~\bibnamefont
  {Romeira}}, \bibinfo {author} {\bibfnamefont {K.}~\bibnamefont {Seunarine}},
  \bibinfo {author} {\bibfnamefont {C.}~\bibnamefont {Ironside}}, \bibinfo
  {author} {\bibfnamefont {A.}~\bibnamefont {Kelly}}, \ and\ \bibinfo {author}
  {\bibfnamefont {J.}~\bibnamefont {Figueiredo}},\ }\bibfield  {title}
  {\enquote {\bibinfo {title} {{A self-synchronized optoelectronic oscillator
  based on an RTD photodetector and a laser diode}},}\ }\href {\doibase
  10.1109/LPT.2011.2154320} {\bibfield  {journal} {\bibinfo  {journal} {IEEE
  Photonics Technology Letters}\ }\textbf {\bibinfo {volume} {23}} (\bibinfo
  {year} {2011}),\ 10.1109/LPT.2011.2154320}\BibitemShut {NoStop}%
\bibitem [{\citenamefont {Romeira}\ \emph
  {et~al.}(2013{\natexlab{c}})\citenamefont {Romeira}, \citenamefont
  {Javaloyes}, \citenamefont {Figueiredo}, \citenamefont {Ironside},
  \citenamefont {Cantu},\ and\ \citenamefont {Kelly}}]{Romeira2013}%
  \BibitemOpen
  \bibfield  {author} {\bibinfo {author} {\bibfnamefont {B.}~\bibnamefont
  {Romeira}}, \bibinfo {author} {\bibfnamefont {J.}~\bibnamefont {Javaloyes}},
  \bibinfo {author} {\bibfnamefont {J.~M.~L.}\ \bibnamefont {Figueiredo}},
  \bibinfo {author} {\bibfnamefont {C.~N.}\ \bibnamefont {Ironside}}, \bibinfo
  {author} {\bibfnamefont {H.~I.}\ \bibnamefont {Cantu}}, \ and\ \bibinfo
  {author} {\bibfnamefont {A.~E.}\ \bibnamefont {Kelly}},\ }\bibfield  {title}
  {\enquote {\bibinfo {title} {{Delayed Feedback Dynamics of Lienard-Type
  Resonant Tunneling-Photo-Detector Optoelectronic Oscillators}},}\ }\href
  {\doibase 10.1109/jqe.2012.2225415} {\bibfield  {journal} {\bibinfo
  {journal} {Ieee Journal of Quantum Electronics}\ }\textbf {\bibinfo {volume}
  {49}},\ \bibinfo {pages} {31--42} (\bibinfo {year}
  {2013}{\natexlab{c}})}\BibitemShut {NoStop}%
\bibitem [{\citenamefont {Mesaritakis}\ \emph {et~al.}(2016)\citenamefont
  {Mesaritakis}, \citenamefont {Kapsalis}, \citenamefont {Bogris},\ and\
  \citenamefont {Syvridis}}]{Mesaritakis2016}%
  \BibitemOpen
  \bibfield  {author} {\bibinfo {author} {\bibfnamefont {C.}~\bibnamefont
  {Mesaritakis}}, \bibinfo {author} {\bibfnamefont {A.}~\bibnamefont
  {Kapsalis}}, \bibinfo {author} {\bibfnamefont {A.}~\bibnamefont {Bogris}}, \
  and\ \bibinfo {author} {\bibfnamefont {D.}~\bibnamefont {Syvridis}},\
  }\bibfield  {title} {\enquote {\bibinfo {title} {Artificial neuron based on
  integrated semiconductor quantum dot mode-locked lasers},}\ }\href@noop {}
  {\bibfield  {journal} {\bibinfo  {journal} {Scientific Reports}\ }\textbf
  {\bibinfo {volume} {6}} (\bibinfo {year} {2016})}\BibitemShut {NoStop}%
\bibitem [{\citenamefont {Romeira}\ \emph {et~al.}(2016)\citenamefont
  {Romeira}, \citenamefont {Av{\'{o}}}, \citenamefont {Figueiredo},
  \citenamefont {Barland},\ and\ \citenamefont {Javaloyes}}]{Romeira2016}%
  \BibitemOpen
  \bibfield  {author} {\bibinfo {author} {\bibfnamefont {B.}~\bibnamefont
  {Romeira}}, \bibinfo {author} {\bibfnamefont {R.}~\bibnamefont {Av{\'{o}}}},
  \bibinfo {author} {\bibfnamefont {J.~M.~L.}\ \bibnamefont {Figueiredo}},
  \bibinfo {author} {\bibfnamefont {S.}~\bibnamefont {Barland}}, \ and\
  \bibinfo {author} {\bibfnamefont {J.}~\bibnamefont {Javaloyes}},\ }\bibfield
  {title} {\enquote {\bibinfo {title} {{Regenerative memory in time-delayed
  neuromorphic photonic resonators}},}\ }\href {\doibase 10.1038/srep19510}
  {\bibfield  {journal} {\bibinfo  {journal} {Scientific Reports}\ }\textbf
  {\bibinfo {volume} {6}},\ \bibinfo {pages} {19510} (\bibinfo {year}
  {2016})},\ \Eprint {http://arxiv.org/abs/1503.07781} {arXiv:1503.07781}
  \BibitemShut {NoStop}%
\bibitem [{\citenamefont {Yanchuk}\ and\ \citenamefont
  {Giacomelli}(2017)}]{YG-JPA-17}%
  \BibitemOpen
  \bibfield  {author} {\bibinfo {author} {\bibfnamefont {S.}~\bibnamefont
  {Yanchuk}}\ and\ \bibinfo {author} {\bibfnamefont {G.}~\bibnamefont
  {Giacomelli}},\ }\bibfield  {title} {\enquote {\bibinfo {title}
  {Spatio-temporal phenomena in complex systems with time delays},}\ }\href
  {http://stacks.iop.org/1751-8121/50/i=10/a=103001} {\bibfield  {journal}
  {\bibinfo  {journal} {Journal of Physics A: Mathematical and Theoretical}\
  }\textbf {\bibinfo {volume} {50}},\ \bibinfo {pages} {103001} (\bibinfo
  {year} {2017})}\BibitemShut {NoStop}%
\bibitem [{\citenamefont {Oshima}\ \emph {et~al.}(2016)\citenamefont {Oshima},
  \citenamefont {Hashimoto}, \citenamefont {Suzuki},\ and\ \citenamefont
  {Asada}}]{Oshima2016}%
  \BibitemOpen
  \bibfield  {author} {\bibinfo {author} {\bibfnamefont {N.}~\bibnamefont
  {Oshima}}, \bibinfo {author} {\bibfnamefont {K.}~\bibnamefont {Hashimoto}},
  \bibinfo {author} {\bibfnamefont {S.}~\bibnamefont {Suzuki}}, \ and\ \bibinfo
  {author} {\bibfnamefont {M.}~\bibnamefont {Asada}},\ }\href {\doibase
  10.1049/el.2016.3120} {\enquote {\bibinfo {title} {{Wireless data
  transmission of 34 Gbit/s at a 500-GHz range using resonant-tunnelling-diode
  terahertz oscillator}},}\ } (\bibinfo {year} {2016})\BibitemShut {NoStop}%
\bibitem [{\citenamefont {Miyamoto}, \citenamefont {Yamaguchi},\ and\
  \citenamefont {Mukai}(2016)}]{Miyamoto2016}%
  \BibitemOpen
  \bibfield  {author} {\bibinfo {author} {\bibfnamefont {T.}~\bibnamefont
  {Miyamoto}}, \bibinfo {author} {\bibfnamefont {A.}~\bibnamefont {Yamaguchi}},
  \ and\ \bibinfo {author} {\bibfnamefont {T.}~\bibnamefont {Mukai}},\
  }\bibfield  {title} {\enquote {\bibinfo {title} {{Terahertz imaging system
  with resonant tunneling diodes}},}\ }\href
  {http://stacks.iop.org/1347-4065/55/i=3/a=032201} {\bibfield  {journal}
  {\bibinfo  {journal} {Japanese Journal of Applied Physics}\ }\textbf
  {\bibinfo {volume} {55}},\ \bibinfo {pages} {32201} (\bibinfo {year}
  {2016})}\BibitemShut {NoStop}%
\bibitem [{\citenamefont {Kamegai}\ \emph {et~al.}(2008)\citenamefont
  {Kamegai}, \citenamefont {Kishimoto}, \citenamefont {Maezawa}, \citenamefont
  {Mizutani}, \citenamefont {Andoh}, \citenamefont {Akamatsu},\ and\
  \citenamefont {Nakata}}]{Kamegai2008}%
  \BibitemOpen
  \bibfield  {author} {\bibinfo {author} {\bibfnamefont {N.}~\bibnamefont
  {Kamegai}}, \bibinfo {author} {\bibfnamefont {S.}~\bibnamefont {Kishimoto}},
  \bibinfo {author} {\bibfnamefont {K.}~\bibnamefont {Maezawa}}, \bibinfo
  {author} {\bibfnamefont {T.}~\bibnamefont {Mizutani}}, \bibinfo {author}
  {\bibfnamefont {H.}~\bibnamefont {Andoh}}, \bibinfo {author} {\bibfnamefont
  {K.}~\bibnamefont {Akamatsu}}, \ and\ \bibinfo {author} {\bibfnamefont
  {H.}~\bibnamefont {Nakata}},\ }\bibfield  {title} {\enquote {\bibinfo {title}
  {{Ultrashort Pulse Generators Using Resonant Tunneling Diodes and Their
  Integration with Antennas on Ceramic Substrates}},}\ }\href
  {http://stacks.iop.org/1347-4065/47/i=4S/a=2833} {\bibfield  {journal}
  {\bibinfo  {journal} {Japanese Journal of Applied Physics}\ }\textbf
  {\bibinfo {volume} {47}},\ \bibinfo {pages} {2833} (\bibinfo {year}
  {2008})}\BibitemShut {NoStop}%
\bibitem [{\citenamefont {Pfenning}\ \emph {et~al.}(2015)\citenamefont
  {Pfenning}, \citenamefont {Hartmann}, \citenamefont {Dias}, \citenamefont
  {Langer}, \citenamefont {Kamp}, \citenamefont {Castelano}, \citenamefont
  {Lopez-Richard}, \citenamefont {Marques}, \citenamefont {H{\"{o}}fling},\
  and\ \citenamefont {Worschech}}]{Pfenning2015}%
  \BibitemOpen
  \bibfield  {author} {\bibinfo {author} {\bibfnamefont {A.}~\bibnamefont
  {Pfenning}}, \bibinfo {author} {\bibfnamefont {F.}~\bibnamefont {Hartmann}},
  \bibinfo {author} {\bibfnamefont {M.~R.~S.}\ \bibnamefont {Dias}}, \bibinfo
  {author} {\bibfnamefont {F.}~\bibnamefont {Langer}}, \bibinfo {author}
  {\bibfnamefont {M.}~\bibnamefont {Kamp}}, \bibinfo {author} {\bibfnamefont
  {L.~K.}\ \bibnamefont {Castelano}}, \bibinfo {author} {\bibfnamefont
  {V.}~\bibnamefont {Lopez-Richard}}, \bibinfo {author} {\bibfnamefont {G.~E.}\
  \bibnamefont {Marques}}, \bibinfo {author} {\bibfnamefont {S.}~\bibnamefont
  {H{\"{o}}fling}}, \ and\ \bibinfo {author} {\bibfnamefont {L.}~\bibnamefont
  {Worschech}},\ }\bibfield  {title} {\enquote {\bibinfo {title}
  {{Photocurrent-voltage relation of resonant tunneling diode
  photodetectors}},}\ }\href {\doibase 10.1063/1.4929424} {\bibfield  {journal}
  {\bibinfo  {journal} {Applied Physics Letters}\ }\textbf {\bibinfo {volume}
  {107}},\ \bibinfo {pages} {81104} (\bibinfo {year} {2015})}\BibitemShut
  {NoStop}%
\bibitem [{\citenamefont {Pfenning}\ \emph {et~al.}(2016)\citenamefont
  {Pfenning}, \citenamefont {Hartmann}, \citenamefont {Langer}, \citenamefont
  {Kamp}, \citenamefont {H{\"{o}}fling},\ and\ \citenamefont
  {Worschech}}]{Pfenning2016}%
  \BibitemOpen
  \bibfield  {author} {\bibinfo {author} {\bibfnamefont {A.}~\bibnamefont
  {Pfenning}}, \bibinfo {author} {\bibfnamefont {F.}~\bibnamefont {Hartmann}},
  \bibinfo {author} {\bibfnamefont {F.}~\bibnamefont {Langer}}, \bibinfo
  {author} {\bibfnamefont {M.}~\bibnamefont {Kamp}}, \bibinfo {author}
  {\bibfnamefont {S.}~\bibnamefont {H{\"{o}}fling}}, \ and\ \bibinfo {author}
  {\bibfnamefont {L.}~\bibnamefont {Worschech}},\ }\bibfield  {title} {\enquote
  {\bibinfo {title} {{Sensitivity of resonant tunneling diode
  photodetectors}},}\ }\href {http://stacks.iop.org/0957-4484/27/i=35/a=355202}
  {\bibfield  {journal} {\bibinfo  {journal} {Nanotechnology}\ }\textbf
  {\bibinfo {volume} {27}},\ \bibinfo {pages} {355202} (\bibinfo {year}
  {2016})}\BibitemShut {NoStop}%
\bibitem [{\citenamefont {Blakesley}\ \emph {et~al.}(2005)\citenamefont
  {Blakesley}, \citenamefont {See}, \citenamefont {Shields}, \citenamefont
  {Kardyna$\backslash$l}, \citenamefont {Atkinson}, \citenamefont {Farrer},\
  and\ \citenamefont {Ritchie}}]{Blakesley2005}%
  \BibitemOpen
  \bibfield  {author} {\bibinfo {author} {\bibfnamefont {J.~C.}\ \bibnamefont
  {Blakesley}}, \bibinfo {author} {\bibfnamefont {P.}~\bibnamefont {See}},
  \bibinfo {author} {\bibfnamefont {A.~J.}\ \bibnamefont {Shields}}, \bibinfo
  {author} {\bibfnamefont {B.~E.}\ \bibnamefont {Kardyna$\backslash$l}},
  \bibinfo {author} {\bibfnamefont {P.}~\bibnamefont {Atkinson}}, \bibinfo
  {author} {\bibfnamefont {I.}~\bibnamefont {Farrer}}, \ and\ \bibinfo {author}
  {\bibfnamefont {D.~A.}\ \bibnamefont {Ritchie}},\ }\bibfield  {title}
  {\enquote {\bibinfo {title} {{Efficient Single Photon Detection by Quantum
  Dot Resonant Tunneling Diodes}},}\ }\href {\doibase
  10.1103/PhysRevLett.94.067401} {\bibfield  {journal} {\bibinfo  {journal}
  {Phys. Rev. Lett.}\ }\textbf {\bibinfo {volume} {94}},\ \bibinfo {pages}
  {67401} (\bibinfo {year} {2005})}\BibitemShut {NoStop}%
\bibitem [{\citenamefont {Li}\ \emph {et~al.}(2008)\citenamefont {Li},
  \citenamefont {Kardyna{\l}}, \citenamefont {Ellis}, \citenamefont {Shields},
  \citenamefont {Farrer},\ and\ \citenamefont {Ritchie}}]{Li2008}%
  \BibitemOpen
  \bibfield  {author} {\bibinfo {author} {\bibfnamefont {H.~W.}\ \bibnamefont
  {Li}}, \bibinfo {author} {\bibfnamefont {B.~E.}\ \bibnamefont {Kardyna{\l}}},
  \bibinfo {author} {\bibfnamefont {D.~J.~P.}\ \bibnamefont {Ellis}}, \bibinfo
  {author} {\bibfnamefont {A.~J.}\ \bibnamefont {Shields}}, \bibinfo {author}
  {\bibfnamefont {I.}~\bibnamefont {Farrer}}, \ and\ \bibinfo {author}
  {\bibfnamefont {D.~A.}\ \bibnamefont {Ritchie}},\ }\bibfield  {title}
  {\enquote {\bibinfo {title} {{Quantum dot resonant tunneling diode single
  photon detector with aluminum oxide aperture defined tunneling area}},}\
  }\href {\doibase 10.1063/1.2978232} {\bibfield  {journal} {\bibinfo
  {journal} {Applied Physics Letters}\ }\textbf {\bibinfo {volume} {93}},\
  \bibinfo {pages} {153503} (\bibinfo {year} {2008})}\BibitemShut {NoStop}%
\bibitem [{\citenamefont {Weng}\ \emph {et~al.}(2015)\citenamefont {Weng},
  \citenamefont {An}, \citenamefont {Zhang}, \citenamefont {Chen},
  \citenamefont {Chen}, \citenamefont {Zhu},\ and\ \citenamefont
  {Lu}}]{Weng2015}%
  \BibitemOpen
  \bibfield  {author} {\bibinfo {author} {\bibfnamefont {Q.}~\bibnamefont
  {Weng}}, \bibinfo {author} {\bibfnamefont {Z.}~\bibnamefont {An}}, \bibinfo
  {author} {\bibfnamefont {B.}~\bibnamefont {Zhang}}, \bibinfo {author}
  {\bibfnamefont {P.}~\bibnamefont {Chen}}, \bibinfo {author} {\bibfnamefont
  {X.}~\bibnamefont {Chen}}, \bibinfo {author} {\bibfnamefont {Z.}~\bibnamefont
  {Zhu}}, \ and\ \bibinfo {author} {\bibfnamefont {W.}~\bibnamefont {Lu}},\
  }\bibfield  {title} {\enquote {\bibinfo {title} {{Quantum dot single-photon
  switches of resonant tunneling current for discriminating-photon-number
  detection}},}\ }\href {http://dx.doi.org/10.1038/srep09389
  http://10.0.4.14/srep09389} {\bibfield  {journal} {\bibinfo  {journal}
  {Scientific Reports}\ }\textbf {\bibinfo {volume} {5}},\ \bibinfo {pages}
  {9389} (\bibinfo {year} {2015})}\BibitemShut {NoStop}%
\bibitem [{\citenamefont {Shimizu}, \citenamefont {Takahashi},\ and\
  \citenamefont {Mitsuaki}(2015)}]{Shimizu2015}%
  \BibitemOpen
  \bibfield  {author} {\bibinfo {author} {\bibfnamefont {M.~N.}\ \bibnamefont
  {Shimizu}}, \bibinfo {author} {\bibfnamefont {T.}~\bibnamefont {Takahashi}},
  \ and\ \bibinfo {author} {\bibnamefont {Mitsuaki}},\ }\bibfield  {title}
  {\enquote {\bibinfo {title} {{Investigating the bistability characteristics
  of GaN/AlN resonant tunneling diodes for ultrafast nonvolatile memory}},}\
  }\href {http://stacks.iop.org/1347-4065/54/i=3/a=034201} {\bibfield
  {journal} {\bibinfo  {journal} {Japanese Journal of Applied Physics}\
  }\textbf {\bibinfo {volume} {54}},\ \bibinfo {pages} {34201} (\bibinfo {year}
  {2015})}\BibitemShut {NoStop}%
\bibitem [{\citenamefont {Klofa{\"{i}}}, \citenamefont {Essimbi},\ and\
  \citenamefont {J{\"{a}}ger}(2015)}]{Klofai2015}%
  \BibitemOpen
  \bibfield  {author} {\bibinfo {author} {\bibfnamefont {Y.}~\bibnamefont
  {Klofa{\"{i}}}}, \bibinfo {author} {\bibfnamefont {B.~Z.}\ \bibnamefont
  {Essimbi}}, \ and\ \bibinfo {author} {\bibfnamefont {D.}~\bibnamefont
  {J{\"{a}}ger}},\ }\bibfield  {title} {\enquote {\bibinfo {title} {{An MMIC
  implementation of FitzHugh–Nagumo neurons using a resonant tunneling diode
  nonlinear transmission line}},}\ }\href
  {http://stacks.iop.org/1402-4896/90/i=2/a=025002} {\bibfield  {journal}
  {\bibinfo  {journal} {Physica Scripta}\ }\textbf {\bibinfo {volume} {90}},\
  \bibinfo {pages} {25002} (\bibinfo {year} {2015})}\BibitemShut {NoStop}%
\bibitem [{\citenamefont {Schulman}, \citenamefont {Santos},\ and\
  \citenamefont {Chow}(1996)}]{Schulman1996}%
  \BibitemOpen
  \bibfield  {author} {\bibinfo {author} {\bibfnamefont {J.~N.}\ \bibnamefont
  {Schulman}}, \bibinfo {author} {\bibfnamefont {H.~J. D.~L.}\ \bibnamefont
  {Santos}}, \ and\ \bibinfo {author} {\bibfnamefont {D.~H.}\ \bibnamefont
  {Chow}},\ }\href {\doibase 10.1109/55.491835} {\enquote {\bibinfo {title}
  {{Physics-based RTD current-voltage equation}},}\ } (\bibinfo {year}
  {1996})\BibitemShut {NoStop}%
\bibitem [{\citenamefont {Hale}()}]{Hale}%
  \BibitemOpen
  \bibfield  {author} {\bibinfo {author} {\bibfnamefont {J.~K.}\ \bibnamefont
  {Hale}},\ }\href@noop {} {\emph {\bibinfo {title} {{Theory of Functional
  Differential Equations}}}}\ (\bibinfo  {publisher} {Springer-Verla},\
  \bibinfo {address} {New York})\BibitemShut {NoStop}%
\bibitem [{\citenamefont {Lins}, \citenamefont {Melo},\ and\ \citenamefont
  {Pugh}(1977)}]{A.LinsW.Melo1977}%
  \BibitemOpen
  \bibfield  {author} {\bibinfo {author} {\bibfnamefont {A.}~\bibnamefont
  {Lins}}, \bibinfo {author} {\bibfnamefont {W.}~\bibnamefont {Melo}}, \ and\
  \bibinfo {author} {\bibfnamefont {C.}~\bibnamefont {Pugh}},\ }\href@noop {}
  {\emph {\bibinfo {title} {{On Lienards Equation}}}}\ (\bibinfo  {publisher}
  {Springer-Verlag},\ \bibinfo {address} {New York},\ \bibinfo {year}
  {1977})\BibitemShut {NoStop}%
\bibitem [{\citenamefont {Figueiredo}\ \emph {et~al.}(2008)\citenamefont
  {Figueiredo}, \citenamefont {Romeira}, \citenamefont {Slight}, \citenamefont
  {Wang}, \citenamefont {Wasige},\ and\ \citenamefont
  {Ironside}}]{Figueiredo2008}%
  \BibitemOpen
  \bibfield  {author} {\bibinfo {author} {\bibfnamefont {J.~M.~L.}\
  \bibnamefont {Figueiredo}}, \bibinfo {author} {\bibfnamefont
  {B.}~\bibnamefont {Romeira}}, \bibinfo {author} {\bibfnamefont {T.~J.}\
  \bibnamefont {Slight}}, \bibinfo {author} {\bibfnamefont {L.}~\bibnamefont
  {Wang}}, \bibinfo {author} {\bibfnamefont {E.}~\bibnamefont {Wasige}}, \ and\
  \bibinfo {author} {\bibfnamefont {C.~N.}\ \bibnamefont {Ironside}},\
  }\bibfield  {title} {\enquote {\bibinfo {title} {Self-oscillation and period
  adding from resonant tunnelling diode-laser diode circuit},}\ }\href
  {\doibase 10.1049/el:20080350} {\bibfield  {journal} {\bibinfo  {journal}
  {Electronics Letters}\ }\textbf {\bibinfo {volume} {44}},\ \bibinfo {pages}
  {876--877} (\bibinfo {year} {2008})}\BibitemShut {NoStop}%
\bibitem [{\citenamefont {Romeira}\ \emph
  {et~al.}(2009{\natexlab{a}})\citenamefont {Romeira}, \citenamefont
  {Figueiredo}, \citenamefont {Slight}, \citenamefont {Wang}, \citenamefont
  {Wasige}, \citenamefont {Ironside}, \citenamefont {Kelly},\ and\
  \citenamefont {Green}}]{Romeira2009a}%
  \BibitemOpen
  \bibfield  {author} {\bibinfo {author} {\bibfnamefont {B.}~\bibnamefont
  {Romeira}}, \bibinfo {author} {\bibfnamefont {J.~M.~L.}\ \bibnamefont
  {Figueiredo}}, \bibinfo {author} {\bibfnamefont {T.~J.}\ \bibnamefont
  {Slight}}, \bibinfo {author} {\bibfnamefont {L.}~\bibnamefont {Wang}},
  \bibinfo {author} {\bibfnamefont {E.}~\bibnamefont {Wasige}}, \bibinfo
  {author} {\bibfnamefont {C.~N.}\ \bibnamefont {Ironside}}, \bibinfo {author}
  {\bibfnamefont {A.~E.}\ \bibnamefont {Kelly}}, \ and\ \bibinfo {author}
  {\bibfnamefont {R.}~\bibnamefont {Green}},\ }\bibfield  {title} {\enquote
  {\bibinfo {title} {{Nonlinear dynamics of resonant tunneling optoelectronic
  circuits for wireless/Optical interfaces}},}\ }\href {\doibase
  10.1109/JQE.2009.2028084} {\bibfield  {journal} {\bibinfo  {journal} {IEEE
  Journal of Quantum Electronics}\ }\textbf {\bibinfo {volume} {45}},\ \bibinfo
  {pages} {1436--1445} (\bibinfo {year} {2009}{\natexlab{a}})}\BibitemShut
  {NoStop}%
\bibitem [{\citenamefont {Romeira}(2008)}]{Romeira2008}%
  \BibitemOpen
  \bibfield  {author} {\bibinfo {author} {\bibfnamefont {B.}~\bibnamefont
  {Romeira}},\ }\bibfield  {title} {{\selectlanguage {English}\enquote
  {\bibinfo {title} {Synchronisation and chaos in a laser diode driven by a
  resonant tunnelling diode},}\ }}\href
  {http://digital-library.theiet.org/content/journals/10.1049/iet-opt_20080024}
  {\bibfield  {journal} {\bibinfo  {journal} {IET Optoelectronics}\ }\textbf
  {\bibinfo {volume} {2}},\ \bibinfo {pages} {211--215(4)} (\bibinfo {year}
  {2008})}\BibitemShut {NoStop}%
\bibitem [{\citenamefont {Romeira}\ \emph
  {et~al.}(2009{\natexlab{b}})\citenamefont {Romeira}, \citenamefont
  {Figueiredo}, \citenamefont {Ironside},\ and\ \citenamefont
  {Slight}}]{Romeira2009}%
  \BibitemOpen
  \bibfield  {author} {\bibinfo {author} {\bibfnamefont {B.}~\bibnamefont
  {Romeira}}, \bibinfo {author} {\bibfnamefont {J.}~\bibnamefont {Figueiredo}},
  \bibinfo {author} {\bibfnamefont {C.}~\bibnamefont {Ironside}}, \ and\
  \bibinfo {author} {\bibfnamefont {T.}~\bibnamefont {Slight}},\ }\bibfield
  {title} {\enquote {\bibinfo {title} {{Chaotic dynamics in resonant tunneling
  optoelectronic voltage controlled oscillators}},}\ }\href {\doibase
  10.1109/LPT.2009.2034129} {\bibfield  {journal} {\bibinfo  {journal} {IEEE
  Photonics Technology Letters}\ }\textbf {\bibinfo {volume} {21}} (\bibinfo
  {year} {2009}{\natexlab{b}}),\ 10.1109/LPT.2009.2034129}\BibitemShut
  {NoStop}%
\bibitem [{\citenamefont {Figueiredo}, \citenamefont {Ironside},\ and\
  \citenamefont {Stanley}(2001)}]{Figueiredo2001}%
  \BibitemOpen
  \bibfield  {author} {\bibinfo {author} {\bibfnamefont {J.~M.~L.}\
  \bibnamefont {Figueiredo}}, \bibinfo {author} {\bibfnamefont {C.~N.}\
  \bibnamefont {Ironside}}, \ and\ \bibinfo {author} {\bibfnamefont {C.~R.}\
  \bibnamefont {Stanley}},\ }\bibfield  {title} {\enquote {\bibinfo {title}
  {Electric field switching in a resonant tunneling diode electroabsorption
  modulator},}\ }\href {\doibase 10.1109/3.970901} {\bibfield  {journal}
  {\bibinfo  {journal} {IEEE Journal of Quantum Electronics}\ }\textbf
  {\bibinfo {volume} {37}},\ \bibinfo {pages} {1547--1552} (\bibinfo {year}
  {2001})}\BibitemShut {NoStop}%
\bibitem [{\citenamefont {Coelho}\ \emph {et~al.}(2004)\citenamefont {Coelho},
  \citenamefont {Martins-Filho}, \citenamefont {Figueiredo},\ and\
  \citenamefont {Ironside}}]{Coelho2004}%
  \BibitemOpen
  \bibfield  {author} {\bibinfo {author} {\bibfnamefont {I.~J.~S.}\
  \bibnamefont {Coelho}}, \bibinfo {author} {\bibfnamefont {J.~F.}\
  \bibnamefont {Martins-Filho}}, \bibinfo {author} {\bibfnamefont {J.~M.~L.}\
  \bibnamefont {Figueiredo}}, \ and\ \bibinfo {author} {\bibfnamefont {C.~N.}\
  \bibnamefont {Ironside}},\ }\bibfield  {title} {\enquote {\bibinfo {title}
  {Modeling of light-sensitive resonant-tunneling-diode devices},}\ }\href
  {\doibase 10.1063/1.1728290} {\bibfield  {journal} {\bibinfo  {journal}
  {Journal of Applied Physics}\ }\textbf {\bibinfo {volume} {95}},\ \bibinfo
  {pages} {8258--8263} (\bibinfo {year} {2004})},\ \Eprint
  {http://arxiv.org/abs/http://dx.doi.org/10.1063/1.1728290}
  {http://dx.doi.org/10.1063/1.1728290} \BibitemShut {NoStop}%
\bibitem [{\citenamefont {Slight}\ and\ \citenamefont
  {Ironside}(2007)}]{Slight2007}%
  \BibitemOpen
  \bibfield  {author} {\bibinfo {author} {\bibfnamefont {T.~J.}\ \bibnamefont
  {Slight}}\ and\ \bibinfo {author} {\bibfnamefont {C.~N.}\ \bibnamefont
  {Ironside}},\ }\bibfield  {title} {\enquote {\bibinfo {title} {Investigation
  into the integration of a resonant tunnelling diode and an optical
  communications laser: Model and experiment},}\ }\href {\doibase
  10.1109/JQE.2007.898847} {\bibfield  {journal} {\bibinfo  {journal} {IEEE
  Journal of Quantum Electronics}\ }\textbf {\bibinfo {volume} {43}},\ \bibinfo
  {pages} {580--587} (\bibinfo {year} {2007})}\BibitemShut {NoStop}%
\bibitem [{\citenamefont {Gardiner}(1995)}]{G-BOOK-95}%
  \BibitemOpen
  \bibfield  {author} {\bibinfo {author} {\bibfnamefont {C.~W.}\ \bibnamefont
  {Gardiner}},\ }\href@noop {} {\emph {\bibinfo {title} {Handbook of stochastic
  methods. 2nd {Ed}.}}}\ (\bibinfo  {publisher} {Springer-Verlag},\ \bibinfo
  {address} {Berlin},\ \bibinfo {year} {1995})\BibitemShut {NoStop}%
\bibitem [{\citenamefont {Hodgkin}\ and\ \citenamefont
  {Huxley}(1952)}]{HH-JOP-52}%
  \BibitemOpen
  \bibfield  {author} {\bibinfo {author} {\bibfnamefont {A.~L.}\ \bibnamefont
  {Hodgkin}}\ and\ \bibinfo {author} {\bibfnamefont {A.~F.}\ \bibnamefont
  {Huxley}},\ }\bibfield  {title} {\enquote {\bibinfo {title} {A quantitative
  description of membrane current and its application to conduction and
  excitation in nerve},}\ }\href {http://stacks.iop.org/0034-4885/61/i=4/a=002}
  {\bibfield  {journal} {\bibinfo  {journal} {Journal of Physiology}\ }\textbf
  {\bibinfo {volume} {117}},\ \bibinfo {pages} {500--544} (\bibinfo {year}
  {1952})}\BibitemShut {NoStop}%
\bibitem [{\citenamefont {Hodgkin}, \citenamefont {Huxley},\ and\ \citenamefont
  {Katz}(1952)}]{HH2-JOP-52}%
  \BibitemOpen
  \bibfield  {author} {\bibinfo {author} {\bibfnamefont {A.~L.}\ \bibnamefont
  {Hodgkin}}, \bibinfo {author} {\bibfnamefont {A.~F.}\ \bibnamefont {Huxley}},
  \ and\ \bibinfo {author} {\bibfnamefont {B.}~\bibnamefont {Katz}},\
  }\bibfield  {title} {\enquote {\bibinfo {title} {Measurement of
  current-voltage relations in the membrane of the giant axon of loligo},}\
  }\href@noop {} {\bibfield  {journal} {\bibinfo  {journal} {The Journal of
  physiology}\ }\textbf {\bibinfo {volume} {116}},\ \bibinfo {pages} {424}
  (\bibinfo {year} {1952})}\BibitemShut {NoStop}%
\bibitem [{\citenamefont {Kuhnert}, \citenamefont {Agladze},\ and\
  \citenamefont {Krinsky}(1989)}]{Kuhnert89}%
  \BibitemOpen
  \bibfield  {author} {\bibinfo {author} {\bibfnamefont {L.}~\bibnamefont
  {Kuhnert}}, \bibinfo {author} {\bibfnamefont {K.~I.}\ \bibnamefont
  {Agladze}}, \ and\ \bibinfo {author} {\bibfnamefont {V.~I.}\ \bibnamefont
  {Krinsky}},\ }\bibfield  {title} {\enquote {\bibinfo {title} {Image
  processing using light-sensitive chemical waves.}}\ }\href@noop {} {\bibfield
   {journal} {\bibinfo  {journal} {Nature}\ }\textbf {\bibinfo {volume}
  {337}},\ \bibinfo {pages} {244--247} (\bibinfo {year} {1989})}\BibitemShut
  {NoStop}%
\bibitem [{\citenamefont {Pedaci}\ \emph {et~al.}(2010)\citenamefont {Pedaci},
  \citenamefont {Huang}, \citenamefont {van Oene}, \citenamefont {Barland},\
  and\ \citenamefont {Dekker}}]{Pedaci10}%
  \BibitemOpen
  \bibfield  {author} {\bibinfo {author} {\bibfnamefont {F.}~\bibnamefont
  {Pedaci}}, \bibinfo {author} {\bibfnamefont {Z.}~\bibnamefont {Huang}},
  \bibinfo {author} {\bibfnamefont {M.}~\bibnamefont {van Oene}}, \bibinfo
  {author} {\bibfnamefont {S.}~\bibnamefont {Barland}}, \ and\ \bibinfo
  {author} {\bibfnamefont {N.~H.}\ \bibnamefont {Dekker}},\ }\bibfield  {title}
  {\enquote {\bibinfo {title} {Excitable particles in an optical torque
  wrench},}\ }\href {\doibase 10.1038/nphys1862} {\bibfield  {journal}
  {\bibinfo  {journal} {Nat Phys}\ }\textbf {\bibinfo {volume} {7}},\ \bibinfo
  {pages} {259--264} (\bibinfo {year} {2010})}\BibitemShut {NoStop}%
\bibitem [{\citenamefont {Samardak}\ \emph {et~al.}(2011)\citenamefont
  {Samardak}, \citenamefont {Nogaret}, \citenamefont {Janson}, \citenamefont
  {Balanov}, \citenamefont {Farrer},\ and\ \citenamefont
  {Ritchie}}]{SNJ-JAP-11}%
  \BibitemOpen
  \bibfield  {author} {\bibinfo {author} {\bibfnamefont {A.}~\bibnamefont
  {Samardak}}, \bibinfo {author} {\bibfnamefont {A.}~\bibnamefont {Nogaret}},
  \bibinfo {author} {\bibfnamefont {N.}~\bibnamefont {Janson}}, \bibinfo
  {author} {\bibfnamefont {A.}~\bibnamefont {Balanov}}, \bibinfo {author}
  {\bibfnamefont {I.}~\bibnamefont {Farrer}}, \ and\ \bibinfo {author}
  {\bibfnamefont {D.}~\bibnamefont {Ritchie}},\ }\bibfield  {title} {\enquote
  {\bibinfo {title} {Spiking computation and stochastic amplification in a
  neuron-like semiconductor microstructure},}\ }\href@noop {} {\bibfield
  {journal} {\bibinfo  {journal} {Journal of Applied Physics}\ }\textbf
  {\bibinfo {volume} {109}} (\bibinfo {year} {2011})}\BibitemShut {NoStop}%
\bibitem [{\citenamefont {Barbay}, \citenamefont {Kuszelewicz},\ and\
  \citenamefont {Yacomotti}(2011)}]{BKY-OL-11}%
  \BibitemOpen
  \bibfield  {author} {\bibinfo {author} {\bibfnamefont {S.}~\bibnamefont
  {Barbay}}, \bibinfo {author} {\bibfnamefont {R.}~\bibnamefont {Kuszelewicz}},
  \ and\ \bibinfo {author} {\bibfnamefont {A.~M.}\ \bibnamefont {Yacomotti}},\
  }\bibfield  {title} {\enquote {\bibinfo {title} {Excitability in a
  semiconductor laser with saturable absorber},}\ }\href@noop {} {\bibfield
  {journal} {\bibinfo  {journal} {Optics letters}\ }\textbf {\bibinfo {volume}
  {36}},\ \bibinfo {pages} {4476--4478} (\bibinfo {year} {2011})}\BibitemShut
  {NoStop}%
\bibitem [{\citenamefont {Goulding}\ \emph {et~al.}(2007)\citenamefont
  {Goulding}, \citenamefont {Hegarty}, \citenamefont {Rasskazov}, \citenamefont
  {Melnik}, \citenamefont {Hartnett}, \citenamefont {Greene}, \citenamefont
  {McInerney}, \citenamefont {Rachinskii},\ and\ \citenamefont
  {Huyet}}]{GHR-PRL-07}%
  \BibitemOpen
  \bibfield  {author} {\bibinfo {author} {\bibfnamefont {D.}~\bibnamefont
  {Goulding}}, \bibinfo {author} {\bibfnamefont {S.~P.}\ \bibnamefont
  {Hegarty}}, \bibinfo {author} {\bibfnamefont {O.}~\bibnamefont {Rasskazov}},
  \bibinfo {author} {\bibfnamefont {S.}~\bibnamefont {Melnik}}, \bibinfo
  {author} {\bibfnamefont {M.}~\bibnamefont {Hartnett}}, \bibinfo {author}
  {\bibfnamefont {G.}~\bibnamefont {Greene}}, \bibinfo {author} {\bibfnamefont
  {J.~G.}\ \bibnamefont {McInerney}}, \bibinfo {author} {\bibfnamefont
  {D.}~\bibnamefont {Rachinskii}}, \ and\ \bibinfo {author} {\bibfnamefont
  {G.}~\bibnamefont {Huyet}},\ }\bibfield  {title} {\enquote {\bibinfo {title}
  {Excitability in a quantum dot semiconductor laser with optical injection},}\
  }\href {\doibase 10.1103/PhysRevLett.98.153903} {\bibfield  {journal}
  {\bibinfo  {journal} {Phys. Rev. Lett.}\ }\textbf {\bibinfo {volume} {98}},\
  \bibinfo {pages} {153903} (\bibinfo {year} {2007})}\BibitemShut {NoStop}%
\bibitem [{\citenamefont {Selmi}\ \emph {et~al.}(2014)\citenamefont {Selmi},
  \citenamefont {Braive}, \citenamefont {Beaudoin}, \citenamefont {Sagnes},
  \citenamefont {Kuszelewicz},\ and\ \citenamefont {Barbay}}]{SBB-PRL-14}%
  \BibitemOpen
  \bibfield  {author} {\bibinfo {author} {\bibfnamefont {F.}~\bibnamefont
  {Selmi}}, \bibinfo {author} {\bibfnamefont {R.}~\bibnamefont {Braive}},
  \bibinfo {author} {\bibfnamefont {G.}~\bibnamefont {Beaudoin}}, \bibinfo
  {author} {\bibfnamefont {I.}~\bibnamefont {Sagnes}}, \bibinfo {author}
  {\bibfnamefont {R.}~\bibnamefont {Kuszelewicz}}, \ and\ \bibinfo {author}
  {\bibfnamefont {S.}~\bibnamefont {Barbay}},\ }\bibfield  {title} {\enquote
  {\bibinfo {title} {Relative refractory period in an excitable semiconductor
  laser},}\ }\href {\doibase 10.1103/PhysRevLett.112.183902} {\bibfield
  {journal} {\bibinfo  {journal} {Phys. Rev. Lett.}\ }\textbf {\bibinfo
  {volume} {112}},\ \bibinfo {pages} {183902} (\bibinfo {year}
  {2014})}\BibitemShut {NoStop}%
\bibitem [{\citenamefont {Desroches}\ \emph {et~al.}(2012)\citenamefont
  {Desroches}, \citenamefont {Guckenheimer}, \citenamefont {Krauskopf},
  \citenamefont {Kuehn}, \citenamefont {Osinga},\ and\ \citenamefont
  {Wechselberger}}]{Desroches2012}%
  \BibitemOpen
  \bibfield  {author} {\bibinfo {author} {\bibfnamefont {M.}~\bibnamefont
  {Desroches}}, \bibinfo {author} {\bibfnamefont {J.}~\bibnamefont
  {Guckenheimer}}, \bibinfo {author} {\bibfnamefont {B.}~\bibnamefont
  {Krauskopf}}, \bibinfo {author} {\bibfnamefont {C.}~\bibnamefont {Kuehn}},
  \bibinfo {author} {\bibfnamefont {H.~M.}\ \bibnamefont {Osinga}}, \ and\
  \bibinfo {author} {\bibfnamefont {M.}~\bibnamefont {Wechselberger}},\
  }\bibfield  {title} {\enquote {\bibinfo {title} {Mixed-mode oscillations with
  multiple time scales},}\ }\href {\doibase 10.1137/100791233} {\bibfield
  {journal} {\bibinfo  {journal} {SIAM Review}\ }\textbf {\bibinfo {volume}
  {54}},\ \bibinfo {pages} {211--288} (\bibinfo {year} {2012})},\ \Eprint
  {http://arxiv.org/abs/http://dx.doi.org/10.1137/100791233}
  {http://dx.doi.org/10.1137/100791233} \BibitemShut {NoStop}%
\bibitem [{\citenamefont {Leo}\ \emph {et~al.}(2010{\natexlab{a}})\citenamefont
  {Leo}, \citenamefont {Coen}, \citenamefont {Kockaert}, \citenamefont {Gorza},
  \citenamefont {Emplit},\ and\ \citenamefont {Haelterman}}]{Leo2010}%
  \BibitemOpen
  \bibfield  {author} {\bibinfo {author} {\bibfnamefont {F.}~\bibnamefont
  {Leo}}, \bibinfo {author} {\bibfnamefont {S.}~\bibnamefont {Coen}}, \bibinfo
  {author} {\bibfnamefont {P.}~\bibnamefont {Kockaert}}, \bibinfo {author}
  {\bibfnamefont {S.-P.}\ \bibnamefont {Gorza}}, \bibinfo {author}
  {\bibfnamefont {P.}~\bibnamefont {Emplit}}, \ and\ \bibinfo {author}
  {\bibfnamefont {M.}~\bibnamefont {Haelterman}},\ }\bibfield  {title}
  {\enquote {\bibinfo {title} {{Temporal cavity solitons in one-dimensional
  Kerr media as bits in an all-optical buffer}},}\ }\href
  {http://dx.doi.org/10.1038/nphoton.2010.120
  http://www.nature.com/nphoton/journal/v4/n7/suppinfo/nphoton.2010.120{\_}S1.html}
  {\bibfield  {journal} {\bibinfo  {journal} {Nat Photon}\ }\textbf {\bibinfo
  {volume} {4}},\ \bibinfo {pages} {471--476} (\bibinfo {year}
  {2010}{\natexlab{a}})}\BibitemShut {NoStop}%
\bibitem [{\citenamefont {Buri\'c}\ and\ \citenamefont
  {Todorovi\'c}(2003)}]{Buri2003}%
  \BibitemOpen
  \bibfield  {author} {\bibinfo {author} {\bibfnamefont {N.}~\bibnamefont
  {Buri\'c}}\ and\ \bibinfo {author} {\bibfnamefont {D.}~\bibnamefont
  {Todorovi\'c}},\ }\bibfield  {title} {\enquote {\bibinfo {title} {{Dynamics
  of FitzHugh-Nagumo excitable systems with delayed coupling}},}\ }\href
  {\doibase 10.1103/PhysRevE.67.066222} {\bibfield  {journal} {\bibinfo
  {journal} {Phys. Rev. E}\ }\textbf {\bibinfo {volume} {67}},\ \bibinfo
  {pages} {66222} (\bibinfo {year} {2003})}\BibitemShut {NoStop}%
\bibitem [{\citenamefont {Stepan}(2009)}]{Stepan2009}%
  \BibitemOpen
  \bibfield  {author} {\bibinfo {author} {\bibfnamefont {G.}~\bibnamefont
  {Stepan}},\ }\bibfield  {title} {\enquote {\bibinfo {title} {{Delay effects
  in brain dynamics}},}\ }\href {\doibase 10.1098/rsta.2008.0279} {\bibfield
  {journal} {\bibinfo  {journal} {Philosophical Transactions of the Royal
  Society of London A: Mathematical, Physical and Engineering Sciences}\
  }\textbf {\bibinfo {volume} {367}},\ \bibinfo {pages} {1059--1062} (\bibinfo
  {year} {2009})}\BibitemShut {NoStop}%
\bibitem [{\citenamefont {Yacomotti}\ \emph {et~al.}(2002)\citenamefont
  {Yacomotti}, \citenamefont {Mindlin}, \citenamefont {Giudici}, \citenamefont
  {Balle}, \citenamefont {Barland},\ and\ \citenamefont
  {Tredicce}}]{Yacomotti2002}%
  \BibitemOpen
  \bibfield  {author} {\bibinfo {author} {\bibfnamefont {A.~M.}\ \bibnamefont
  {Yacomotti}}, \bibinfo {author} {\bibfnamefont {G.~B.}\ \bibnamefont
  {Mindlin}}, \bibinfo {author} {\bibfnamefont {M.}~\bibnamefont {Giudici}},
  \bibinfo {author} {\bibfnamefont {S.}~\bibnamefont {Balle}}, \bibinfo
  {author} {\bibfnamefont {S.}~\bibnamefont {Barland}}, \ and\ \bibinfo
  {author} {\bibfnamefont {J.}~\bibnamefont {Tredicce}},\ }\bibfield  {title}
  {\enquote {\bibinfo {title} {{Coupled optical excitable cells}},}\ }\href
  {\doibase 10.1103/PhysRevE.66.036227} {\bibfield  {journal} {\bibinfo
  {journal} {Phys. Rev. E}\ }\textbf {\bibinfo {volume} {66}},\ \bibinfo
  {pages} {36227} (\bibinfo {year} {2002})}\BibitemShut {NoStop}%
\bibitem [{\citenamefont {Sch{\"{o}}ll}\ \emph {et~al.}(2009)\citenamefont
  {Sch{\"{o}}ll}, \citenamefont {Hiller}, \citenamefont {H{\"{o}}vel},\ and\
  \citenamefont {Dahlem}}]{Scholl2009}%
  \BibitemOpen
  \bibfield  {author} {\bibinfo {author} {\bibfnamefont {E.}~\bibnamefont
  {Sch{\"{o}}ll}}, \bibinfo {author} {\bibfnamefont {G.}~\bibnamefont
  {Hiller}}, \bibinfo {author} {\bibfnamefont {P.}~\bibnamefont {H{\"{o}}vel}},
  \ and\ \bibinfo {author} {\bibfnamefont {M.~A.}\ \bibnamefont {Dahlem}},\
  }\bibfield  {title} {\enquote {\bibinfo {title} {{Time-delayed feedback in
  neurosystems}},}\ }\href {\doibase 10.1098/rsta.2008.0258} {\bibfield
  {journal} {\bibinfo  {journal} {Philosophical Transactions of the Royal
  Society of London A: Mathematical, Physical and Engineering Sciences}\
  }\textbf {\bibinfo {volume} {367}},\ \bibinfo {pages} {1079--1096} (\bibinfo
  {year} {2009})}\BibitemShut {NoStop}%
\bibitem [{\citenamefont {Kelleher}\ \emph {et~al.}(2010)\citenamefont
  {Kelleher}, \citenamefont {Bonatto}, \citenamefont {Skoda}, \citenamefont
  {Hegarty},\ and\ \citenamefont {Huyet}}]{Kelleher2010}%
  \BibitemOpen
  \bibfield  {author} {\bibinfo {author} {\bibfnamefont {B.}~\bibnamefont
  {Kelleher}}, \bibinfo {author} {\bibfnamefont {C.}~\bibnamefont {Bonatto}},
  \bibinfo {author} {\bibfnamefont {P.}~\bibnamefont {Skoda}}, \bibinfo
  {author} {\bibfnamefont {S.~P.}\ \bibnamefont {Hegarty}}, \ and\ \bibinfo
  {author} {\bibfnamefont {G.}~\bibnamefont {Huyet}},\ }\bibfield  {title}
  {\enquote {\bibinfo {title} {{Excitation regeneration in delay-coupled
  oscillators}},}\ }\href {\doibase 10.1103/PhysRevE.81.036204} {\bibfield
  {journal} {\bibinfo  {journal} {Phys. Rev. E}\ }\textbf {\bibinfo {volume}
  {81}},\ \bibinfo {pages} {36204} (\bibinfo {year} {2010})}\BibitemShut
  {NoStop}%
\bibitem [{\citenamefont {Weicker}\ \emph {et~al.}(2014)\citenamefont
  {Weicker}, \citenamefont {Erneux}, \citenamefont {Keuninckx},\ and\
  \citenamefont {Danckaert}}]{Weicker2014}%
  \BibitemOpen
  \bibfield  {author} {\bibinfo {author} {\bibfnamefont {L.}~\bibnamefont
  {Weicker}}, \bibinfo {author} {\bibfnamefont {T.}~\bibnamefont {Erneux}},
  \bibinfo {author} {\bibfnamefont {L.}~\bibnamefont {Keuninckx}}, \ and\
  \bibinfo {author} {\bibfnamefont {J.}~\bibnamefont {Danckaert}},\ }\bibfield
  {title} {\enquote {\bibinfo {title} {{Analytical and experimental study of
  two delay-coupled excitable units}},}\ }\href {\doibase
  10.1103/PhysRevE.89.012908} {\bibfield  {journal} {\bibinfo  {journal} {Phys.
  Rev. E}\ }\textbf {\bibinfo {volume} {89}},\ \bibinfo {pages} {12908}
  (\bibinfo {year} {2014})}\BibitemShut {NoStop}%
\bibitem [{\citenamefont {Ikeda}(1979)}]{Ikeda1979}%
  \BibitemOpen
  \bibfield  {author} {\bibinfo {author} {\bibfnamefont {K.}~\bibnamefont
  {Ikeda}},\ }\bibfield  {title} {\enquote {\bibinfo {title} {{Multiple-valued
  stationary state and its instability of the transmitted light by a ring
  cavity system}},}\ }\href {\doibase
  http://dx.doi.org/10.1016/0030-4018(79)90090-7} {\bibfield  {journal}
  {\bibinfo  {journal} {Optics Communications}\ }\textbf {\bibinfo {volume}
  {30}},\ \bibinfo {pages} {257--261} (\bibinfo {year} {1979})}\BibitemShut
  {NoStop}%
\bibitem [{\citenamefont {Neyer}\ and\ \citenamefont
  {Voges}(1982)}]{Neyer1982}%
  \BibitemOpen
  \bibfield  {author} {\bibinfo {author} {\bibfnamefont {A.}~\bibnamefont
  {Neyer}}\ and\ \bibinfo {author} {\bibfnamefont {E.}~\bibnamefont {Voges}},\
  }\href {\doibase 10.1109/JQE.1982.1071487} {\enquote {\bibinfo {title}
  {{Dynamics of electrooptic bistable devices with delayed feedback}},}\ }
  (\bibinfo {year} {1982})\BibitemShut {NoStop}%
\bibitem [{\citenamefont {Aida}\ and\ \citenamefont {Davis}(1992)}]{Aida1992}%
  \BibitemOpen
  \bibfield  {author} {\bibinfo {author} {\bibfnamefont {T.}~\bibnamefont
  {Aida}}\ and\ \bibinfo {author} {\bibfnamefont {P.}~\bibnamefont {Davis}},\
  }\href {\doibase 10.1109/3.124994} {\enquote {\bibinfo {title} {{Oscillation
  modes of laser diode pumped hybrid bistable system with large delay and
  application to dynamical memory}},}\ } (\bibinfo {year} {1992})\BibitemShut
  {NoStop}%
\bibitem [{\citenamefont {Pyragas}(1992)}]{P-PLA-92}%
  \BibitemOpen
  \bibfield  {author} {\bibinfo {author} {\bibfnamefont {K.}~\bibnamefont
  {Pyragas}},\ }\bibfield  {title} {\enquote {\bibinfo {title} {Continuous
  control of chaos by self-controlling feedback},}\ }\href {\doibase
  http://dx.doi.org/10.1016/0375-9601(92)90745-8} {\bibfield  {journal}
  {\bibinfo  {journal} {Physics Letters A}\ }\textbf {\bibinfo {volume}
  {170}},\ \bibinfo {pages} {421 -- 428} (\bibinfo {year} {1992})}\BibitemShut
  {NoStop}%
\bibitem [{\citenamefont {Giacomelli}\ and\ \citenamefont
  {Politi}(1996)}]{Giacomelli1996}%
  \BibitemOpen
  \bibfield  {author} {\bibinfo {author} {\bibfnamefont {G.}~\bibnamefont
  {Giacomelli}}\ and\ \bibinfo {author} {\bibfnamefont {A.}~\bibnamefont
  {Politi}},\ }\bibfield  {title} {\enquote {\bibinfo {title} {{Relationship
  between Delayed and Spatially Extended Dynamical Systems}},}\ }\href
  {\doibase 10.1103/PhysRevLett.76.2686} {\bibfield  {journal} {\bibinfo
  {journal} {Phys. Rev. Lett.}\ }\textbf {\bibinfo {volume} {76}},\ \bibinfo
  {pages} {2686--2689} (\bibinfo {year} {1996})}\BibitemShut {NoStop}%
\bibitem [{\citenamefont {Umbanhowar}, \citenamefont {Melo},\ and\
  \citenamefont {Swinney}(1996)}]{Umbanhowar1996}%
  \BibitemOpen
  \bibfield  {author} {\bibinfo {author} {\bibfnamefont {P.~B.}\ \bibnamefont
  {Umbanhowar}}, \bibinfo {author} {\bibfnamefont {F.}~\bibnamefont {Melo}}, \
  and\ \bibinfo {author} {\bibfnamefont {H.~L.}\ \bibnamefont {Swinney}},\
  }\bibfield  {title} {\enquote {\bibinfo {title} {{Localized excitations in a
  vertically vibrated granular layer}},}\ }\href
  {http://dx.doi.org/10.1038/382793a0} {\bibfield  {journal} {\bibinfo
  {journal} {Nature}\ }\textbf {\bibinfo {volume} {382}},\ \bibinfo {pages}
  {793--796} (\bibinfo {year} {1996})}\BibitemShut {NoStop}%
\bibitem [{\citenamefont {Astrov}\ and\ \citenamefont
  {Purwins}(2001)}]{Astrov2001}%
  \BibitemOpen
  \bibfield  {author} {\bibinfo {author} {\bibfnamefont {Y.~A.}\ \bibnamefont
  {Astrov}}\ and\ \bibinfo {author} {\bibfnamefont {H.-G.}\ \bibnamefont
  {Purwins}},\ }\bibfield  {title} {\enquote {\bibinfo {title} {{Plasma spots
  in a gas discharge system: birth, scattering and formation of molecules}},}\
  }\href {\doibase http://doi.org/10.1016/S0375-9601(01)00257-2} {\bibfield
  {journal} {\bibinfo  {journal} {Physics Letters A}\ }\textbf {\bibinfo
  {volume} {283}},\ \bibinfo {pages} {349--354} (\bibinfo {year}
  {2001})}\BibitemShut {NoStop}%
\bibitem [{\citenamefont {Lee}\ \emph {et~al.}(1994)\citenamefont {Lee},
  \citenamefont {McCormick}, \citenamefont {Pearson},\ and\ \citenamefont
  {Swinney}}]{Lee1994}%
  \BibitemOpen
  \bibfield  {author} {\bibinfo {author} {\bibfnamefont {K.-J.}\ \bibnamefont
  {Lee}}, \bibinfo {author} {\bibfnamefont {W.~D.}\ \bibnamefont {McCormick}},
  \bibinfo {author} {\bibfnamefont {J.~E.}\ \bibnamefont {Pearson}}, \ and\
  \bibinfo {author} {\bibfnamefont {H.~L.}\ \bibnamefont {Swinney}},\
  }\bibfield  {title} {\enquote {\bibinfo {title} {{Experimental observation of
  self-replicating spots in a reaction-diffusion system}},}\ }\href
  {http://dx.doi.org/10.1038/369215a0} {\bibfield  {journal} {\bibinfo
  {journal} {Nature}\ }\textbf {\bibinfo {volume} {369}},\ \bibinfo {pages}
  {215--218} (\bibinfo {year} {1994})}\BibitemShut {NoStop}%
\bibitem [{\citenamefont {Wu}, \citenamefont {Keolian},\ and\ \citenamefont
  {Rudnick}(1984)}]{Wu1994}%
  \BibitemOpen
  \bibfield  {author} {\bibinfo {author} {\bibfnamefont {J.}~\bibnamefont
  {Wu}}, \bibinfo {author} {\bibfnamefont {R.}~\bibnamefont {Keolian}}, \ and\
  \bibinfo {author} {\bibfnamefont {I.}~\bibnamefont {Rudnick}},\ }\bibfield
  {title} {\enquote {\bibinfo {title} {{Observation of a Nonpropagating
  Hydrodynamic Soliton}},}\ }\href {\doibase 10.1103/PhysRevLett.52.1421}
  {\bibfield  {journal} {\bibinfo  {journal} {Phys. Rev. Lett.}\ }\textbf
  {\bibinfo {volume} {52}},\ \bibinfo {pages} {1421--1424} (\bibinfo {year}
  {1984})}\BibitemShut {NoStop}%
\bibitem [{\citenamefont {Moses}, \citenamefont {Fineberg},\ and\ \citenamefont
  {Steinberg}(1987)}]{Moses1987}%
  \BibitemOpen
  \bibfield  {author} {\bibinfo {author} {\bibfnamefont {E.}~\bibnamefont
  {Moses}}, \bibinfo {author} {\bibfnamefont {J.}~\bibnamefont {Fineberg}}, \
  and\ \bibinfo {author} {\bibfnamefont {V.}~\bibnamefont {Steinberg}},\
  }\bibfield  {title} {\enquote {\bibinfo {title} {{Multistability and confined
  traveling-wave patterns in a convecting binary mixture}},}\ }\href {\doibase
  10.1103/PhysRevA.35.2757} {\bibfield  {journal} {\bibinfo  {journal} {Phys.
  Rev. A}\ }\textbf {\bibinfo {volume} {35}},\ \bibinfo {pages} {2757--2760}
  (\bibinfo {year} {1987})}\BibitemShut {NoStop}%
\bibitem [{\citenamefont {Barland}\ \emph {et~al.}(2002)\citenamefont
  {Barland}, \citenamefont {Tredicce}, \citenamefont {Brambilla}, \citenamefont
  {Lugiato}, \citenamefont {Balle}, \citenamefont {Giudici}, \citenamefont
  {Maggipinto}, \citenamefont {Spinelli}, \citenamefont {Tissoni},
  \citenamefont {Knodl}, \citenamefont {Miller},\ and\ \citenamefont
  {Jager}}]{Barland2002}%
  \BibitemOpen
  \bibfield  {author} {\bibinfo {author} {\bibfnamefont {S.}~\bibnamefont
  {Barland}}, \bibinfo {author} {\bibfnamefont {J.~R.}\ \bibnamefont
  {Tredicce}}, \bibinfo {author} {\bibfnamefont {M.}~\bibnamefont {Brambilla}},
  \bibinfo {author} {\bibfnamefont {L.~A.}\ \bibnamefont {Lugiato}}, \bibinfo
  {author} {\bibfnamefont {S.}~\bibnamefont {Balle}}, \bibinfo {author}
  {\bibfnamefont {M.}~\bibnamefont {Giudici}}, \bibinfo {author} {\bibfnamefont
  {T.}~\bibnamefont {Maggipinto}}, \bibinfo {author} {\bibfnamefont
  {L.}~\bibnamefont {Spinelli}}, \bibinfo {author} {\bibfnamefont
  {G.}~\bibnamefont {Tissoni}}, \bibinfo {author} {\bibfnamefont
  {T.}~\bibnamefont {Knodl}}, \bibinfo {author} {\bibfnamefont
  {M.}~\bibnamefont {Miller}}, \ and\ \bibinfo {author} {\bibfnamefont
  {R.}~\bibnamefont {Jager}},\ }\bibfield  {title} {\enquote {\bibinfo {title}
  {{Cavity solitons as pixels in semiconductor microcavities}},}\ }\href
  {http://dx.doi.org/10.1038/nature01049} {\bibfield  {journal} {\bibinfo
  {journal} {Nature}\ }\textbf {\bibinfo {volume} {419}},\ \bibinfo {pages}
  {699--702} (\bibinfo {year} {2002})}\BibitemShut {NoStop}%
\bibitem [{\citenamefont {Marino}, \citenamefont {Giacomelli},\ and\
  \citenamefont {Barland}(2014)}]{Marino2014}%
  \BibitemOpen
  \bibfield  {author} {\bibinfo {author} {\bibfnamefont {F.}~\bibnamefont
  {Marino}}, \bibinfo {author} {\bibfnamefont {G.}~\bibnamefont {Giacomelli}},
  \ and\ \bibinfo {author} {\bibfnamefont {S.}~\bibnamefont {Barland}},\
  }\bibfield  {title} {\enquote {\bibinfo {title} {{Front Pinning and Localized
  States Analogues in Long-Delayed Bistable Systems}},}\ }\href {\doibase
  10.1103/PhysRevLett.112.103901} {\bibfield  {journal} {\bibinfo  {journal}
  {Phys. Rev. Lett.}\ }\textbf {\bibinfo {volume} {112}},\ \bibinfo {pages}
  {103901} (\bibinfo {year} {2014})}\BibitemShut {NoStop}%
\bibitem [{\citenamefont {Marconi}\ \emph {et~al.}(2014)\citenamefont
  {Marconi}, \citenamefont {Javaloyes}, \citenamefont {Balle},\ and\
  \citenamefont {Giudici}}]{MJB-PRL-14}%
  \BibitemOpen
  \bibfield  {author} {\bibinfo {author} {\bibfnamefont {M.}~\bibnamefont
  {Marconi}}, \bibinfo {author} {\bibfnamefont {J.}~\bibnamefont {Javaloyes}},
  \bibinfo {author} {\bibfnamefont {S.}~\bibnamefont {Balle}}, \ and\ \bibinfo
  {author} {\bibfnamefont {M.}~\bibnamefont {Giudici}},\ }\bibfield  {title}
  {\enquote {\bibinfo {title} {How lasing localized structures evolve out of
  passive mode locking},}\ }\href {\doibase 10.1103/PhysRevLett.112.223901}
  {\bibfield  {journal} {\bibinfo  {journal} {Phys. Rev. Lett.}\ }\textbf
  {\bibinfo {volume} {112}},\ \bibinfo {pages} {223901} (\bibinfo {year}
  {2014})}\BibitemShut {NoStop}%
\bibitem [{\citenamefont {Garbin}\ \emph {et~al.}(2015)\citenamefont {Garbin},
  \citenamefont {Javaloyes}, \citenamefont {Tissoni},\ and\ \citenamefont
  {Barland}}]{Garbin2015}%
  \BibitemOpen
  \bibfield  {author} {\bibinfo {author} {\bibfnamefont {B.}~\bibnamefont
  {Garbin}}, \bibinfo {author} {\bibfnamefont {J.}~\bibnamefont {Javaloyes}},
  \bibinfo {author} {\bibfnamefont {G.}~\bibnamefont {Tissoni}}, \ and\
  \bibinfo {author} {\bibfnamefont {S.}~\bibnamefont {Barland}},\ }\bibfield
  {title} {\enquote {\bibinfo {title} {{Topological solitons as addressable
  phase bits in a driven laser}},}\ }\href
  {http://dx.doi.org/10.1038/ncomms6915 http://10.0.4.14/ncomms6915
  http://www.nature.com/articles/ncomms6915{\#}supplementary-information}
  {\bibfield  {journal} {\bibinfo  {journal} {Nature Communications}\ }\textbf
  {\bibinfo {volume} {6}},\ \bibinfo {pages} {5915} (\bibinfo {year}
  {2015})}\BibitemShut {NoStop}%
\bibitem [{\citenamefont {Marconi}\ \emph {et~al.}(2015)\citenamefont
  {Marconi}, \citenamefont {Javaloyes}, \citenamefont {Barland}, \citenamefont
  {Balle},\ and\ \citenamefont {Giudici}}]{MJB-NAP-15}%
  \BibitemOpen
  \bibfield  {author} {\bibinfo {author} {\bibfnamefont {M.}~\bibnamefont
  {Marconi}}, \bibinfo {author} {\bibfnamefont {J.}~\bibnamefont {Javaloyes}},
  \bibinfo {author} {\bibfnamefont {S.}~\bibnamefont {Barland}}, \bibinfo
  {author} {\bibfnamefont {S.}~\bibnamefont {Balle}}, \ and\ \bibinfo {author}
  {\bibfnamefont {M.}~\bibnamefont {Giudici}},\ }\bibfield  {title} {\enquote
  {\bibinfo {title} {Vectorial dissipative solitons in vertical-cavity
  surface-emitting lasers with delays},}\ }\href {\doibase DOI:
  10.1038/NPHOTON.2015.92} {\bibfield  {journal} {\bibinfo  {journal} {Nature
  Photonics}\ }\textbf {\bibinfo {volume} {9}},\ \bibinfo {pages} {450--455}
  (\bibinfo {year} {2015})}\BibitemShut {NoStop}%
\bibitem [{\citenamefont {Javaloyes}, \citenamefont {Ackemann},\ and\
  \citenamefont {Hurtado}(2015)}]{JAH-PRL-15}%
  \BibitemOpen
  \bibfield  {author} {\bibinfo {author} {\bibfnamefont {J.}~\bibnamefont
  {Javaloyes}}, \bibinfo {author} {\bibfnamefont {T.}~\bibnamefont {Ackemann}},
  \ and\ \bibinfo {author} {\bibfnamefont {A.}~\bibnamefont {Hurtado}},\
  }\bibfield  {title} {\enquote {\bibinfo {title} {Arrest of domain coarsening
  via anti-periodic regimes in delay systems},}\ }\href {\doibase
  10.1103/PhysRevLett.112.223901} {\bibfield  {journal} {\bibinfo  {journal}
  {Phys. Rev. Lett.}\ }\textbf {\bibinfo {volume} {115}},\ \bibinfo {pages}
  {223901} (\bibinfo {year} {2015})}\BibitemShut {NoStop}%
\bibitem [{\citenamefont {Coullet}, \citenamefont {Riera},\ and\ \citenamefont
  {Tresser}(2004)}]{Coullet2004}%
  \BibitemOpen
  \bibfield  {author} {\bibinfo {author} {\bibfnamefont {P.}~\bibnamefont
  {Coullet}}, \bibinfo {author} {\bibfnamefont {C.}~\bibnamefont {Riera}}, \
  and\ \bibinfo {author} {\bibfnamefont {C.}~\bibnamefont {Tresser}},\
  }\bibfield  {title} {\enquote {\bibinfo {title} {{A new approach to data
  storage using localized structures}},}\ }\href {\doibase 10.1063/1.1642311}
  {\bibfield  {journal} {\bibinfo  {journal} {Chaos: An Interdisciplinary
  Journal of Nonlinear Science}\ }\textbf {\bibinfo {volume} {14}},\ \bibinfo
  {pages} {193--198} (\bibinfo {year} {2004})}\BibitemShut {NoStop}%
\bibitem [{\citenamefont {Engelborghs}, \citenamefont {Luzyanina},\ and\
  \citenamefont {Roose}(2002)}]{DDEBT}%
  \BibitemOpen
  \bibfield  {author} {\bibinfo {author} {\bibfnamefont {K.}~\bibnamefont
  {Engelborghs}}, \bibinfo {author} {\bibfnamefont {T.}~\bibnamefont
  {Luzyanina}}, \ and\ \bibinfo {author} {\bibfnamefont {D.}~\bibnamefont
  {Roose}},\ }\bibfield  {title} {\enquote {\bibinfo {title} {Numerical
  bifurcation analysis of delay differential equations using dde-biftool},}\
  }\href {\doibase 10.1145/513001.513002} {\bibfield  {journal} {\bibinfo
  {journal} {ACM Trans. Math. Softw.}\ }\textbf {\bibinfo {volume} {28}},\
  \bibinfo {pages} {1--21} (\bibinfo {year} {2002})}\BibitemShut {NoStop}%
\bibitem [{\citenamefont {Leo}\ \emph {et~al.}(2010{\natexlab{b}})\citenamefont
  {Leo}, \citenamefont {Coen}, \citenamefont {Kockaert}, \citenamefont {Gorza},
  \citenamefont {Emplit},\ and\ \citenamefont {Haelterman}}]{LCK-NAP-10}%
  \BibitemOpen
  \bibfield  {author} {\bibinfo {author} {\bibfnamefont {F.}~\bibnamefont
  {Leo}}, \bibinfo {author} {\bibfnamefont {S.}~\bibnamefont {Coen}}, \bibinfo
  {author} {\bibfnamefont {P.}~\bibnamefont {Kockaert}}, \bibinfo {author}
  {\bibfnamefont {S.}~\bibnamefont {Gorza}}, \bibinfo {author} {\bibfnamefont
  {P.}~\bibnamefont {Emplit}}, \ and\ \bibinfo {author} {\bibfnamefont
  {M.}~\bibnamefont {Haelterman}},\ }\bibfield  {title} {\enquote {\bibinfo
  {title} {Temporal cavity solitons in one-dimensional kerr media as bits in an
  all-optical buffer},}\ }\href {\doibase 10.1038/nphoton.2010.120} {\bibfield
  {journal} {\bibinfo  {journal} {Nat Photon}\ }\textbf {\bibinfo {volume}
  {4}},\ \bibinfo {pages} {471--476} (\bibinfo {year}
  {2010}{\natexlab{b}})}\BibitemShut {NoStop}%
\bibitem [{\citenamefont {Camelin}\ \emph {et~al.}(2016)\citenamefont
  {Camelin}, \citenamefont {Javaloyes}, \citenamefont {Marconi},\ and\
  \citenamefont {Giudici}}]{CJM-PRA-16}%
  \BibitemOpen
  \bibfield  {author} {\bibinfo {author} {\bibfnamefont {P.}~\bibnamefont
  {Camelin}}, \bibinfo {author} {\bibfnamefont {J.}~\bibnamefont {Javaloyes}},
  \bibinfo {author} {\bibfnamefont {M.}~\bibnamefont {Marconi}}, \ and\
  \bibinfo {author} {\bibfnamefont {M.}~\bibnamefont {Giudici}},\ }\bibfield
  {title} {\enquote {\bibinfo {title} {Electrical addressing and temporal
  tweezing of localized pulses in passively-mode-locked semiconductor
  lasers},}\ }\href {\doibase 10.1103/PhysRevA.94.063854} {\bibfield  {journal}
  {\bibinfo  {journal} {Phys. Rev. A}\ }\textbf {\bibinfo {volume} {94}},\
  \bibinfo {pages} {063854} (\bibinfo {year} {2016})}\BibitemShut {NoStop}%
\bibitem [{\citenamefont {Romeira}(2012)}]{Romeira2012}%
  \BibitemOpen
  \bibfield  {author} {\bibinfo {author} {\bibfnamefont {B.}~\bibnamefont
  {Romeira}},\ }\emph {\bibinfo {title} {{Dynamics of resonant tunneling diode
  optoelectronic oscillators}}},\ \href@noop {} {Ph.D. thesis},\ \bibinfo
  {school} {Universidade do Algarve} (\bibinfo {year} {2012})\BibitemShut
  {NoStop}%
\bibitem [{\citenamefont {Romeira}\ \emph
  {et~al.}(2014{\natexlab{c}})\citenamefont {Romeira}, \citenamefont
  {Figueiredo}, \citenamefont {Javaloyes}, \citenamefont {Piro},\ and\
  \citenamefont {Balle}}]{Romeira2014a}%
  \BibitemOpen
  \bibfield  {author} {\bibinfo {author} {\bibfnamefont {B.}~\bibnamefont
  {Romeira}}, \bibinfo {author} {\bibfnamefont {J.}~\bibnamefont {Figueiredo}},
  \bibinfo {author} {\bibfnamefont {J.}~\bibnamefont {Javaloyes}}, \bibinfo
  {author} {\bibfnamefont {O.}~\bibnamefont {Piro}}, \ and\ \bibinfo {author}
  {\bibfnamefont {S.}~\bibnamefont {Balle}},\ }\bibfield  {title} {\enquote
  {\bibinfo {title} {{Mixed mode oscillations in a forced optoelectronic
  circuit for pattern and random bit generation}},}\ }in\ \href {\doibase
  10.1109/NUSOD.2014.6935378} {\emph {\bibinfo {booktitle} {Proceedings of the
  International Conference on Numerical Simulation of Optoelectronic Devices,
  NUSOD}}}\ (\bibinfo {year} {2014})\BibitemShut {NoStop}%
\end{thebibliography}%

%

\end{document}